\documentclass[aps,prd,10pt,tightenlines,notitlepage,showpacs,nofootinbib,superscriptaddress,twocolumn,eqsecnum,raggedbottom,preprintnumbers,longbibliography]{revtex4-1}

\usepackage{bbm,lmodern,amsmath,graphicx,epsfig,amssymb,amsfonts,dsfont,mathtools}
\usepackage{amsmath,amssymb,mathrsfs}
\usepackage{rotating}
\usepackage{bigstrut}
\usepackage{morefloats}
\usepackage{multirow}
\usepackage{longtable}
\usepackage{dcolumn}
\usepackage{placeins}
\usepackage[normalem]{ulem}
\usepackage[dvipsnames]{xcolor}
\usepackage[utf8]{inputenc}
\usepackage{nicefrac} 

\usepackage{hyperref}
\hypersetup{
	colorlinks	=true,
	urlcolor	=blue,
	linkcolor	=blue,
	citecolor	=blue,
	pdftitle	={},
	pdfauthor	={D.~R\"onchen, M.~D\"oring, U.-G.~Mei\ss ner, },
	pdfsubject	={$\mathbf{K\Sigma}$ photoproduction...}
	}

\allowdisplaybreaks[1]

\newcommand{\be}{\begin{equation}}
\newcommand{\ee}{\end{equation}}
\newcommand{\ba}{\begin{eqnarray}}
\newcommand{\ea}{\end{eqnarray}}

\begin{document}

\preprint{JLAB-THY-22-3667}

\title{Light baryon resonances from a coupled-channel study including  $\mathbf{K\Sigma}$ photoproduction}

\author{D.~R\"onchen}
\email{d.roenchen@fz-juelich.de}
\affiliation{Institute for Advanced Simulation and J\"ulich Center for Hadron Physics, Forschungszentrum J\"ulich, 
52425 J\"ulich, Germany}

\author{M.~D\"oring}
\email{doring@gwu.edu}
\affiliation{Institute for Nuclear Studies and Department of Physics, The George Washington University,
Washington, DC 20052, USA}
\affiliation{Thomas Jefferson National Accelerator Facility, 12000 Jefferson Avenue, Newport News,VA, USA}

\author{Ulf-G.~Mei\ss ner}
\email{meissner@hiskp.uni-bonn.de}
\affiliation{Helmholtz-Institut f\"ur Strahlen- und Kernphysik (Theorie) and Bethe Center for Theoretical
Physics,  Universit\"at Bonn, 53115 Bonn, Germany}
\affiliation{Institute for Advanced Simulation and J\"ulich Center for Hadron Physics, Forschungszentrum J\"ulich, 
52425 J\"ulich, Germany}
\affiliation{Tbilisi State University, 0186 Tbilisi, Georgia}

\author{Chao-Wei Shen}
\affiliation{Institute for Advanced Simulation and J\"ulich Center for Hadron Physics, Forschungszentrum J\"ulich, 
52425 J\"ulich, Germany}

\begin{abstract}
The J\"ulich-Bonn dynamical coupled-channel approach is extended to include $K\Sigma$ photoproduction off the proton. Differential cross section and (double) polarization data for $K^+\Sigma^0$ and $K^0\Sigma^+$ are analysed simultaneously with the pion- and photon-induced production of $\pi N$, $\eta N$, $K\Lambda$, and $K\Sigma$ final states, totaling almost {72,000} data points for center-of-mass energies $W<2.4$~GeV. Based on the fit results the spectrum of $N^*$ and $\Delta$ resonances is extracted in terms of pole positions and residues. We discuss the impact of the $\gamma p\to K\Sigma$ channels in detail and investigate the influence of recent polarization data for $\eta p$ photoproduction.
\end{abstract}


\maketitle

\section{Introduction}

The photoproduction of the $K\Sigma$ final states provides ample opportunity to refine our current picture of the light baryon resonance spectrum at medium energies. On the one hand, complementing $K\Lambda$ as a strangeness channel, it holds the promise to reveal states that couple only weakly to $\pi N$ or $\eta N$. Many more states have been predicted at energies beyond 1.7 GeV, e.g. by quark models~\cite{Capstick:1993kb,Ronniger:2011td}, Dyson-Schwinger calculations~\cite{Qin:2019hgk, Eichmann:2016hgl}, or lattice simulations~\cite{Edwards:2012fx,Engel:2013ig,Lang:2016hnn,Andersen:2017una,Silvi:2021uya}, than have been observed by traditional partial-wave analyses~\cite{Hoehler1,Cutkosky:1979fy,Arndt:2006bf} of $\pi N$  scattering. 
Recently, the use of meson and baryon operators in lattice QCD~\cite{Lang:2016hnn} has enabled the extractions of phase shifts encoding baryonic resonance widths~\cite{Silvi:2021uya}. For a recent review,
see Ref.~\cite{Mai:2022eur}.
In contrast to $K\Lambda$, however, $K\Sigma$ is mixed-isospin, and holds valuable information pertaining $\Delta$ resonances. Compared to the  $N^*$ states, fewer $\Delta$ states with a four or three star rating -- meaning the existence of the state is certain or very likely -- are listed by the Particle Data Group~\cite{Workman:2022ynf}.  While in the past experimental information on the $K\Sigma$ channel were limited to pion-induced reactions, with a partially problematic data base~\cite{Ronchen:2012eg,Anisovich:2013vpa}, in the last decades photoproduction experiments like ELSA, JLab, MAMI, SPring-8 and ELPH have contributed significantly to a much better foundation of the light baryon spectrum with a large number of high-quality data sets for several hadronic final states, see, e.g., Refs.~\cite{Ireland:2019uwn, Thiel:2022xtb} for {reviews} or Ref.~\cite{Jude:2019qqd} for prospects of upcoming strangeness production experiments at BGOOD. Among the different photoproduction channels, the $\gamma N\to KY$ reactions are unique candidates for a ``complete experiment"~\cite{Barker:1975bp}, as the self-analyzing weak decay of the hyperons facilitates the determination of recoil polarization observables. A ``complete experiment" describes a set of eight specific observables that are, in principle, sufficient for an unambiguous determination of the pseudoscalar meson photoproduction amplitude, up to an overall phase~\cite{Chiang:1996em,Keaton:1996pe}, see also Refs.~\cite{Sandorfi:2010uv,Ireland:2010bi,Nys:2015kqa}. Such a set always includes recoil polarization measurements {and all observables have to be measured in the same angular and energy range. In contrast, there are also truncated-partial wave complete experiments that allow for measurements at different angles~\cite{Tiator:2017cde}}. In Ref.~\cite{Wunderlich:2020umg} complete sets of polarization observables were also studied for pion-nucleon scattering and electroproduction of pseudoscalar meson, see also Refs.~\cite{Workman:2016irf, Wunderlich:2021xhp}.   

Theoretical approaches to $K\Sigma$ photoproduction are ranging from chiral perturbation theory~\cite{Steininger:1996xw,Mai:2009ce} or chiral unitarized frameworks~\cite{Kaiser:1996js,Borasoy:2007ku} for studying the threshold region, over quark model calculations~\cite{Golli:2016dlj}, to effective Lagrangian approaches~\cite{Maxwell:2015psa, Wei:2022nqp}, isobar models~\cite{Mart:2017mwj, Mart:2019fau, Clymton:2021wof, Luthfiyah:2021yqe, Egorov:2021scy}
including Kaon-MAID~\cite{Lee:1999kd, Tiator:2018pjq},
and Regge-plus-resonance parametrizations~\cite{Corthals:2006nz}. 

Since the threshold energy of the $K\Sigma$ channel is $W\simeq 1686\,$MeV, where many other meson-baryon channels are already open, a more comprehensive picture of the resonance spectrum can be obtained from coupled-channel frameworks. The Bonn-Gatchina multichannel partial-wave analysis comprises a large data base for many different pion- and photon-induced reactions~\cite{Anisovich:2011fc}, see Ref.~\cite{CBELSATAPS:2019ylw} for the recent solution BnGa2019. The Gie\ss en group applied a multi-channel $K$-matrix formalism with a microscopic background to analyse $K\Sigma$ photoproduction in Ref.~\cite{Cao:2013psa}. Dynamical coupled-channel (DCC) models are based on effective Lagrangians and represent an especially capable and theoretically well-founded approach to extract the resonance spectrum in terms of complex poles and residues, as they include or at least approximate theoretical constraints of the $S$-matrix like analyticity, left-hand cuts and complex branch points, and two or three-body unitarity. The ANL/Osaka group analysed Kaon photoproduction in a DCC framework in Refs.~\cite{Kamano:2013iva,Kamano:2016bgm} { with updated results in Ref.~\cite{Kamano:2019gtm} available also at a web site~\cite{anlweb}}. In the present study we apply the J\"ulich-Bonn (J\"uBo) DCC model, an approach that was developed and refined over more than two decades~\cite{Schutz:1994ue,Doring:2009bi,Doring:2009yv,Doring:2010ap,Ronchen:2012eg,Ronchen:2014cna,Ronchen:2015vfa}. In its most recent version the framework was extended to $K\Lambda$ photoproducion~\cite{Ronchen:2018ury}. Here, we add $K\Sigma$ photoproduction off the proton to the {list of analyzed reactions, while the corresponding $K\Sigma$ channel and its pion-induced production were introduced before~\cite{Ronchen:2012eg}. 
We stress that this analysis is currently the only one taking the changed $\Lambda$ decay parameter $\alpha_-$~\cite{BESIII:2018cnd, Ireland:2019uja, BESIII:2022qax} into account that affects a substantial part of polarization data with $KY$ final states.}

Moreover, in Ref.~\cite{Mai:2021vsw} the J\"uBo formalism was extended to virtual photons {(``J\"ulich-Bonn-Washington'' (JBW) approach)}. A first-ever coupled-channel analysis of pion and eta electroproduction was performed in Ref.~\cite{Mai:2021aui}. The present extension of J\"uBo represents a prerequisite for an analysis of $K\Sigma$ electroproduction since the J\"uBo photoproduction amplitude enters the electroproduction potential as a boundary condition at $Q^2=0$. A wealth of electroproduction data for different hadronic final states is already available, including the $K\Sigma$ final state, see, e.g., Ref.~\cite{CLAS:2022yzd}. Even more data are expected soon from experiments with the CLAS12 detector at the 12 GeV upgrade of JLab~\cite{Carman:2019lkk}.

In Refs.~\cite{Shen:2017ayv,Wang:2022oof} the J\"uBo framework was applied in a study of heavy meson -- heavy baryon reactions with hidden charm.

The paper is organized as follows: in Sec.~\ref{sec:formalism} we give a short overview of the theoretical foundations of the formalism. Sec.~\ref{sec:results} contains details pertaining the numerical fit and the fit results for $K\Sigma$ photproduction are presented. In Sec.~\ref{sec:spectrum} we discuss the extracted resonance spectrum. The appendix includes further selected fit results and the $K\Sigma$ multipole amplitudes. 

\section{Formalism}
\label{sec:formalism}

In this section we outline the basic features of the J\"ulich-Bonn DCC approach. For further details the reader is referred to Refs.~\cite{Ronchen:2012eg,Ronchen:2014cna,Ronchen:2015vfa}.

The (hadronic) meson-baryon
interaction is described by the $T$-matrix  $T_{\mu\nu}$ and obeys the Lippmann-Schwinger equation
\begin{multline}
T_{\mu\nu}(q,p',W)=V_{\mu\nu}(q,p',W) \\
+\sum_{\kappa}\int\limits_0^\infty dp\,
 p^2\,V_{\mu\kappa}(q,p,W)G^{}_\kappa(p,W)\,T_{\kappa\nu}(p,p',W) \ ,
\label{eq:LS}
\end{multline}
where $W$ is the scattering energy in the center-of-mass system, $q$ ($p,p'$) is the modulus of the outgoing (intermediate, incoming) momentum and $G^{}_\kappa(p,W)$ the meson-baryon propagator. Eq.~(\ref{eq:LS}) is formulated in partial-wave basis (partial-wave indices suppressed) with a maximum total spin of $J=9/2$. The scattering matrix $T$ with the final and initial channels $\mu$ and $\nu$ enters the calculation of observables that can then be fit to experiment.
The model includes the channels $\kappa = \pi N$, $\eta N$, $K\Lambda$, $K\Sigma$, $\sigma N$,
$\rho N$ and $\pi \Delta$. The latter three channels account effectively for the three-body $\pi\pi N$ channel. They
 are included in our model by fitting the corresponding $\pi\pi$ and $\pi N$ phase shifts~\cite{Schutz:1994ue,Doring:2009yv}{, but the actual  data of the reaction $\pi N\to\pi\pi N$ are not yet included in the analysis}. Recently, the channel space was increased by $\pi N\to \omega N$~\cite{Wang:2022osj}.
 
 In Eq.~(\ref{eq:LS}), $V_{\mu\nu}$ stands for the driving transition amplitude from the initial meson-baryon channel $\nu$ to the final meson-baryon channel $\mu$. This scattering potential is derived from an effective Lagrangian using time-ordered perturbation theory (TOPT) and is iterated in Eq.~(\ref{eq:LS}), which automatically ensures two-body unitarity. Two-to-three and three-to-three body unitarity is approximately fulfilled. For a comprehensive discussion, see Ref.~\cite{Mai:2017vot}. 
 
 The potential $V_{\mu\nu}$ is constructed of $t$- and $u$-channel exchanges of known mesons and baryons, $s$-channel pole terms that account for genuine resonances and phenomenological contact diagrams, which are included to absorb physics beyond the explicit processes.  The role of contact diagrams is discussed in Ref.~\cite{Ronchen:2018ury}. Details on the explicit form of $V_{\mu\nu}$ can be found in Refs.~\cite{Doring:2010ap,Ronchen:2012eg} and in the appendix of Ref.~\cite{Wang:2022osj}.  An important feature of the approach is that the unitarization of Eq.~(\ref{eq:LS}) allows the dynamical generation of poles without  explicit $s$-channel terms. The decomposition into a pole ($s$-channel terms) and non-pole part or {\it background}  ($t$-, $u$-channels and contact terms) versus \emph{dressed reosnances} is, thus, not unique~\cite{Doring:2009bi} and we do not attribute any physical meaning to bare masses or couplings that enter the $s$-channel diagrams. The pole positions and residues of the full amplitude are the only relevant and physically well-defined quantities. 
 
The photon is coupled to the hadronic final-state interaction in the semi-phenomenological approach of Ref.~\cite{Ronchen:2014cna} with the electric or magnetic photoproduction multipole amplitude ${\cal M}$ given by 
\begin{multline}
{\cal M}_{\mu\gamma}(q,W)=V_{\mu\gamma}(q,W) \\+\sum_{\kappa}\int\limits_0^\infty dp\,p^2\,
T_{\mu\kappa}(q,p,W)G^{}_\kappa(p,W)V_{\kappa\gamma}(p,W)\ .
\label{eq:photo}
\end{multline}
The index $\gamma$ denotes the initial $\gamma N$ channel with a real ($Q^2=0$) photon, and $\mu$ ($\kappa$) denotes the final (intermediate) meson-baryon pair. In the present study this channel space is extended to $K\Sigma$ photoproduction, i.e. it now includes also $\mu=K\Sigma$ besides $\pi N$, $\eta N$ and $K\Lambda$. $T_{\mu\kappa}$ is the hadronic half-off-shell matrix of Eq.~(\ref{eq:LS}) with the off-shell momentum $p$ and the on-shell momentum $q$, and $V_{\mu\gamma}$ represents the driving photoproduction amplitude. As in Eq.~(\ref{eq:LS}), $G_\kappa$ denotes the meson-baryon two-body propagator with $\kappa=\pi N$, $\eta N$, $K\Lambda$, $K\Sigma$ and $\pi\Delta$. As we do not yet analyze $\pi\pi N$ photoproduction reactions, we only allow for photoexcitation of the $\pi\Delta$ state out of the three effective $\pi\pi N$ states.
Equation~(\ref{eq:photo}) can be straightforwardly extended to virtual photons, i.e. to electroproduction processes, as done in Refs.~\cite{Mai:2021vsw,Mai:2021aui}.

The driving photoproduction potential $V_{\mu\gamma}$ in Eq.~(\ref{eq:photo}) is written as
\begin{equation}
    V_{\mu\gamma}(p,W)=\alpha^{NP}_{\mu\gamma}(p,W)+\sum_i\frac{\gamma^a_{\mu;i}(p)\gamma_{\gamma;i}(W)}{W-m^b_i}\,,
\label{eq:Vphoto1}
\end{equation}
where $\gamma^a_{\mu\, i}$  denotes the (real) meson-baryon-resonance vertex function and ${\gamma}_{\gamma;i}$ the resonance-photon-baryon vertex function. The summation runs over the number of resonances per partial wave, 
specified by the index $i$.  While the hadronic vertex function $\gamma^a_{\mu;i}$ is precisely the same as that employed in the field-theoretical description of the hadronic reactions in Eq.~(\ref{eq:LS}), the photon vertex ${\gamma}_{\gamma;i}$ is parameterized phenomenologically as a polynomial function in energy $W$ of the system and includes free parameters for each genuine $s$-channel state to be determined by a fit to the data. 
The non-pole part in Eq.~(\ref{eq:Vphoto1}), $\alpha^{NP}_{\mu\gamma}$, is {also parameterized by energy-dependent polynomials with additional fit parameters depending on the partial wave and the hadronic final state.} This polynomial parameterization is of numerical advantage as the evaluation of the amplitude is much faster than in a field-theoretical description as, e.g., in Ref.~\cite{Huang:2011as}. The explicit forms of ${\gamma}_{\gamma;i}$ and $\alpha^{NP}_{\mu\gamma}$ are given in Ref.~\cite{Ronchen:2014cna}.


\section{Results}
\label{sec:results}


\subsection{Data base}
\label{sec:data}

The data analyzed in the present study are listed in Tab.~\ref{tab:data}. {A detailed discussion of the pion-induced data that are included in the fits can be found in Ref.~\cite{Ronchen:2012eg}. In case of the elastic $\pi N$ channel we do not fit to actual data but to the partial-wave amplitudes of the GWU/SAID WI08 analysis~\cite{Workman:2012hx}. We include the energy-dependent solution in steps of 5 MeV from $\pi N$ threshold up to $W=2400$~MeV. However, higher partial waves of the WI08 solution are essentially zero at lower energies and we adapt the fitted energy range accordingly. Taking into account that the amplitude consists of a real and an imaginary part, the number of fitted points for the elastic $\pi N$ channel amounts to the number quoted in Tab.~\ref{tab:data}.  } 

A substantial part of the {photoproduction} data sets was obtained from the GWU/SAID~\cite{SAID} and BnGa webpages~\cite{BnGa_web}. A full list of references to data sets for  reactions other than $\gamma p\to K\Sigma$ is provided in supplementary material online~\cite{Juelichmodel:online}. {Note that first measurements of  $K^+\Sigma^-$ and $K^0\Sigma^0$ final states produced on quasifree neutrons begin to emerge~\cite{CLAS:2021hex, BGOOD:2021oxp, A2:2018doh} that will be analyzed in the future.} 

In total, the data base for $K^+\Sigma^0$ photoproduction comprises more than 10 times the number of data points of $K^0\Sigma^+$. An increase in the amount of data and in the availability of different polarization observables for the latter channel is urgently needed to disentangle the isospin content in the $K\Sigma$ channel. Especially since the quantity and quality of the data base for pion-induced $K\Sigma$ production is not comparable to the photoproduction measurements, see Refs.~\cite{Doring:2010ap, Ronchen:2012eg} for a representation of the former. 

For $\gamma p\to K^+\Sigma^0$, we do not fit the SAPHIR differential cross section data from Refs.~\cite{Bockhorst:1994jf,SAPHIR:1998fev,Glander:2003jw} due to inconsistencies at backward angles with more recent data sets. See the discussion in Ref.~\cite{CLAS:2010aen}, where those inconsistencies are attributed to possible overall normalization issues. Data for the recoil polarization $P$ from SAPHIR are, however, included in our fit, except for Ref.~\cite{SAPHIR:1998fev}.
Moreover, differential cross sections at forward angles of older data sets between $W=1680$~MeV and $2000$~MeV are not taken into account. To avoid inconsistencies, we fit instead the very recent measurement from the BGOOD experiment at ELSA~\cite{Jude:2020byj}. The BGOOD data on $d\sigma/d\Omega$ and $P$ at forward angles for $K^+\Lambda$ photoproduction~\cite{Alef:2020yul} are also included. Compared to the J\"uBo2017 analysis~\cite{Ronchen:2018ury}, a number of recent data sets were additionally included in the fit, such as new data in $\eta p$ photoproduction for the beam asymmetry $\Sigma$~\cite{CBELSATAPS:2020cwk} and $T$, $E$, $P$, $H$, and $G$~\cite{CBELSATAPS:2019ylw} by the CBELSA/TAPS Collaboration. 

The BES-III Collaboration recently reported on a new value for the weak decay parameter $\alpha_-$ of the $\Lambda$ baryon~\cite{BESIII:2018cnd,BESIII:2022qax}. This parameter is a crucial quantity in the extraction process of polarization observables for $K^+\Lambda$, but also for $K^+\Sigma^0$ photoproduction, since the decay chain $K^+\Sigma^0\to K^+\gamma\Lambda\to K^+\gamma p\pi^-$ is utilized in the experiments. That the new value of $\alpha_-$ is significantly larger than the previous PDG value of $\alpha_-=0.642(13)$, was confirmed in a re-analsyis of CLAS $K^+\Lambda$ photoproduction data in Ref.~\cite{Ireland:2019uja} where a value of $\alpha_-=0.721(6)(5)$ was obtained. This implies that all polarization observables affected by $\alpha_-$, i.e. $P$, $T$, $C_{x,z}$ and $O_{x,z}$, are actually about 17~\% smaller. Accordingly, we scale all affected polarization data in $\gamma p\to K^+\Lambda$, $K^+\Sigma^0$, but also in $\pi^-p\to K^0\Lambda$, $K^0\Sigma^0$ by a factor of 0.642/0.721. The present analysis is currently the only one taking this change into account.
Analytic expressions of the observables in terms of partial waves, CGNL amplitudes and multipoles are given in Refs.~\cite{Ronchen:2014cna,Ronchen:2018ury}.

\begin{table*}
\caption{Data included in the fit. A full list of references to the different experimental publications can be found online~\cite{Juelichmodel:online}.}
\begin{center}
\renewcommand{\arraystretch}{1.9}
\begin {tabular}{l|l|r} 
\hline\hline
Reaction & Observables ($\#$ data points) & $\#$ data p./channel \\ \hline
$\pi N\to\pi N$ & {PWA  GW-SAID WI08 \cite{Workman:2012hx} (ED solution) } & {8,396}\\
$\pi^-p\to\eta n$ &{$d\sigma/d\Omega$ (676), $P$ (79) }
& 755\\
$\pi^-p\to K^0 \Lambda$ &{$d\sigma/d\Omega$ (814), $P$ (472), $\beta$ (72) }
& 1,358\\
$\pi^-p\to K^0 \Sigma^0$ &{$d\sigma/d\Omega$ (470), $P$ (120) }
& 590\\
$\pi^-p\to K^+ \Sigma^-$ &{$d\sigma/d\Omega$ (150)}
& 150\\
$\pi^+p\to K^+ \Sigma^+$ &{$d\sigma/d\Omega$ (1124), $P$ (551) , $\beta$ (7)}
& 1,682\\ 
 \hline
$\gamma p\to \pi^0p$ & $d\sigma/d\Omega$ (18721), $\Sigma$ (3287), $P$ (768), $T$ (1404), $\Delta\sigma_{31}$ (140),& \\
& $G$ (393), $H$ (225), $E$ (1227), $F$ (397), $C_{x^\prime_\text {L}}$ (74), $C_{z^\prime_\text{L}}$ (26)
&26,662 \\
$\gamma p\to \pi^+n$ & $d\sigma/d\Omega$ (5670), $\Sigma$ (1456), $P$ (265), $T$ (718), $\Delta\sigma_{31}$ (231), &\\
& $G$ (86), $H$ (128), $E$ (903)
& 9,457 \\
$\gamma p\to \eta p$ &$d\sigma/d\Omega$ (9112), $\Sigma$ (535), $P$ (63), $T$ (291), $F$ (144), $E$ (306), $G$ (47), $H$ (56)
& 10,554\\
$\gamma p\to K^+\Lambda$ & 
$d\sigma/d\Omega$ (2563), 
$P$ (1663), 
$\Sigma$ (459), 
$T$ (383), 
 &\\ &
$C_{x^\prime}$ (121), 
$C_{z^\prime}$ (123), 
$O_{x^\prime}$ (66), 
$O_{z^\prime}$ (66), 
$O_x$ (314), 
$O_z$ (314), 
&6,072 \\ 
$\gamma p\to K^+\Sigma^0$ & {\footnotesize $d\sigma/d\Omega$ (4381)~\cite{Jude:2020byj,Brody:1960zz,Anderson:1962za,Fujii:1970gn,Bleckmann:1970kb,Feller:1972ph,CLAS:2003zrd,CLAS:2005lui,LEPS:2005hji,Kohri:2006yx,CLAS:2010aen,CrystalBallatMAMI:2013iig}, $P$ (402)~\cite{Bockhorst:1994jf,CLAS:2010aen,Glander:2003jw,CLAS:2003zrd,LEPS:2005hji,Kohri:2006yx,Lleres:2007tx}, $\Sigma$ (280)~\cite{Zegers:2003ux,Lleres:2007tx,Paterson:2016vmc} } & \\
& \footnotesize{$T$ (127)~\cite{Paterson:2016vmc}, $C_{x'}$ (94)~\cite{Bradford:2006ba}, $C_{z'}$ (94)~\cite{Bradford:2006ba}, $O_{x}$ (127)~\cite{Paterson:2016vmc}, $O_{z}$ (127)~\cite{Paterson:2016vmc}  } & 5,632 \\ 
$\gamma p\to K^0\Sigma^+$ & {\footnotesize $d\sigma/d\Omega$ (281)~\cite{Carnahan:2003mk,CBELSATAPS:2007oqn,CBELSATAPS:2011gly,A2:2013cqk,A2:2018doh}, $P$ (167)~\cite{SAPHIR:1999wfu,A2:2013cqk,CBELSATAPS:2007oqn,CLAS:2013owj} } & 448 \\ \hline
& \multicolumn{1}{r}{in total} & {71,756}  
\\
\hline\hline
\end {tabular}
\end{center}
\label{tab:data}
\end{table*}


\subsection{Numerical details and fit parameters}
\label{sec:numerics}

The free parameters of the approach are adjusted to the experimental data in a $\chi^2$ minimization using MINUIT on the supercomputer JURECA-DC at the J\"ulich Supercomputing Center~\cite{Thornig2021-hh}. In the current study we consider partial waves of a total spin up to $J=9/2$ and  the following types of parameters were fitted:

    \textbf{$\bullet$ $s$-channel or ``pole parameters":} one bare mass and several coupling constants to the different hadronic channels ($\pi N$, $\rho N$, $\eta N$, $\pi\Delta$, $K\Lambda$, $K\Sigma$) as allowed by isospin for each of the 12 genuine isospin $I=1/2$ poles and the 10 $I=3/2$ poles amount to 134 fit parameters. Note that the nucleon is included in the J\"uBo approach as an $s$-channel state in the $P_{11}$ partial wave. Its bare mass and coupling to the $\pi N$ channel are renormalized as described in Ref.~\cite{Ronchen:2015vfa} to match the physical values of $W=m_N=938$~MeV and $f_{\pi NN}=0.964$~\cite{Baru:2011bw}. The cut-off parameter in the form factor of the nucleon $s$-channel diagram is treated as a free parameter.
    
    \textbf{$\bullet$ phenomenological contact term parameters:} we include one contact term in each partial wave with couplings to $\pi N$, $\eta N$, $K\Lambda$ and $K\Sigma$. In case of the $P_{13}$ partial wave, the contact term also couples to the $\pi\Delta$ channel. This amounts to 61 fit parameters.
    
    \textbf{$\bullet$ $t$-, $u$-channel or ``background parameters":} while almost all coupling constants for these diagrams are fixed from SU(3) flavour symmetry, the cut-off values in the corresponding  form factors are treated as free parameters of the model. Our framework comprises 68 fit parameters of this type. The numerical evaluation of the background terms is much more time-consuming than for the other building blocks of the amplitude and we therefore refrained from an in-depth re-fit of the parameters connected to them. However, due to a recent change in sign convention for a smaller number of exchange diagrams~\cite{Wang:2022osj}, the background parameters were also slightly re-adjusted. 
    
    \textbf{$\bullet$ ``photo parameters":} the parameters directly related to the photoproduction kernel are the couplings in the energy-dependent polynomials that are used to parameterize $V_{\mu\gamma}$ as described below Eq.~(\ref{eq:Vphoto1}). In the current study, polynomials of order 4 or less are suffiecent to achieve a good description of the data. We have 764 fit parameters of this type. Note that the photoproduction amplitude also depends on the ``pole", contact term, and ``background" parameters via the hadronic final state interaction in Eqs.~(\ref{eq:photo}) and (\ref{eq:Vphoto1}).
    
    The numerical evaluation of a model as complex as the present one is not straightforward and while all data are fitted simultaneously, not all parameters can be adjusted in the same run. Moreover, not all free parameters are indispensable from a physical point of view. This applies especially to the ``photo parameters" in the polynomial parameterization. But also the number of genuine $s$-channel states is not predetermined by symmetry constraints or the like.
    A systematic study regarding those points using model selection tools such as the LASSO method to reduce the number of parameters~\cite{Landay:2016cjw} and/or minimize the resonance content~\cite{Landay:2018wgf} are planned for the future. On the other hand, the large number of fit parameters connected to the polynomial parameterization can also be regarded as an advantage since it prevents the inclusion of superfluous $s$-channel states to improve the fit result.
  
If provided, experimental systematic errors of more recent measurements are taken into account as angle-independent normalization factors, following the SAID approach~(see, e.g., Ref.~\cite{Doring:2016snk}), 
i.e. they are not added in quadrature to the statistical ones because systematic errors cannot necessarily be considered as Gaussian and induce correlations between data. However, for older data sets we add a general 5~\% uncertainty to the statistical ones to account for unspecified systematic errors.  
  
The amount of available data points varies significantly for the different reaction channels and observables, c.f. Tab.~\ref{tab:data}. We therefore apply weighting factors to the different data sets in the $\chi^2$ minimization in order to achieve a good description also of those sets the fit would otherwise ignore due to the limited number of data points. Such a weighting procedure is standard in the field for this type of analyses~\cite{Arndt:1995bj, Anisovich:2011fc, Briscoe:2020qat, CLAS:2015ykk} and inevitable in situations as in $K\Sigma$ photoproduction where much more data are available for one of the two possible isospin channels. The weights applied for $\gamma p\to K^+\Sigma^0$, $K^0\Sigma^+$ in the present study can be found in Tab.~\ref{tab:weightskps0} in Appendix~\ref{app:weights}. {Note that due to the strong variations in the number of available data points, the values of those weights can be very different.  }

\subsection{Uncertainty estimation}

A rigorous statistical error analysis of the extracted baryon spectrum  would include the study of the propagation of statistical and systematic uncertainties from experimental data to the resonance parameters, as well as an application of model selection tools. For the elastic $\pi N$ channel, it would also require the transition from fitting mere partial-wave amplitudes to including the covariance matrices and performing a correlated $\chi^2$ fit~\cite{Doring:2016snk}. Such an analysis is beyond the scope of the present study and has, so far, never been fully carried out in any analysis efforts in the field.

Instead, we estimate the uncertainties of the resonance parameters from re-fits with modified model parameterization, where the difference lies in the number of $s$-channel states. For each partial wave with only one $s$-channel resonance, a second genuine state is included and an extensive re-fit of all ``pole", contact term, and ``photo parameters" is performed. This amounts to 16 re-fits. The uncertainties for the resonance parameters quoted in Tables~\ref{tab:poles1} to \ref{tab:photo} denote the maximal deviation of the original resonance values from the ones extracted from the re-fits. In none of the re-fits the data description is significantly improved compared to the original fit. We thus conclude that {there is no need of adding additional $s$-channel states in the considered partial waves }.

This procedure represents a qualitative estimation of uncertainties from statistical and systematical sources and allows to determine the relative size of uncertainty among the different resonances. The absolute size of the errors, however, is not well determined. We still regard this as a valuable compromise since a rigorous error analysis is not feasible in the current work. {In this  context, the electroproduction fits of Refs.~\cite{Mai:2021vsw, Mai:2021aui} were carried out with up to eight radically different fit strategies and parameter starting points, leading to a better exploration of different local minima in parameter space. Our current effort to estimate uncertainties goes in the same direction, but the data base in this study is much larger and more heterogeneous. In any case, it is clear that systematic uncertainties and amplitude ambiguities dominate the error bars of any extracted quantity.}

\subsection{Fit results}

\begin{figure}
\begin{center}
\includegraphics[width=1.\linewidth]{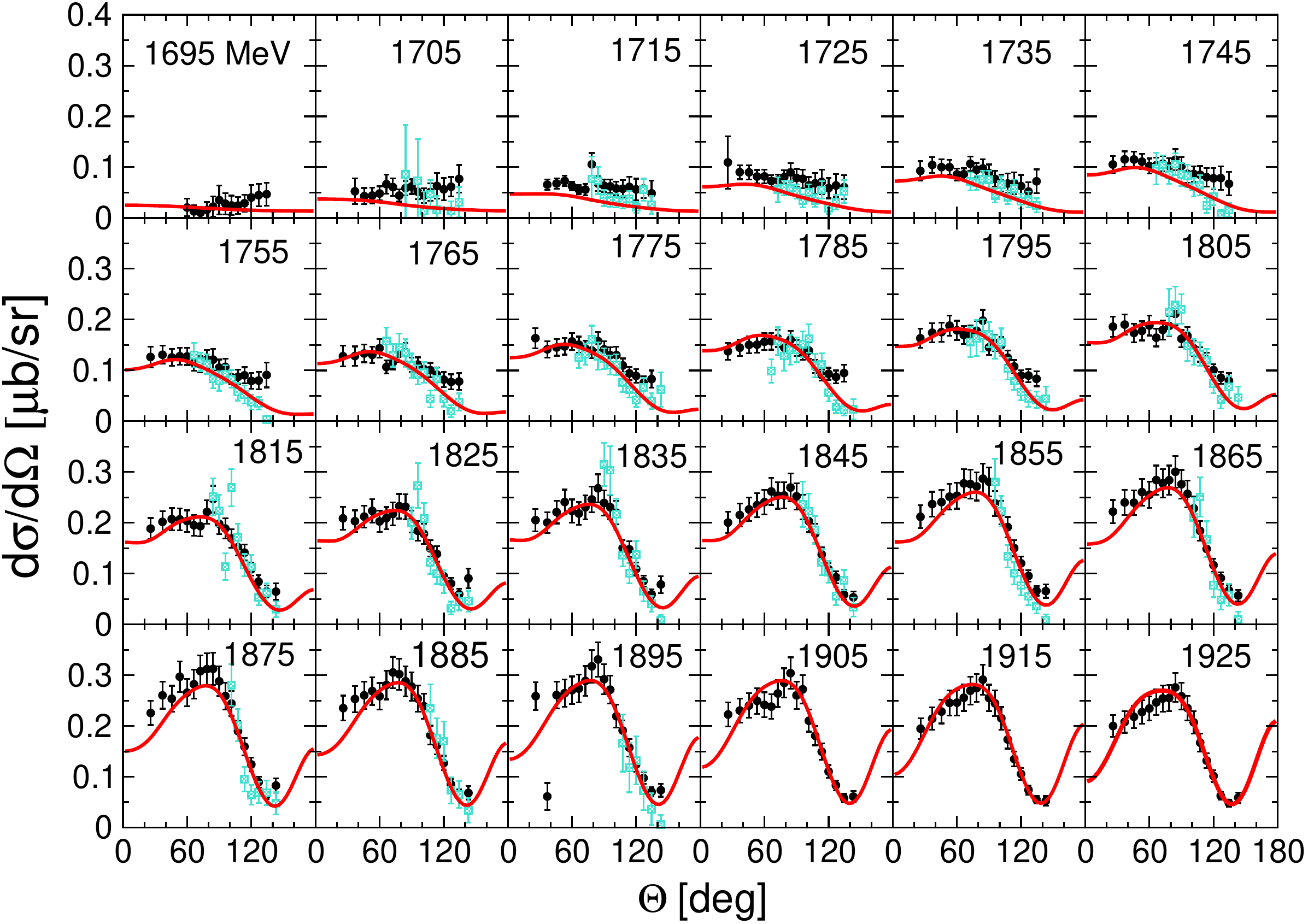} 
\includegraphics[width=1\linewidth]{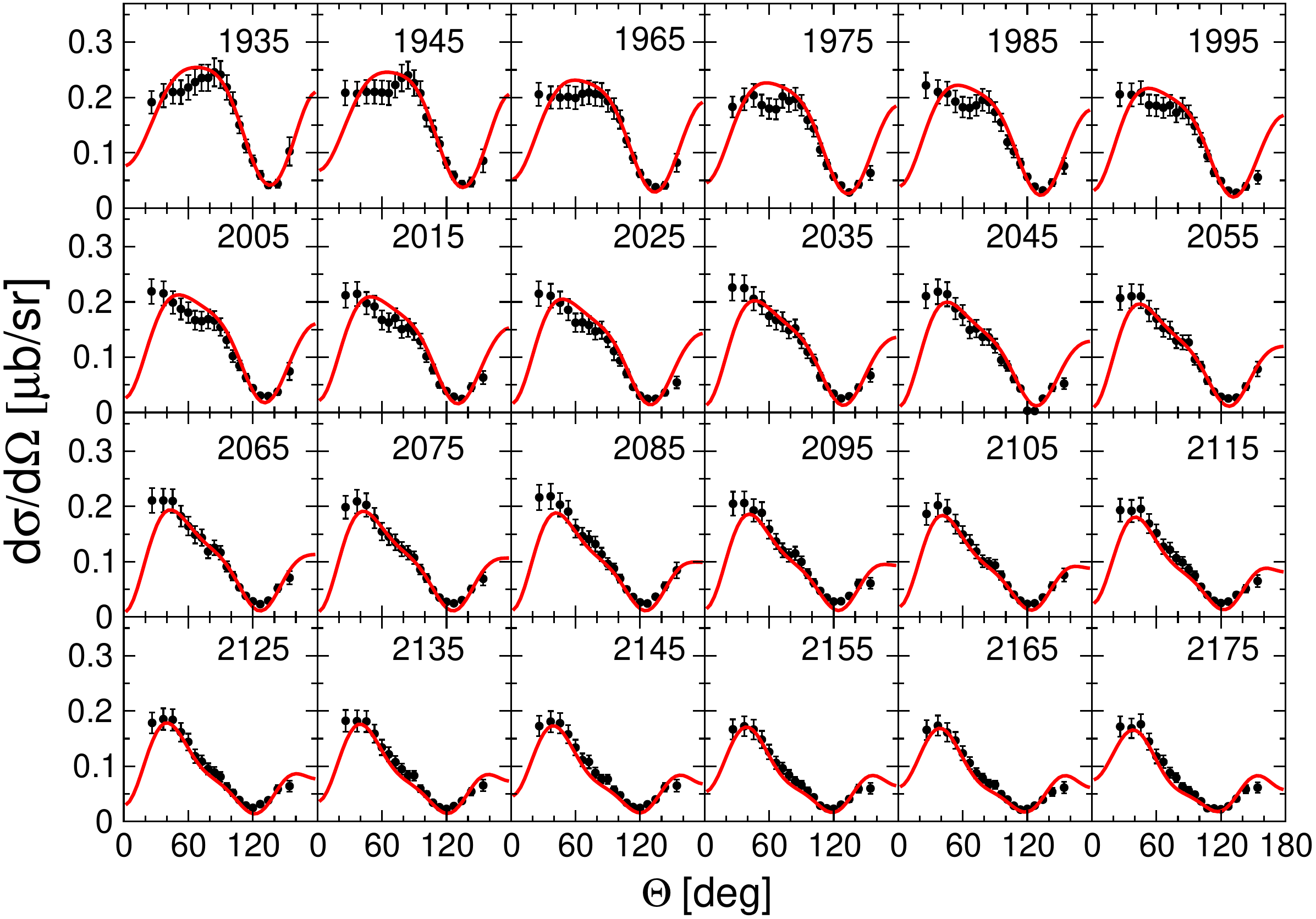} 
\includegraphics[width=1\linewidth]{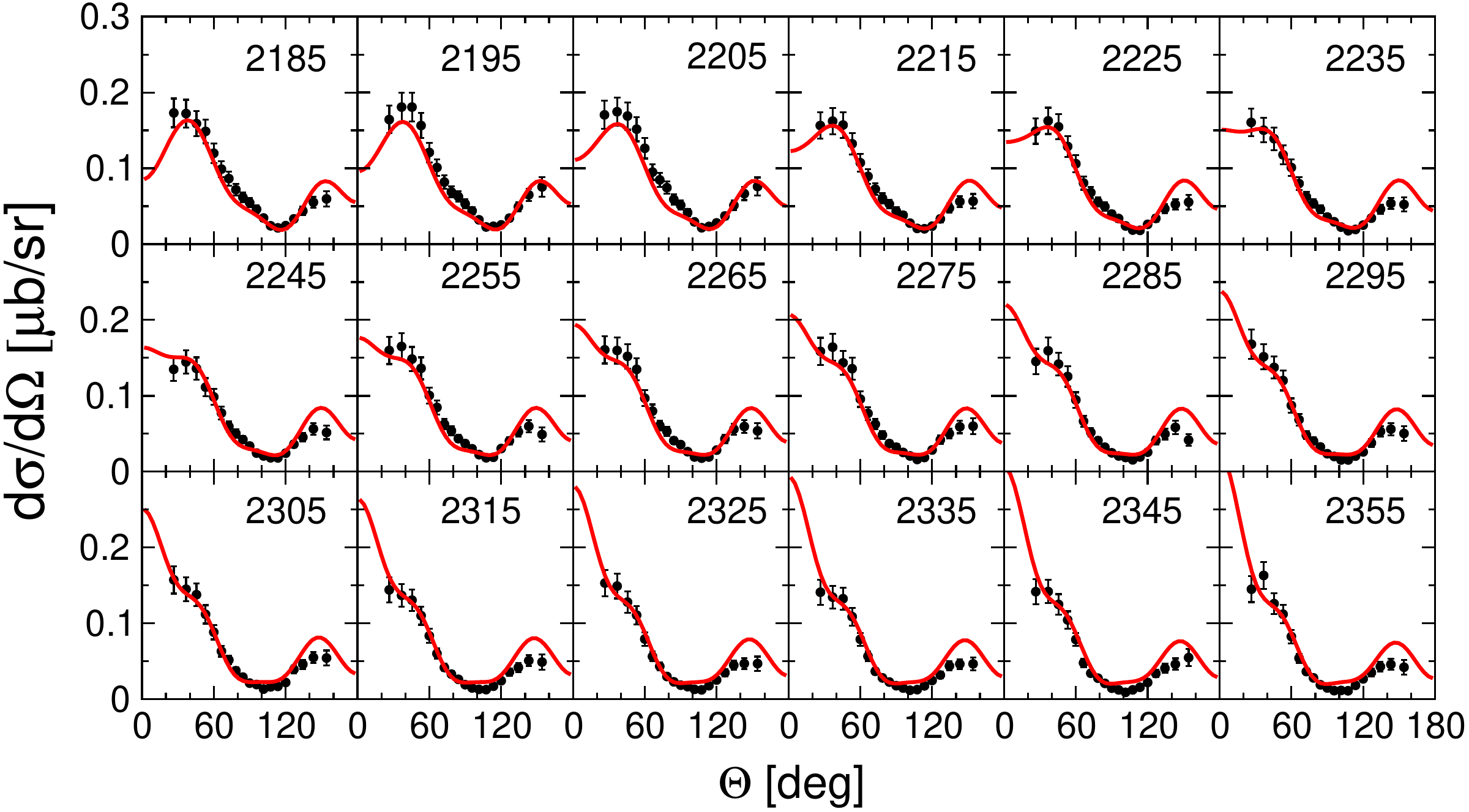} 
\end{center}
\caption{Selected fit results (solid (red) lines) for the differential cross section of the reaction $\gamma p\to K^+\Sigma^0$. Data: (black) circles: CLAS (Dey {\it et al.}~\cite{CLAS:2010aen}); (cyan) squares: MAMI (Jude {\it et al.}~\cite{CrystalBallatMAMI:2013iig}). The numbers in the individual panels in this and all subsequent figures denote the scattering energy $W$ in the center-of-mass system in MeV.}
\label{fig:dsdok+s0}
\end{figure}

\begin{figure}
\begin{center}
\includegraphics[width=1.\linewidth]{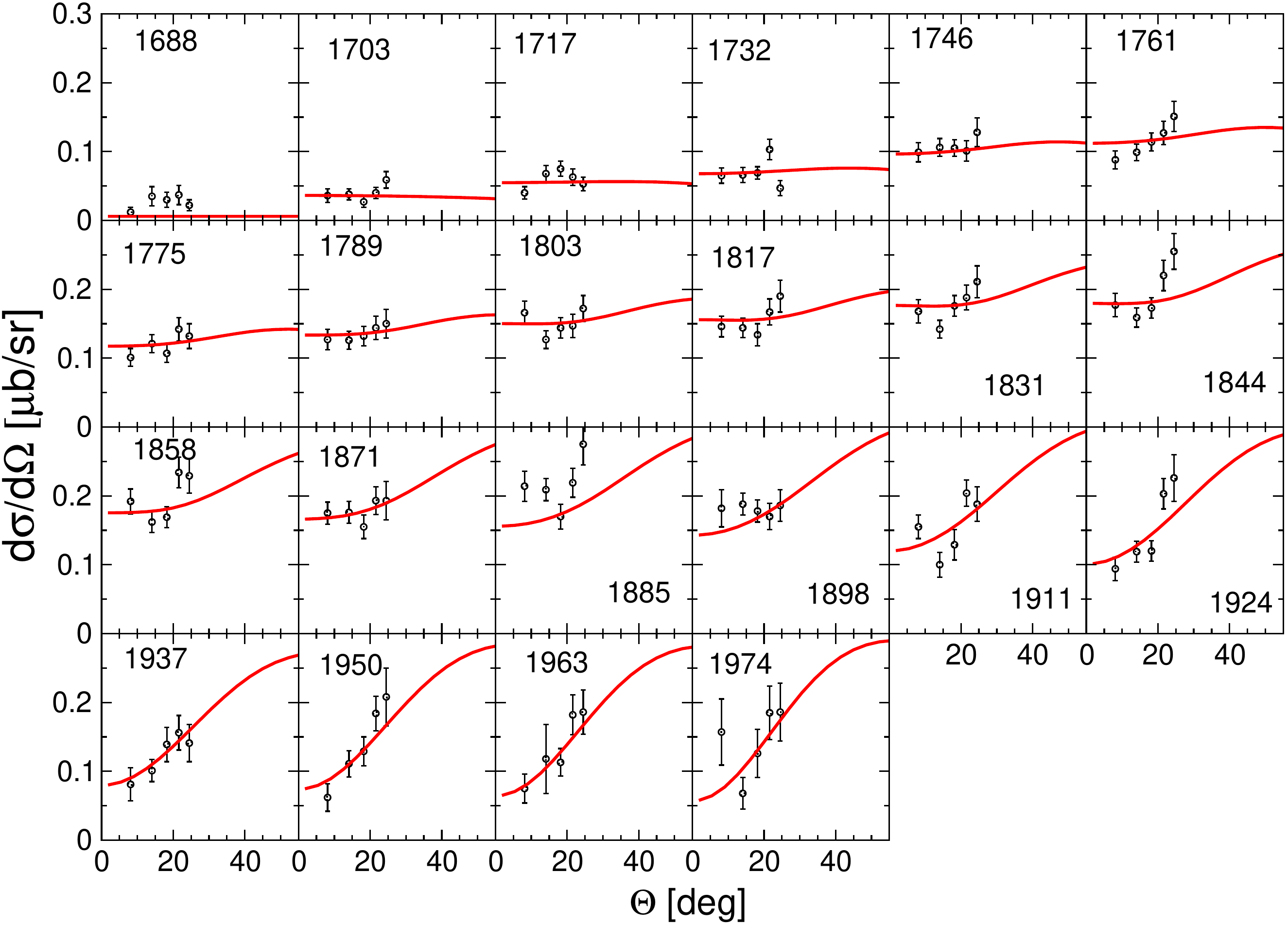} 
\end{center}
\caption{Fit results (solid (red) lines) for the differential cross section in forward direction of the reaction $\gamma p\to K^+\Sigma^0$. Data: BGOOD (Jude {\it et al.}~\cite{Jude:2020byj}).}
\label{fig:dsdok+s0BGOOD}
\end{figure}

\begin{figure}
\begin{center}
\includegraphics[width=1.\linewidth]{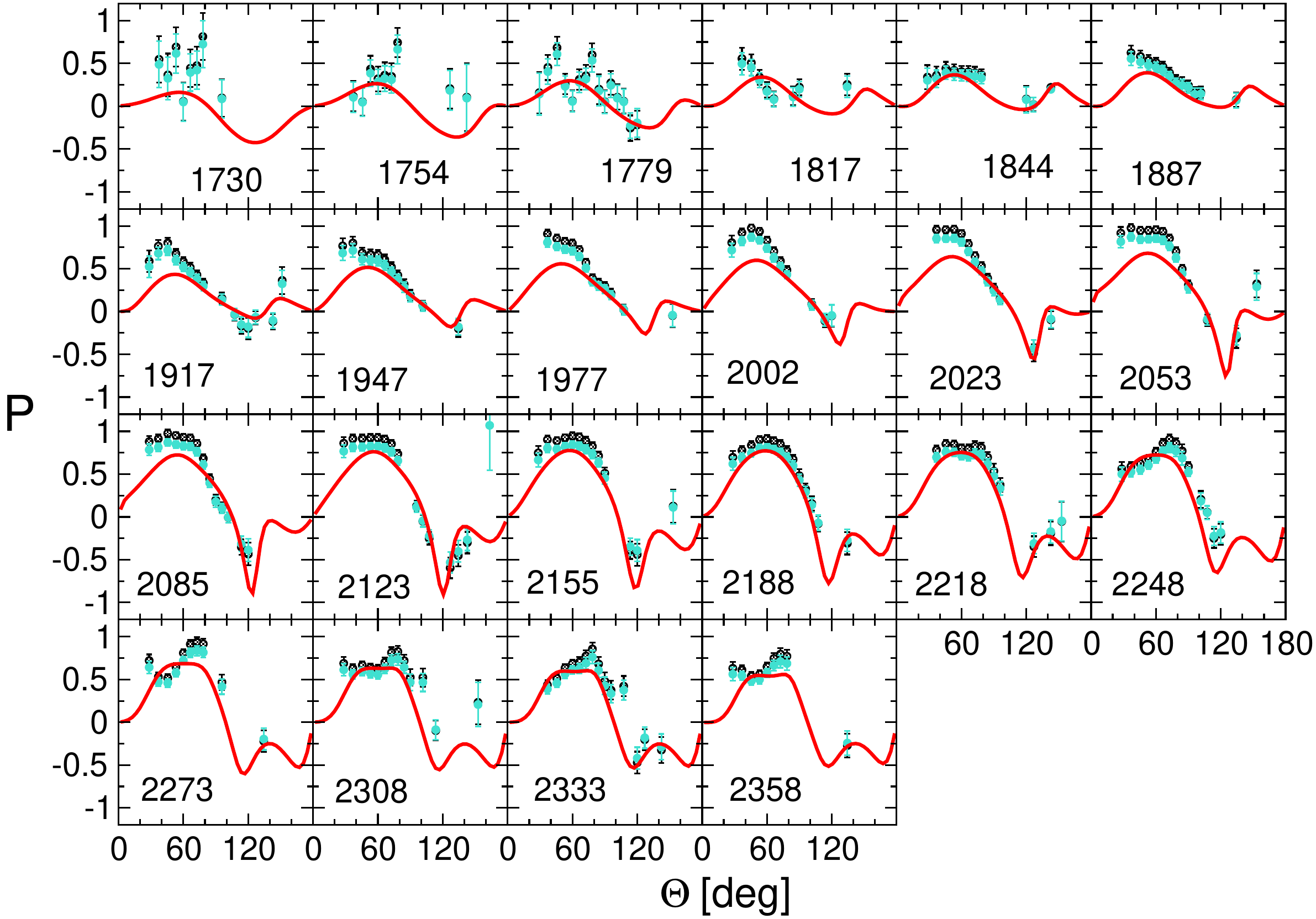} 
\end{center}
\caption{Selected fit results (solid (red) lines) for the recoil polarization of the reaction $\gamma p\to K^+\Sigma^0$. Data: (black) circles: CLAS (Dey {\it et al.}~\cite{CLAS:2010aen}); (turquoise) circles: same data but scaled by the new value of the $\Lambda$ decay parameter $\alpha_-$~\cite{Ireland:2019uja}. Data are shown accumulated in energy bins of up to 30 MeV.}
\label{fig:polak+s0}
\end{figure}

\begin{figure}
\begin{center}
\includegraphics[width=1.\linewidth]{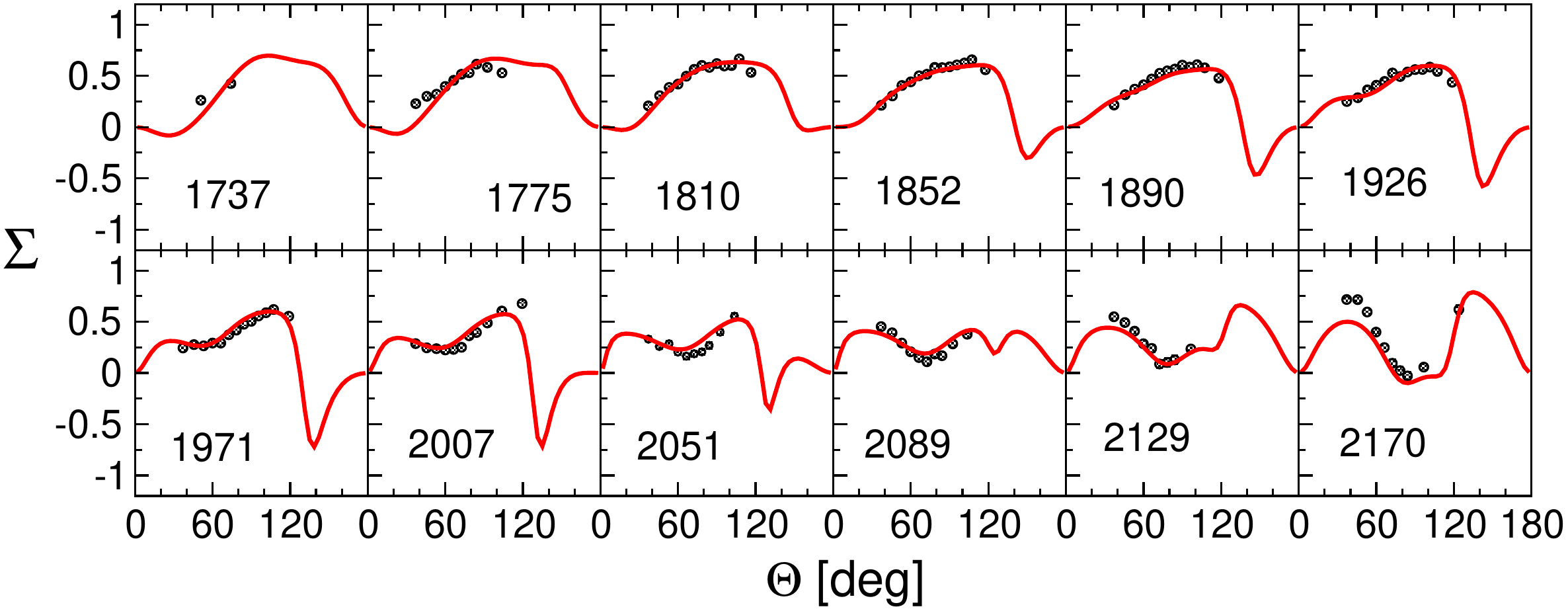} 
\includegraphics[width=1.\linewidth]{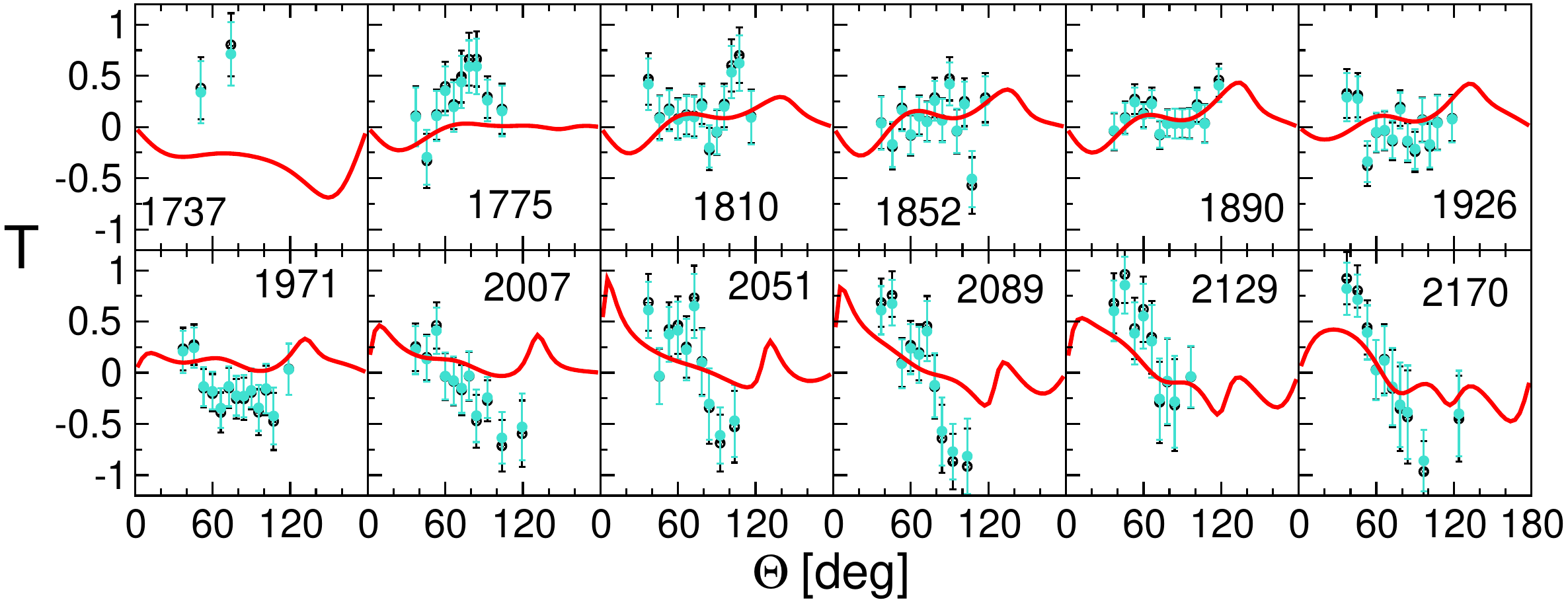} 
\end{center}
\caption{Fit results (solid (red) lines) for the beam asymmetry $\Sigma$ and the target asymmetry $T$ of the reaction $\gamma p\to K^+\Sigma^0$. Data: (black) circles: CLAS (Paterson {\it et al.}~\cite{Paterson:2016vmc}); (turquoise) circles: same data but scaled by the new value of the $\Lambda$ decay parameter $\alpha_-$~\cite{Ireland:2019uja}. }
\label{fig:STk+s0}
\end{figure}

\begin{figure}
\begin{center}
\includegraphics[width=1.\linewidth]{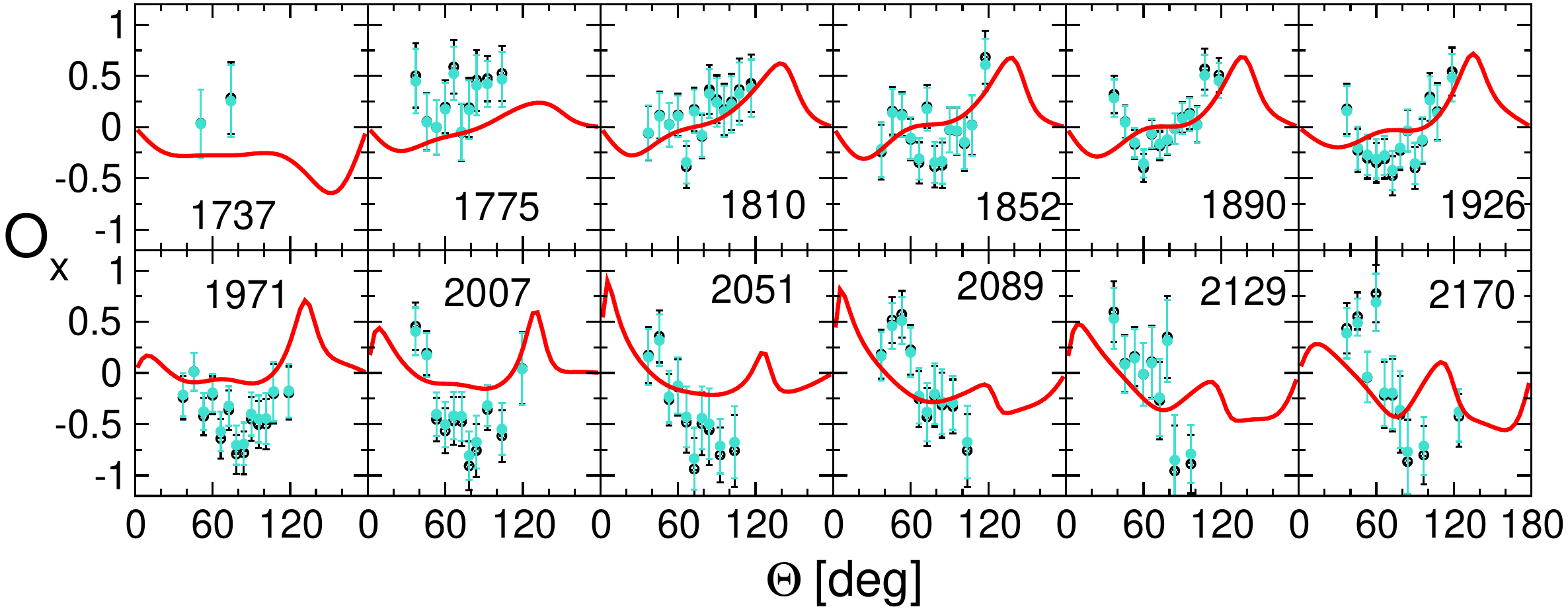} 
\includegraphics[width=1.\linewidth]{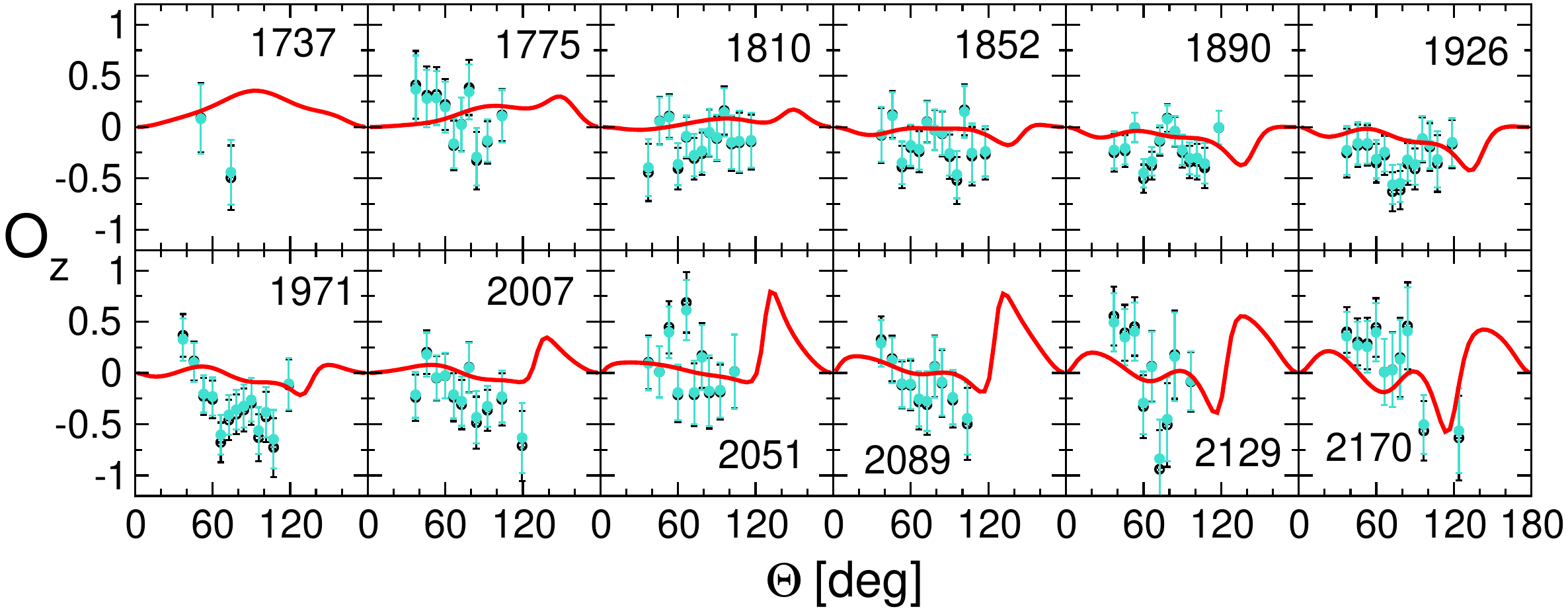} 
\end{center}
\caption{Fit results (solid (red) lines) for the beam recoil polarizations $O_x$ and $O_z$ of the reaction $\gamma p\to K^+\Sigma^0$. Data: (black) circles: CLAS (Paterson {\it et al.}~\cite{Paterson:2016vmc}); (turquoise) circles: same data but scaled by the new value of the $\Lambda$ decay parameter $\alpha_-$~\cite{Ireland:2019uja}. }
\label{fig:OxOzk+s0}
\end{figure}

\begin{figure}
\begin{center}
\includegraphics[width=1.\linewidth]{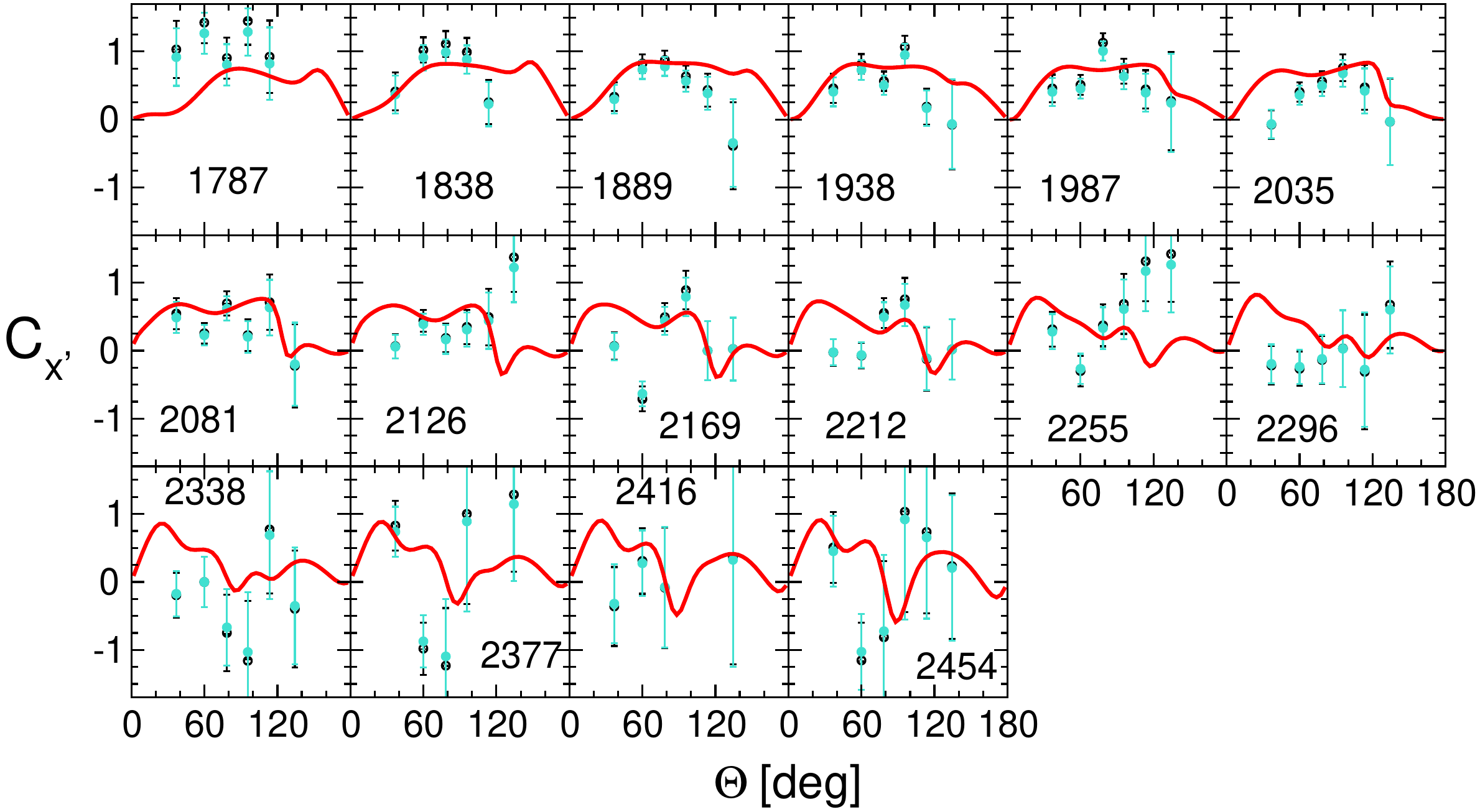} 
\includegraphics[width=1.\linewidth]{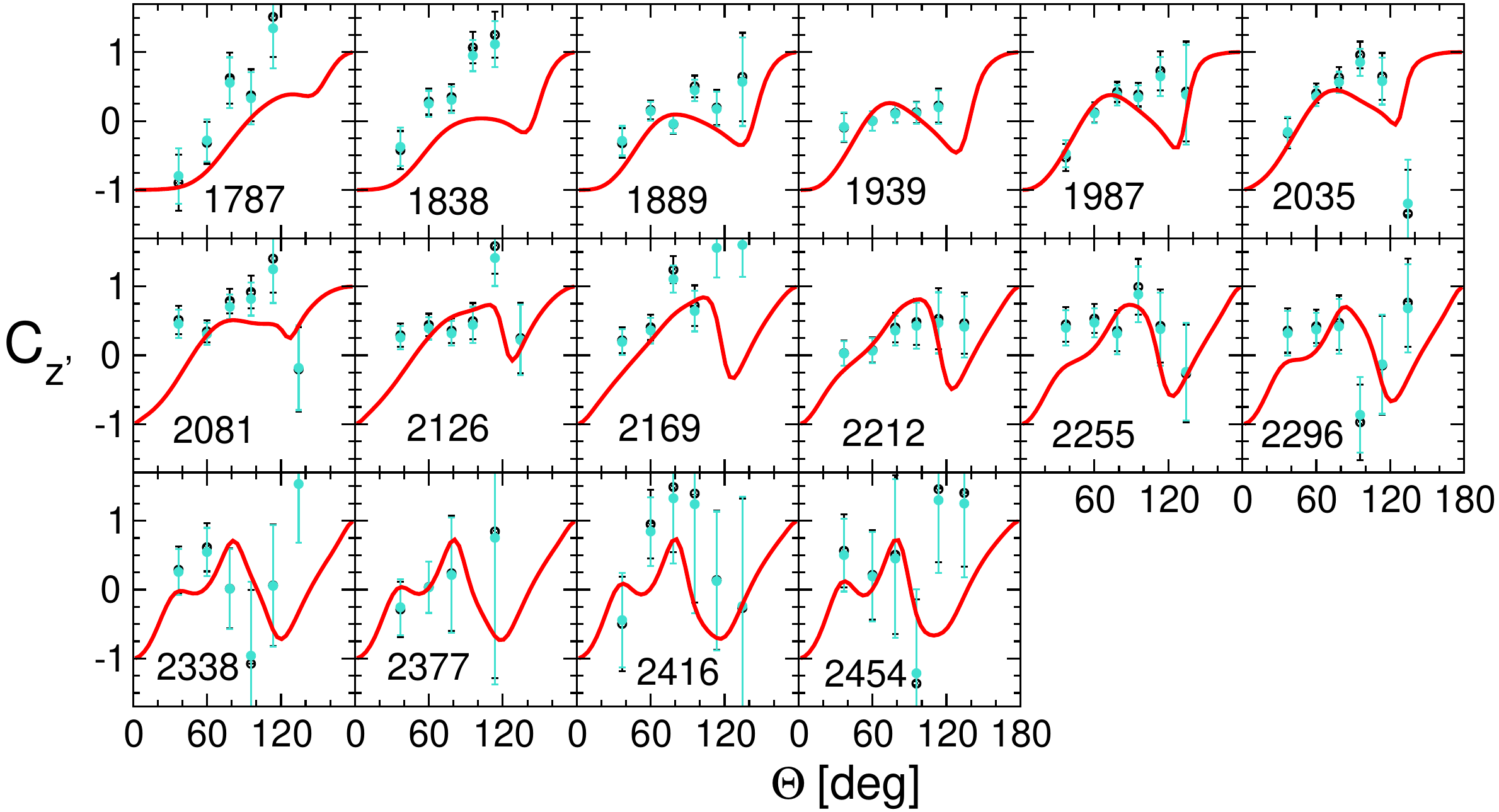} 
\end{center}
\caption{Fit results (solid (red) lines) for the beam recoil polarizations $C_{x^\prime}$ and $C_{z^\prime}$ of the reaction $\gamma p\to K^+\Sigma^0$. Data: (black) circles: CLAS (Bradford {\it et al.}~\cite{Bradford:2006ba}); (turquoise) circles: same data but scaled by the new value of the $\Lambda$ decay parameter $\alpha_-$~\cite{Ireland:2019uja}. }
\label{fig:CxCzk+s0}
\end{figure}

\begin{figure}
\begin{center}
\includegraphics[width=1.\linewidth]{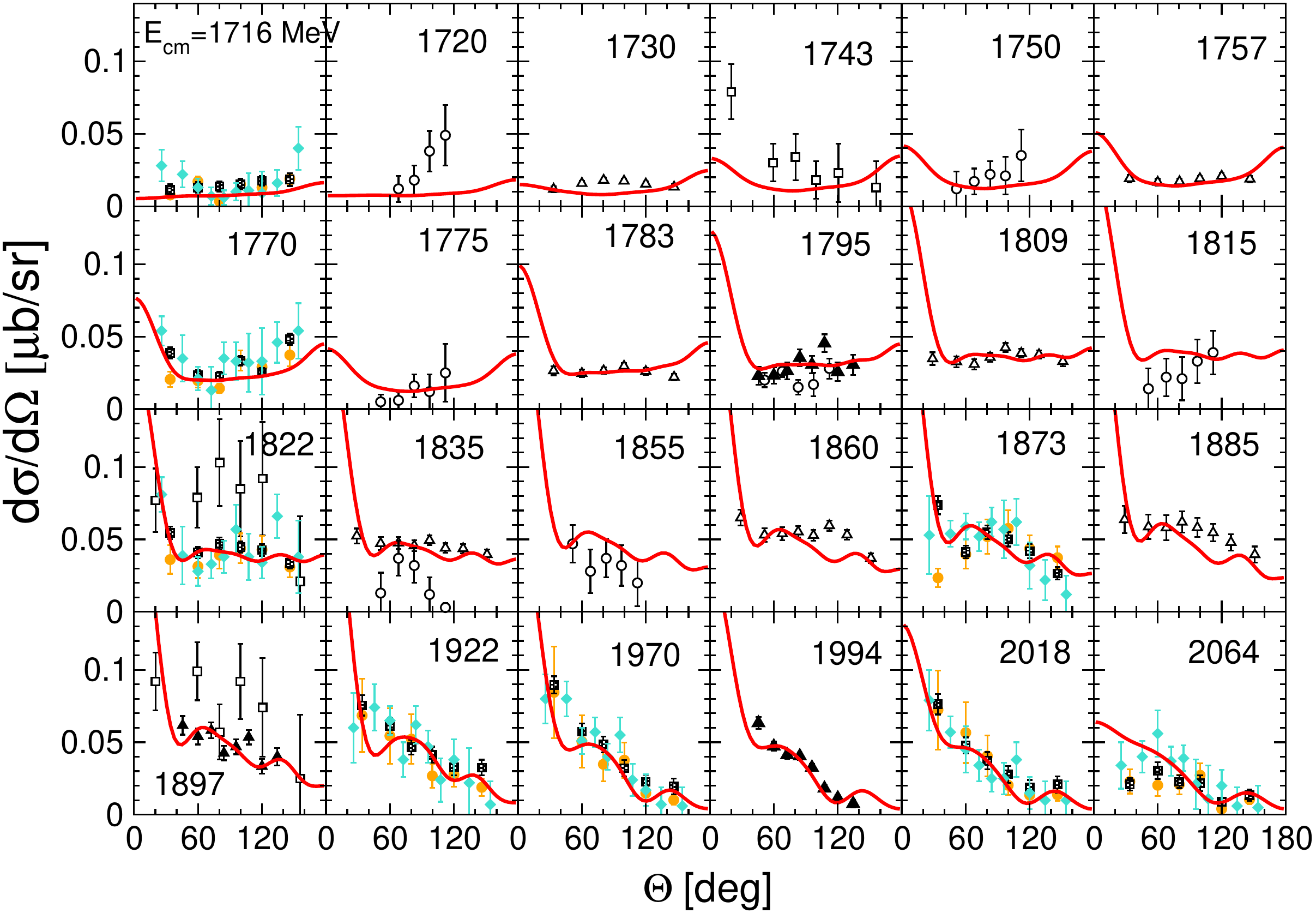} 
\includegraphics[width=1.\linewidth]{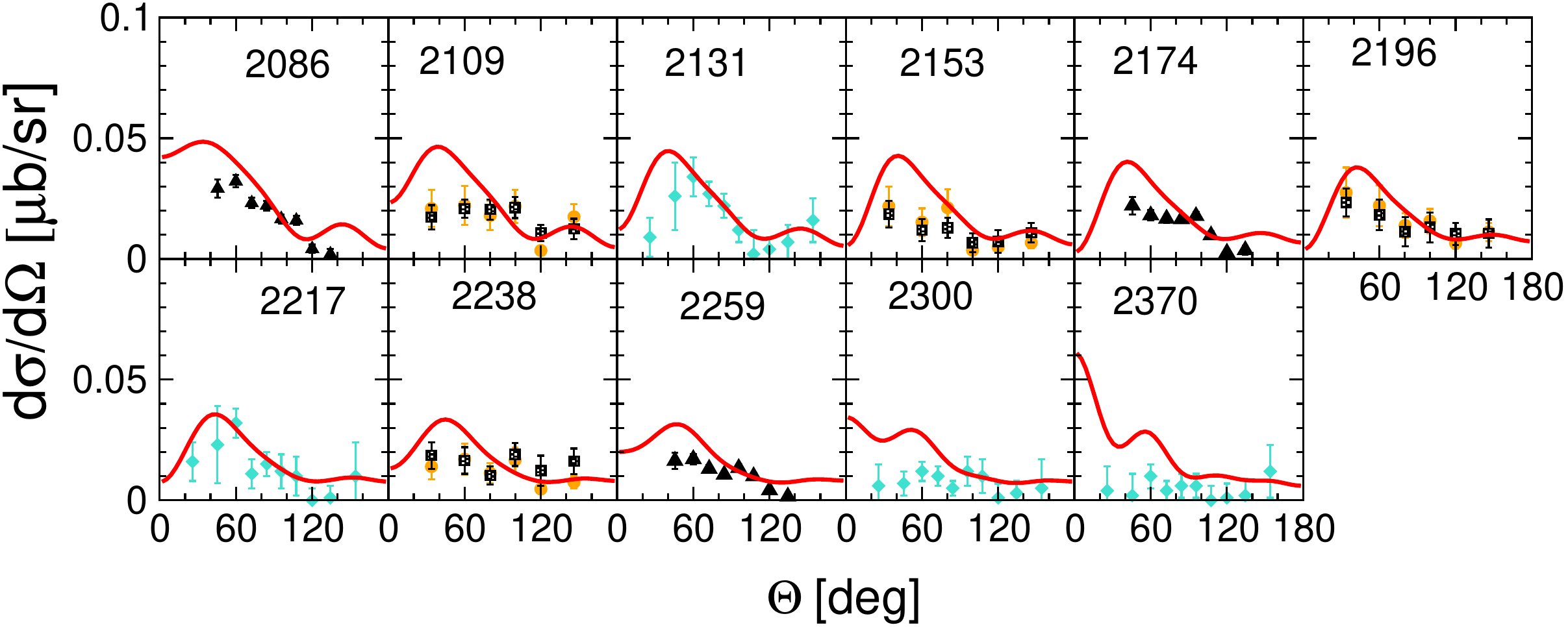} 
\end{center}
\caption{Fit results (solid (red) lines) for the differential cross section of the reaction $\gamma p\to K^0\Sigma^+$. Data: filled (orange) circles: CBELSA/TAPS 2007~\cite{CBELSATAPS:2007oqn}; filled (black) squares: CBELSA/TAPS 2011~\cite{CBELSATAPS:2011gly}; filled (turquoise) diamonds: SAPHIR~\cite{Lawall:2005np} ; open circles: A2 MAMI 2018~\cite{A2:2018doh}; open triangles: A2 MAMI 2013~\cite{A2:2013cqk}; open squares: SAPHIR 1999~\cite{SAPHIR:1999wfu} (omitted from fit); filled (black) triangles: JLab Hall B 2003~\cite{Carnahan:2003mk};  }
\label{fig:dsdok0s+}
\end{figure}

\begin{figure}
\begin{center}
\includegraphics[width=1.\linewidth]{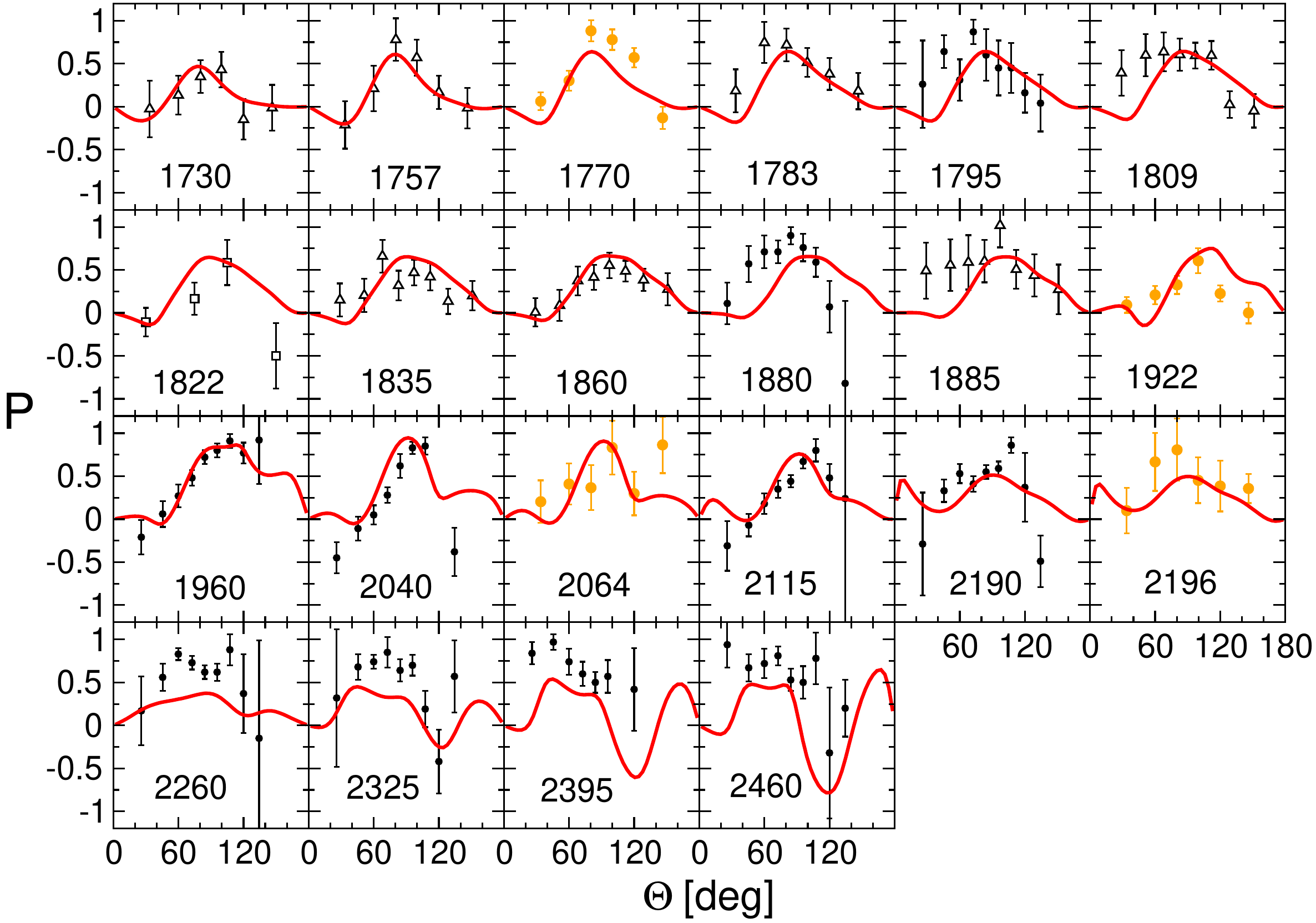} 
\end{center}
\caption{Fit results (solid (red) lines) for the recoil polarization of the reaction $\gamma p\to K^0\Sigma^+$. Data: open triangles: A2 MAMI 2013~\cite{A2:2013cqk}; filled (orange) circles: CBELSA/TAPS 2007~\cite{CBELSATAPS:2007oqn}; filled (black) circles: CLAS 2013~\cite{CLAS:2013owj}; open squares: SAPHIR 1999~\cite{SAPHIR:1999wfu}. }
\label{fig:polak0s+}
\end{figure}

In Figs.~\ref{fig:dsdok+s0} to \ref{fig:polak0s+} selected fit results for the reactions $\gamma p\to K^+\Sigma^0$ and $K^0\Sigma^+$ are shown. Fit results for other pion- and photon-induced reactions can be found online~\cite{Juelichmodel:online}. The $\chi^2$ values for the $\gamma p\to K\Sigma$ channels obtained in the current fit are given in Tab.~\ref{tab:chi2KSphoto}. These numbers were obtained with all weights set to one (cf. Appendix~\ref{app:weights}).

For $\gamma p\to K^+\Sigma^0$ we achieve an overall good fit result of the data. Exceptions are, to a certain extend, the beam-recoil observables $C_{x'}$ and $C_{z'}$ in Fig.~\ref{fig:CxCzk+s0}, where the fit fails to give a very accurate description of the data. Regarding the large uncertainties of the data and a number of points with unphysical values larger than one, we consider major modifications of the model, such as the inclusion of new genuine $s$-channel resonances, unjustified if they are just based on $C_{x'}$ and $C_{z'}$ alone, at least without an in-depth statistical analysis, which is beyond the scope of the present study. {Note that while a $\chi^2\sim 2$ might be large from a pure statistical point of view, this value is not unusual for coupled-channel fits that include many different reactions and data sets from different measurements, c.f. the $\chi^2$ values obtained in Ref.~\cite{Anisovich:2011fc}.}

Due to the limited data base it is much harder to achieve a good description of the $K^0\Sigma^+$ data. Moreover, the available data are not entirely consistent as can be seen in Fig.~\ref{fig:dsdok0s+}.  Accordingly, the $\chi^2$ is worse than for $K^+\Sigma^0$. More data, especially for polarization observables would be helpful to better determine the amplitude for the $K^0\Sigma^+$ channel.

The extracted multipole amplitudes for $\gamma p\to K^+\Sigma^0$ and $K^0\Sigma^+$ can be found in Appendix~\ref{app:multipoles}. One of the striking features of the $K\Sigma$ photoproduction data is the appearance of relatively sharp drops in cross section of both charge final states, especially for $K^+\Sigma^0$~\cite{Jude:2020byj}. This issue is discussed in the next section.

\begin{table}
\caption{The $\chi^2$ for the reactions $\gamma p\to K^+\Sigma^0$ and $K^0\Sigma^+$.}
\begin{center}
\renewcommand{\arraystretch}{1.9}
\begin {tabular}{l|c|c} 
\hline\hline
Reaction & Observable ($\#$ data points) & $\chi^2$/data point \\ \hline
$\gamma p\to K^+\Sigma^0$ & $d\sigma/d\Omega$ (4381) & 1.52 \\
  & $P$ (402) & 2.60\\
  & $\Sigma$ (280) & 1.88 \\
  & $T$ (127) & 1.50\\
  & $C_{x'}$ (94) & 3.11\\
  & $C_{z'}$ (94) & 2.65\\
  & $O_{x}$ (127) & 1.87\\
  & $O_{z}$ (127) & 1.39\\
  & in total & 1.66\\ \hline
$\gamma p\to K^0\Sigma^+$  & $d\sigma/d\Omega$ (281) & 3.53\\
 & $P$ (167) & 2.54\\
 & in total &  3.16\\
\hline \hline

\end{tabular}
\end{center}
\label{tab:chi2KSphoto}
\end{table}

\setlength{\unitlength}{\textwidth}

\section{Resonance Spectrum}
\label{sec:spectrum}

As required for a reliable determination of the resonance spectrum, in the present approach a resonance state is defined in terms of a pole in the complex energy plane on the unphysical, second Riemann sheet of the full scattering matrix $T_{\mu\nu}$. In Ref.~\cite{Doring:2009yv} the analytic properties, the sheet structure and cuts of our model are described in detail as well as the analytic continuation of the amplitude to the second sheet. The latter is achieved by a contour deformation of the momentum integration; see Ref.~\cite{Sadasivan:2021emk} for an updated and simplified discussion. The relevance of including complex branch points for channels with unstable particles ($\pi\Delta$, $\rho N$ and $\sigma N$ in the present case) in order to avoid a false resonance signal, is further discussed in Ref.~\cite{Ceci:2011ae}.  

The coupling strength of the individual states to the different hadronic channels is parameterized by the \textit{normalized residues}. Our definition of this quantity is in agreement with that of the Particle Data Group~\cite{Workman:2022ynf} and can be found in Ref.~\cite{Ronchen:2012eg}. In the present study, we apply the method described in the appendix of Ref.~\cite{Doring:2010ap} to calculate the residues of the complex poles. Following the PDG convention, the coupling of the $\gamma N$ channel to the resonances is defined by the so-called \textit{photocouplings at the pole}. See Ref.~\cite{Ronchen:2014cna} for an explicit definition and its decomposition into electric and magnetic multipoles. Note that those photocouplings at the pole are independent of the hadronic final state. 

In Tabs.~\ref{tab:poles1} and \ref{tab:poles2} we list the pole positions and residues of established states found in the present study. The corresponding photocouplings at the pole are given in Tab.~\ref{tab:photo}. {All poles in Tabs.~\ref{tab:poles1} and \ref{tab:poles2} lie on the second Riemann sheet, which we define as the unphysical sheet closest to the respective section of the physical axis. The physical implications of poles on higher sheets are in general limited and those poles are not listed here.} We compare the present results to the J\"uBo2017 analyis~\cite{Ronchen:2018ury}, where $K\Lambda$ but not $K\Sigma$ photoproducion was already included, besides the other channels listed in Tab.~\ref{tab:data}. We also give the estimates of resonance properties by the Particle Data Group~\cite{Workman:2022ynf}  in a shortened form. For example, the expression for the real part of the pole position of the $N(1680)5/2^+$, ``1665 to 1680 ($\approx$ 1675) OUR ESTIMATE", is quoted as ``$1675^{+5}_{-10}$". In cases where the PDG does not provide an estimate, i.e. for states with less than three stars, we average the entries from ``above the line" to have a point of comparison. The PDG values for the normalized residues of inelastic channels all originate from studies of the Bonn-Gatchina group~\cite{Anisovich:2011fc,Sokhoyan:2015fra}. 

We find all 4-star $I=1/2$ and $3/2$ resonances with $J\le 9/2$  except for the $N(1895)1/2^+$, which is not needed in the present study to achieve a good fit result. A number of states rated with less than 4 stars are also observed.
In addition to the resonances in Tabs.~\ref{tab:poles1} and \ref{tab:poles2}, we see indications for other states that are not listed in the tables since further evidence for their existence is needed. Those states and the properties of selected resonances are discussed in the following.

\begin{table*} \caption{Properties of the $I=1/2$ resonances: Pole positions $W_0$ ($\Gamma_{\rm tot}$ defined as -2Im$W_0$), elastic $\pi N$ residues $(|r_{\pi N}|,\theta_{\pi
N\to\pi
N})$, and the normalized residues $(\sqrt{\Gamma_{\pi N}\Gamma_\mu}/\Gamma_{\rm tot},\theta_{\pi N\to \mu})$ of the inelastic reactions $\pi
N\to \mu$ with $\mu=\eta N$, $K\Lambda$, $K\Sigma$. Resonances with italic numbers in the parentheses are not identified with a PDG state; subscript (a): dynamically generated  in the present study. We show the results of the present study J\"uBo2022 (``2022") and for comparison the results of the J\"uBo2017 analysis~\cite{Ronchen:2018ury} (``2017") and the estimates of and from the Particle Data Group~\cite{Workman:2022ynf} (``PDG"), if available, as well as the PDG star rating. See text for further explanations regarding states not listed here.
 }
\begin{center}
\renewcommand{\arraystretch}{1.3}
\resizebox{\textwidth}{!}{
\begin {tabular}{ll|ll|ll|ll|ll|ll} 
\hline\hline
&&\multicolumn{1}{|l}{Re $W_0$ \hspace*{0.5cm} }
& \multicolumn{1}{l|}{$-$2Im $W_0$\hspace*{0.1cm} }
& \multicolumn{1}{l}{$|r_{\pi N}|$\hspace*{0.2cm}} 
& \multicolumn{1}{l}{$\theta_{\pi N\to\pi N}$ } 
& \multicolumn{1}{|l}{$\displaystyle{\frac{\Gamma^{1/2}_{\pi N}\Gamma^{1/2}_{\eta N}}{\Gamma_{\rm tot}}}$}
& \multicolumn{1}{l|}{$\theta_{\pi N\to\eta N}$\hspace*{0.1cm}}
& \multicolumn{1}{l}{$\displaystyle{\frac{\Gamma^{1/2}_{\pi N}\Gamma^{1/2}_{K\Lambda}}{\Gamma_{\rm tot}}}$} 
& \multicolumn{1}{l|}{$\theta_{\pi N\to K\Lambda}$\hspace*{0.1cm}}
& \multicolumn{1}{l}{$\displaystyle{\frac{\Gamma^{1/2}_{\pi N}\Gamma^{1/2}_{K\Sigma}}{\Gamma_{\rm tot}}}$} 
& \multicolumn{1}{l}{$\theta_{\pi N\to K\Sigma}$}
\bigstrut[t]\\[0.2cm]
&&\multicolumn{1}{|l}{[MeV]} & \multicolumn{1}{l|}{[MeV]} & \multicolumn{1}{l}{[MeV]} & \multicolumn{1}{l}{[deg]} 
& \multicolumn{1}{|l}{[\%]}  & \multicolumn{1}{l|}{[deg]} & \multicolumn{1}{l}{[\%]}  & \multicolumn{1}{l|}{[deg]} &\multicolumn{1}{l}{[\%]} & \multicolumn{1}{l}{[deg]} \\
		  & fit &&&&&&&&
\bigstrut[t]\\
\hline

$N (1535)$ 1/2$^-$ 
& 2022 & 1504(0)& 74 (1)& 18 (1)& $-37(3)$ & 50(3) & 118(3) & 26(2) & $-67(3)$ & 28(2) & 92(3) \\
& 2017 &  1495$(2)$  & 112$(1)$ & 23$(1)$ & $-52 (4)$ & 51$(1)$ & 105$(3)$ & 6.0$(1.5)$ & $-44(30)$ & 5.7$(1.6)$ & $-86(6)$   \\
 ****	&PDG& $1510\pm 10$ & $130\pm 20$ & $25\pm 10$& $-15\pm 15$	&$43\pm 3$&$-76\pm 5$&---&---&---&---		\\ \hline
	
 $N (1650)$ 1/2$^-$ 
 & 2022 & 1678(3) & 127(3) &59(21) & $-18(46)$ & 34(12) & 71(45) & 26(10) &  $-40(46)$ & 41(15) & $-21(47)$\\
& 2017& 1674$(3)$ & 130$(9)$ & 29$(4)$ & $-53(7)$ & 18$(3)$ & 28$(5)$ & 17$(1)$ & $-59(3)$ &21$(2)$ & $-67(4)$    \\
  ****&PDG &$1655\pm 15$ &$135\pm 35 $ & $45^{+10}_{-20} $ & $-70^{+20}_{-10}$ & $29\pm 3$&$134\pm 10$&---&---&---&---\\
				\hline
				
 $N (1440)$ 1/2$^+_{(a)}$ 
 & 2022 & 1353(1) & 203(3) &  59(2) & $-104(4)$ & 8.4(0.4) & $-28(4)$ & 2.5(0.9) & $-92(86)$ & 0.2(0.5) & $-32(154)$\\
 & 2017& 1353$(4)$ & 213$(2)$ & 62$(2)$ & $-100(7)$ & 8.6$(0.9)$ & $-29(7)$ & 4.8$(0.4)$ & 129$(6)$ & 2.1$(0.4)$ & 87$(22)$ \\
  **** &PDG & $1370\pm 10$ &$175\pm 15$ &$50\pm 4$ & $-90\pm 10$ &---&---&---&---&---&---\\
				\hline


$N (1710)$  1/2$^+_{(a)}$ 
& 2022 &1605(14) & 115(9) & 5.5(4.7) & $-114(57)$ & 28(26) & 91(63) & 20(19)& $-144(77)$ & 5.5(4.8) & 162(305) \\
 & 2017& 1731(7) & 157$(6)$ & 1.5$(0.1)$ & 178$(9)$ & 1.6$(0.4)$ & $-137(46)$ & 10$(1)$ & 52$(5)$ & 1.4$(0.1)$ & $-79(24)$ \\
****  &PDG &$1700\pm 20$ & $ 120\pm 40$ &$7\pm 3$ &$190\pm 70$ &$12\pm 4$ &$0\pm 45$ &---&---&---&---\\
				\hline
 
 $N (\textit{1750})$  1/2$^+_{(a)}$ 
 & 2022 &  not seen &&&&&&&\\
 & 2017& 1750$(2)$ & 318$(3)$ & 2.9$(2.8)$ & 100$(29)$ &0.7$(0.5)$ & $-31(30)$ & 1.0$(0.2)$ & 164$(19)$ & 3.2$(0.6)$ & 29$(15)$ \\
				\hline

 $N (1720)$ 3/2$^+$ 
& 2022 & 1726(8) & 185(12) & 15(2) & $-60(5)$ & 4.9(0.9) & 64(10) & 3.4(0.4) & $-101(8)$ &5.9(1) & 82(6) \\
 & 2017& 1689$(4)$ & 191$(3)$ & 2.3$(1.5)$ & $-57(22)$ & 0.3$(0.2)$ & 139$(35)$ & 1.5$(0.9)$ & $-66(30)$ & 0.6$(0.4)$ & 26$(58)$ \\
 **** &PDG &$1675\pm 15$ &$250^{+150}_{-100} $ &$15^{+10}_{-5}$ & $-130\pm 30$ &$3\pm 2$ &---&$6\pm 4$ &$-150\pm 45$&---&---\\
				\hline

 $N (1900)$ 3/2$^+$ 
 & 2022 & 1905(3) & 93(4) & 1.6(0.3) & 44(21) & 1.0(0.3) & 55(29)&2.9(0.6) & 5.4(18.6)&1.3(0.3) & $-40(18)$\\
& 2017& 1923$(2)$ & 217$(23)$ & 1.6$(1.2)$ & $-61(121)$ & 1.1$(0.7)$ & $-10(79)$ & 2.1$(1.4)$ & 1.7$(86)$ & 10$(7)$ & $-34(74)$ \\
****  &PDG &$1920\pm 20$ & $150\pm 50$ &$4\pm 2$ &$-20\pm 30$&$5\pm 2$ &$70\pm 60$ &$3\pm 2$&$90\pm 40$&$4\pm 2$&$110\pm 30$\\
				\hline
 
 $N (1520)$ 3/2$^-$ 
 & 2022 & 1482(6) & 126(18)& 27(21) & $-36(48)$ & 2.1(1.8) &34(53) & 2.6(1.9) & 127(47) & 1.0(1.2) & 94(68) \\
 & 2017& 1509$(5)$ & 98$(3)$ & 33$(6)$ & $-16(23)$ & 3.7$(0.6)$ & 85$(18)$ & 0.8$(0.3)$ & 83$(43)$ & 3.0$(1.0)$ &$-28(21)$\\
 ****	 &PDG &$1510\pm 5$ &$110^{+10}_{-5}$ &$35\pm 3$ &$-10\pm 5$ &--- &--- &---&---&---&---\\
				 \hline

 $N (1675)$ 5/2$^-$ 
 & 2022 & 1652(3) & 119(1) & 22(1) & $-17(2)$ & 6.3(0.9) & $-39(2)$ & $<0.1(0.2)$ & 174(161) & 2.4(0.2) &$-166(5)$ \\
& 2017&  1647$(8)$ & 135$(9)$ & 28$(2)$ & $-22(3)$ & 9.1$(1.8)$ & $-45(3)$ & 0.7$(0.2)$ & $-91(6)$ & 2.3$(0.2)$ & $-175(10)$\\
****  &PDG &$ 1660\pm 5$ &$135^{+15}_{-10} $ &$28\pm 5$ &$-25\pm 5 $ &--- &--- &---&---&---&---\\
				\hline

 
 $N (1680)$ 5/2$^+$ 
 & 2022 & 1657(3) & 120(2) &36(1) & $-31(1)$ & 0.6(0.7)& 118(2)& 0.6(0.1)& $-119(3)$ & $<0.1(0.2)$ & $-46(29)$\\
& 2017& 1666$(4)$ & 81$(2)$ & 29$(1)$ & $-12(1)$ & 1.7$(0.5)$ & 145$(1)$ & 0.9$(0.1)$ & $-77(2)$ & $<0.1$ & $-33(161)$  \\
****  &PDG &$1675^{+5}_{-10}$ &$120^{+15}_{-10}$ &$40\pm 5 $ &$-5\pm 15 $ &--- &--- &---&---&---&---\\
				\hline

 $N (1990)$ 7/2$^+$ 
 & 2022 & 1861(9) & 72(5) & 0.16(0.01) & $-119(4)$ & 4.8(0.2) & $-43(4)$ & 0.4(0.1) &133(4)&  1.0(0.3) & $-54(4)$ \\
& 2017& 2152$(12)$ & 225$(20)$ & 0.2$(0)$ & 92$(10)$ & 0.4$(0.2)$ & $-9.1(5.5)$ & 1.4$(0.3)$ & $-13(5)$ & 1.5$(0.3)$ & $-18(6)$  \\
**&PDG &$1965\pm 80$ &$250\pm 60$ &--- &--- &--- &--- &---&---&---&---\\
				\hline

 $N (2190)$  7/2$^-$ 
 & 2022 & 1965(12) & 287(66) & 18(7) & $-45(27)$ & 2.1(1) & $-65(29)$ & 2.6(1.4) & $-78(30)$ & 0.5(0.2) & $-92(31)$ \\
& 2017& 2084$(7)$ & 281$(6)$ & 20$(2)$ & $-31(1)$ & 1.2$(0.6)$ & 140$(1)$ & 3.7$(0.3)$ & $-47(1)$ & 0.3$(1.1)$ & 124$(2)$\\
****  &PDG &$2100\pm 50 $ &$400\pm 100 $ &$50^{+20}_{-25}$ &$0^{+30}_{-30}$ &--- &--- &$3\pm 1$&$20\pm 15$&---&---\\
				\hline
 
 $N (2250)$ 9/2$^-$ 
 & 2022 & 2095(20) & 422(26) & 14(2) & $-67(17)$ & 1.8(0.2) & $-89(9)$ & 0.3(0.1) & 80(9) & 0.4(0.4) & $-111(9)$ \\
& 2017& 1910$(53)$ & 243$(73)$ & 0.4$(0.1)$ & $-56(25)$ & 0.9$(0.2)$ & $-80(21)$ & $<0.1$ & $-96(21)$ & 0.2$(0.2)$ & $-110(19)$  \\
****  &PDG &$2200\pm 50 $ & $420^{+80}_{-70} $ &$25\pm 5 $ &$-40\pm 20 $ &--- &--- &---&---&---&---\\
				\hline
 
 $N (2220)$ 9/2$^+$ 
 & 2022 & 2131(12) & 388(12) & 48(10)  & $-13(3)$ & 4.2(1.1)   & $-48(4)  $ & 2.0(0.5)  & $-60(4)  $ & 0.3(1.6) &$-70(4) $ \\
 & 2017& 2207$(89)$ & 659$(140)$ & 91$(47)$ & $-68(16)$ & 0.3$(0.4)$ & $-109(17)$ & $<0.1$ & 31$(150)$ & 1.0$(0.9)$ & 44$(19)$ \\
 ****  &PDG &$2170^{+30}_{-40}$ &$400^{+ 80}_{-40}$ &$45^{+15}_{-10} $ &$-50^{+20}_{-10} $ &--- &--- &---&---&---&---\\
\hline\hline
\end {tabular}
}
\end{center}
\label{tab:poles1}
\end{table*}

\begin{table*}
\caption{Properties of the $I=3/2$ resonances: Pole positions $W_0$ ($\Gamma_{\rm tot}$ defined as -2Im$W_0$), elastic $\pi N$ residues $(|r_{\pi N}|,\theta_{\pi N\to\pi N})$, and
the
normalized residues  $(\sqrt{\Gamma_{\pi N}\Gamma_\mu}/\Gamma_{\rm tot},\theta_{\pi N\to \mu})$ of the inelastic reactions $\pi N\to K\Sigma$
and $\pi N\to\pi\Delta$ with the number in brackets indicating $L$ of the $\pi\Delta$ state. Subscript (a): dynamically generated  in the present study. We show the results of the present study J\"uBo2022 (``2022") and for comparison the results of the J\"uBo2017 analysis~\cite{Ronchen:2018ury} (``2017") and the estimates of and from the Particle Data Group~\cite{Workman:2022ynf} (``PDG"), if available, as well as the PDG star rating. See text for further explanations regarding states not listed here.
}
\begin{center}
\renewcommand{\arraystretch}{1.3}
\resizebox{\textwidth}{!}{
\begin {tabular}{ll|ll|ll|ll|ll|ll}  \hline\hline
&&\multicolumn{2}{|l}{Pole position}
 &\multicolumn{2}{|l}{$\pi N$ Residue}
 &\multicolumn{2}{|l}{$K\Sigma$ channel} 
 &\multicolumn{2}{|l|}{$\pi\Delta$, channel (6)}
 &\multicolumn{2}{|l}{$\pi\Delta$, channel (7)}
\bigstrut[t]\\[0.1cm]
&&\multicolumn{1}{|l}{Re $W_0$ \hspace*{0.5cm} }
& \multicolumn{1}{l|}{$-$2Im $W_0$\hspace*{0.1cm} }
& \multicolumn{1}{l}{$|r_{\pi N}|$\hspace*{0.cm}} 
& \multicolumn{1}{l}{$\theta_{\pi N\to\pi N}$ } 
& \multicolumn{1}{|l}{$\displaystyle{\frac{\Gamma^{1/2}_{\pi N}\Gamma^{1/2}_{K\Sigma}}{\Gamma_{\rm tot}}}$}
& \multicolumn{1}{l|}{$\theta_{\pi N\to K\Sigma}$\hspace*{0.1cm}}
& \multicolumn{1}{l}{$\displaystyle{\frac{\Gamma^{1/2}_{\pi N}\Gamma^{1/2}_{\pi\Delta}}{\Gamma_{\rm tot}}}$} 
& \multicolumn{1}{l|}{$\theta_{\pi N\to \pi\Delta}$\hspace*{0.1cm}}
& \multicolumn{1}{l}{$\displaystyle{\frac{\Gamma^{1/2}_{\pi N}\Gamma^{1/2}_{\pi\Delta}}{\Gamma_{\rm tot}}}$} 
& \multicolumn{1}{l}{$\theta_{\pi N\to \pi\Delta}$}
\\
&&\multicolumn{1}{|l}{[MeV]} & \multicolumn{1}{l|}{[MeV]} & \multicolumn{1}{l}{[MeV]} & \multicolumn{1}{l}{[deg]} 
& \multicolumn{1}{|l}{[\%]}  & \multicolumn{1}{l|}{[deg]} & \multicolumn{1}{l}{[\%]}  & \multicolumn{1}{l|}{[deg]} &\multicolumn{1}{l}{[\%]} & \multicolumn{1}{l}{[deg]} \\
		  & fit &&&&&&&&
\bigstrut[t]\\
\hline

 $\Delta(1620)$	1/2$^-$ &
  2022 & 1607(4) & 85(5) & 12(2) & $126(4)$ & 11(2) & $-120(5)$ & -- &--& 32(2) {\footnotesize (D)} & 81(2) \\
&2017& 1601$(4)$ & 66$(7)$ & 16$(3)$ & $-106(3)$ & 31$(6)$ & $-103(2)$ & --- & --- & $57(4)$ {\footnotesize (D)} & 103$(1)$  \\
****& PDG &$1600\pm 10$ &$ 120\pm 20$ & $17^{+3}_{-2}$ &$-100\pm 20$& ---&---&---&---&$42\pm 6$&$-90\pm 20$ \\
				\hline
 	 			
$\Delta(1910)$ 1/2$^+$  &
 2022 & 1802(11) & 550(22) &35(25) & 93(14) &0.2(0.4) & 138(19)&  24(18) {\footnotesize (P)}& $-42(14)$ &--- &---\\
 &2017&  1798$(5)$ & 621$(35)$ &  81$(68)$ & $-87(18)$ & 5.1$(2.2)$ & $-96(58)$ & $53(42)$ {\footnotesize (P)} & 126$(15)$ & ---& ---  	 \\ 
 ****	& PDG & $1860\pm 30$& $300\pm 100 $ &$25\pm 5$&$130\pm 50$&$7\pm 2$&$-110\pm 30$&$24\pm 10$&$85\pm 35$&---&--- \\
				\hline
 	 			

$\Delta(1232)$ 3/2$^+$ &
  2022 & 1215(2) & 93(1)& 50(2) & $-39(1)$ &&& & \\
 	& 2017& 1215$(4)$ & 97$(2)$ & 48$(1)$ & $-40(2)$ &  &&&& \\
 ****	& PDG &$1210\pm1$ &$ 100\pm 2$ &$50^{+2}_{-1}$ & $-46^{+1}_{-2}$&&&&& \\
				\hline

 $\Delta(1600)$ 3/2$^+$&
  2022 & 1590(1) & 136(1) & 11(1) & $-106(2)$ & 14(1) & 14(2) & 30(3) {\footnotesize (P)}  & 87(3) & 0.4(0.04) {\footnotesize (F)}  &$-62(9)$\\
 & 2017&  1579$(17)$ & 180$(30)$ & 11$(6)$ & $-162(41)$ & 13$(7)$ & $-21(40)$ & $31(16)$ {\footnotesize (P)} & 37$(40)$ & $0.6(0.9)$ {\footnotesize (F)} & $-56(117)$	 \\
 ****	& PDG &$1510\pm 50 $ &$270\pm 70 $ &$25\pm15 $ &$180\pm30 $&---&---&$15\pm 4$&$30\pm 35$&
$1\pm 0.5$&---\\
				\hline

$\Delta(1920)$	3/2$^+_{(a)}$ &
  2022 & 1883(4) & 844(10) & 41(5) & 11(7) & 20(2) & 104(4) & 5.7(0.5) {\footnotesize(P)} & $-48(5)$ & 2.0(0.3) {\footnotesize (F)} & 147(7)\\
&2017& 1939$(141)$ & 838$(38)$ & 26$(9)$ & 96$(35)$ & 14$(3)$ & 146$(18)$ & $2.7(1.0)$ {\footnotesize(P)} & 31$(16)$ & $0.6(0.4)$ {\footnotesize (F)} & $-115(86)$   	 \\
***	& PDG &$1900\pm 50$ &$300\pm 100 $ &$16\pm 8$&$-100\pm 50$&$9\pm 3$&$80\pm 40$&$20\pm 8$&$-105\pm 25$&$37\pm 10$&$-90\pm 20$ \\
				\hline

$\Delta(1700)$ 3/2$^-$ &
  2022 & 1637(64) & 295(58) & 15(23) & $-13(147)$ & 0.7(1.5) & $-176(320)$ &3.8(7.8)  {\footnotesize (D)}& 127(254) & 20(29) {\footnotesize (S)}  & 146(266)  \\
 &2017& 1667$(28)$ & 305$(45)$ & 22$(6)$& $-8.6(32.1)$ & 0.7$(1.8)$ & 176$(152)$ & $4.8(2.0)$ {\footnotesize (D)} & 169$(26)$ & $38(14)$ {\footnotesize (S)} & 146$(30)$   \\
 ****& PDG &$1665\pm 25 $ &$250\pm 50 $ &$25\pm 15$&$-20\pm20 $&---&---&$12\pm 6$&$-160\pm 30$&$25\pm 12$&$135\pm 45$ \\
				\hline

 $\Delta(1930)$	5/2$^-$ &
  2022 & 1821(4)& 447(13) & 15(3)  & $-108(9) $ & 1.0(0.2)  & 49(9)  & 12(3)  {\footnotesize (D)}  & 64(7)  & 0.8(0.2) {\footnotesize (G)} & 148(4) \\
 &2017& 1663$(43)$ & 263$(76)$ & 5.1$(2.4)$ & $-112(23)$ & 2.5$(0.9)$ & $-27(18)$ & $17(5)$ {\footnotesize (D)} & 68$(17)$ & $0.2(0.2)$ {\footnotesize (G)} & $-134(48)$  \\
 *** & PDG &$1880\pm 40 $ &$280\pm 50 $ &$14\pm 6 $&$-30^{+20}_{-10} $&---&---&---&---&---&--- \\
				\hline
 	 			
 $\Delta(1905)$	5/2$^+$ &
  2022 & 1707(1) & 127(8) & 3.7(1.0) & $-92(12)$ & 0.2(0.03) & 154(11) & 1.7(0.3)  {\footnotesize (F)} & 18(15) & 10(1) {\footnotesize (P)}& $-109(14)$\\
&2017& 1733$(47)$ & 435$(264)$ & 21$(20)$ & 110$(93)$ & 0.5$(0.5)$ & $-4.3(345)$ & $3.6(3.4)$ {\footnotesize (F)} & $-117(309)$ & $15(15)$ {\footnotesize (P)} & $-61(230)$  	  \\
 ****	& PDG &$ 1800\pm 30$ &$300 \pm 40 $ &$20\pm 5$&$-50^{+20}_{-70} $&---&---&---&---&$19\pm 7$&$10\pm 30$ \\
				 \hline

 $\Delta(1950)$	7/2$^+$ &
  2022 & 1875(1) & 166(3) &  27(2) & 1.1(2.0) & 2.0(0.3) & $-40(7)$ & 30(54) {\footnotesize (F)} & $166(2)$ & 5.1(0.7) {\footnotesize (H)}& $-11(2)$ \\
&2017& 1850$(37)$ & 259$(61)$ & 34$(20)$ & $-48(46)$ & 1.4$(1.4)$ & $-106(50)$ & $35(25)$ {\footnotesize (F)} & 119$(46)$ & $1.7(1.0)$ {\footnotesize (H)} & $-103(59)$   \\
**** 	& PDG &$1880\pm 10$ &$240\pm 20 $ &$52\pm 8 $ &$ -32\pm8$ &$5\pm 1$&$-65\pm 25$&$12\pm 4$&---&---&--- \\
				\hline

$\Delta (2200)$ 7/2$^-$ &
  2022 & 1963(2) & 328(3) & 6.8(0.6) & $-80(2)$ & $<0.1(0.03)$ & $-123(2)$ & 0.3(0.1) {\footnotesize (G)} & $152(5)$ & 16(1) {\footnotesize (D)}& 100(2)\\
 & 2017& 2290$(132)$ & 388$(204)$ & 33$(92)$ & $-32(138)$ & 1.0$(1.0)$ & 118$(165)$ & $7.0(21.1)$ {\footnotesize (G)} & $-103(328)$ & $53(124)$ {\footnotesize (D)} & 137$(132)$ \\
 ***		& PDG &$2100\pm 50$ &$340\pm 80$ &$8\pm 3$&$-70\pm 40$&---&---&---&---&---&--- \\
				 \hline

 $\Delta (2400)$ 9/2$^-$ &
  2022 & 2458(3) & 280(2) & 5.4(5) & 8.4(33) & 0.4(0.6) & 17(30) & 10(11){\footnotesize (G)} & 17(23) & 1.9(0.5) {\footnotesize (I)}& $-120(49)$ \\
&2017& 1783$(86)$ & 244$(194)$ & 7.2$(8.6)$ & $-78(30)$ & 0.5$(0.6)$ & 9.1$(9.0)$ & $19(9)$ {\footnotesize (G)} & $-95(36)$ & $1.6(1.0)$ {\footnotesize (I)} & $-18(90)$  	\\
** 	& PDG &$2260\pm 60$&$320\pm 160$ &$8\pm 4$&$-25\pm 15$&---&---&---&---&---&--- \\

\hline\hline
\end {tabular}
}
\end{center}
\label{tab:poles2}
\end{table*}

\begin{table*}
\caption{Properties of the $I=1/2$ (left) and $I=3/2$ resonances (right): 
photocouplings at the pole ($A^h_{pole}$, $\vartheta^h$). Subscript (a): dynamically generated in the present study. We show the results of the present study J\"uBo2022 (``2022") and for comparison the results of fit B of the J\"uBo2017 analysis~\cite{Ronchen:2018ury} (``2017"). 
The uncertainties quoted in parentheses provide a rather rough estimate as explained in the text.
}
\begin{center}
\renewcommand{\arraystretch}{1.5}
\resizebox{2.07\columnwidth}{!}{
\begin {tabular}{ll| cc|cc || l l |cc|cc} 
\hline\hline
& & \multicolumn{1}{c}{$\mathbf{A^{1/2}_{pole}}$\hspace*{0.2cm}} 
& \multicolumn{1}{c|}{$\mathbf{\vartheta^{1/2}}$ } 
& \multicolumn{1}{c}{$\mathbf{A^{3/2}_{pole}}$\hspace*{0.2cm}} 
& \multicolumn{1}{c||}{$\mathbf{\vartheta^{3/2}}$ } 
& & & \multicolumn{1}{c}{$\mathbf{A^{1/2}_{pole}}$\hspace*{0.2cm}} 
& \multicolumn{1}{c|}{$\mathbf{\vartheta^{1/2}}$ } 
& \multicolumn{1}{c}{$\mathbf{A^{3/2}_{pole}}$\hspace*{0.2cm}} 
& \multicolumn{1}{c}{$\mathbf{\vartheta^{3/2}}$ } 
\bigstrut[t]\\[0.2cm]
&& \multicolumn{1}{c}{{\footnotesize[$10^{-3}$ GeV$^{-\nicefrac{1}{2}}$]}} &\multicolumn{1}{c|}{\footnotesize[deg]} & \multicolumn{1}{c}{\footnotesize[$10^{-3}$ GeV$^{-\nicefrac{1}{2}}$]} & \multicolumn{1}{c||}{\footnotesize[deg]} 
&&& \multicolumn{1}{c}{\footnotesize[$10^{-3}$ GeV$^{-\nicefrac{1}{2}}$]} &\multicolumn{1}{c|}{\footnotesize[deg]} & \multicolumn{1}{c}{\footnotesize[$10^{-3}$ GeV$^{-\nicefrac{1}{2}}$]} & \multicolumn{1}{c}{\footnotesize[deg]} 
 \\
		  & fit 		&&&&& &fit &	&
\bigstrut[t]\\
\hline
$N (1535)$ 1/2$^-$ & 2022 & 84(5) &$-12(3)$ &&& $\Delta(1620)$	1/2$^-$ & 2022 & 11(4)& $57(24)$ \\
   & 2017&106$(3)$  &$-1.6(2.1)$          &&& & 2017&  19$(9)$ &15$(7)$  	\\ 
 \hline
 
 $N (1650)$ 1/2$^-$  & 2022 &39(10) &$-0.2(27)$&&&	$\Delta(1910)$ 1/2$^+$ & 2022 & $-446(72)$ & $-70(21)$\\ 			
  & 2017&  51$(3)$ & $-1.4(3.9)$      &&&		& 2017&$-238(149)$ &$-87(35)$  	\\
 \hline

 $N (1440)$ 1/2$^+_{(a)}$ & 2022 & $-90(13)$ & $-30(5)$ &&&  $\Delta(1232)$ 3/2$^+$  & 2022& $-126(4) $ & $-18(3) $ & $-245(7) $ & $-0.7(1.7) $ \\
& 2017&$-90(13)$ &$-33(18)$  &&&  & 2017&$ -120(5)$ &$-14(3)$  &$-236(6)$ & 0.5$(1.1)$  	 \\     
  \hline

 $N (1710)$  1/2$^+_{(a)}$ & 2022 & $-18(19)$ & $40(109) $ && & $\Delta(1600)$ 3/2$^+$& 2022  & 25(10)& 0.5(5.9)& $-6.0(2.6)$ & $62(63)$ \\ 				 
 & 2017&$-14(2)$  &$-23(188)$ &&& & 2017& $54(25)$ &$144(31)$   &$-46(19)$  &$-8.5(36)$  \\
  \hline

 $N (1720)$ 3/2$^+$ & 2022 & 39(7) & 60(10) & $-25(7)$ & $-5.7(13)$ &
 $\Delta(1920)$	3/2$^+_{(a)}$ & 2022 & 138(12) & $-8.9(3.9)$ & 252(14) & 14(3) \\
 & 2017& 48$(24)$&30$(24)$ &$-27(19)$ & $-11(29)$ &  &
   2017& 35$(15)$  	& $-89(44)$    &77$(17)$  &$-26(39)$ \\
   \hline

$N (1900)$ 3/2$^+$ & 2022 & 9.1(2.7)& 80(23)& $-7.7(3.4) $ & $-42(23) $ &
$\Delta(1700)$ 3/2$^-$ & 2022 & 163(120) & $-4.4(78)$  & 221(185)  & $-12(79)  $ \\			
 & 2017& 34$(13)$ &$-20(65)$  &109$(64)$  &12$(23)$  &  		& 
 2017& 191$(43)$	& 14$(36)$ &244$(58)$  &$-5.8(32)$ \\ 
\hline

$N (1520)$ 3/2$^-$ & 2022 & $-43(25)$ & $-47(20)$ & 112(64) &	1.8(37)	&  			$\Delta(1930)$	5/2$^-$ & 2022 &104(18)& 129(16) & 322(44) & 142(7) \\ 
 & 2017&$-35(10)$ &$-10(7)$ &77$(17)$ &8.6$(13.1)$  	&  & 
 2017& 159$(133)$	&8.7$(26.5)$  & 97$(32)$  &69$(30)$   \\
\hline

 $N (1675)$ 5/2$^-$ & 2022& 25(4) & $-1.2(7.8)$ & 51(4) & $-1.0(3.7)$&		$\Delta(1905)$	5/2$^+$ & 2022 & 55(8)& $-159(3)$ & $-168(40)$ & $172(1.7)$  \\
  & 2017&38$(3)$ &17$(10)$  &52$(23)$ &$-11(7)$  & & 
 2017& 59$(181)$ &11$(235)$  &$-125(295)$ & 28$(195)$    	 \\
\hline

 $N (1680)$ 5/2$^+$ & 2022 & $-17(6)$ & 70(14) & 95(6)& $-57(7)$ & 				  $\Delta(1950)$	7/2$^+$ & 2022& $-31(4)$ & $-81(7)$ & $-45(4)$ & $-89(4)$ \\
 & 2017&$-8.0(1.8)$&$-42(35)$  &95$(6)$ &$-28(11)$  &  & 
 2017&$ -68(29)$ & $-49(35)$ &$-95(43)$ &$-53(46)$  \\
\hline


  $N (1990)$ 7/2$^+$ & 2022 & $-30(16)$ & $-135(25)$ & $-18(11)$ & 53(32)  &  		 
 $\Delta (2200)$ 7/2$^-$ & 2022 & 104(22) & $-139(3)$ & 21(25) & $-180(39)$\\	 
  &  2017& $-22(48)$ &13$(236)$  & $-41(69)$ &11$(233)$  & 	  & 
 2017& 110$(146)$ & 49$(94)$ &57$(69)$  &$-84(64)$  \\
 \hline

 $N (2190)$  7/2$^-$ & 2022 & $-15(8)$ & 111(17)& 62(22) & 179(26) &
 $\Delta (2400)$ 9/2$^-$ & 2022 & 21(14) & $-67(23)$ & 22(14) & 122(14) \\	
  & 2017&$-23(13)$ & 70$(40)$  &53$(10)$  &$-82(12)$  & & 
 2017&  14$(84)$	&58$(66)$ & 22$(41)$ &89$(82)$    \\ 
  \hline

				 
 $N (2250)$ 9/2$^-$ & 2022 & $-108(14)$ & 112(7) &50(22) & 69(16)  \\
 & 2017&$-41(11)$ &$-20(68)$  &20$(15)$  & $-74(60)$& 	\\
 \hline
				 
 $N (2220)$ 9/2$^+$ & 2022 & 357(39) & $-91(7)$ & $-273(50)$ & $-102(6)$ \\
 & 2017& 536$(435)$	& 69$(62)$ & $-445(355)$ & $82(44)$  & \\

\hline\hline
\end {tabular}
}
\end{center}
\label{tab:photo}
\end{table*}


\subsection{Discussion of specific resonances }

In the $S_{11}$ partial wave two bare $s$-channel poles are included which correspond to the $N(1535)1/2^-$ and the $N(1650)1/2^-$. The $N(1535)1/2^-$  is narrower than in previous J\"uBo studies~\cite{Ronchen:2018ury,Ronchen:2015vfa}. Note that in an earlier SAID solution~\cite{Arndt:2006bf} this resonance was also found to be rather narrow ($\Gamma=95$~MeV) but the width varies in different analyses~\cite{Workman:2022ynf}. The pole position of the $N(1650)1/2^-$, on the other hand, is only slightly different from our previous anlysis~\cite{Ronchen:2018ury} and very close to the PDG values. The residues are less stable in general. While the coupling of the $N(1535)1/2^-$ to $\pi N$ is smaller than in J\"uBo2017, the $N(1650)1/2^-$ now couples stronger to this channel. 

The $\eta N$ coupling to the $N(1535)1/2^-$ remains large and takes a very similar value as in previous J\"uBo studies. The $\eta N$ residue of the $N(1650)1/2^-$, however, changes significantly: the new value is almost twice as large as in JüBo2017. We ascribe this change to the recently published polarization data  for $\eta p$ photoproduction by the CBELSA/TAPS Collaboration~\cite{CBELSATAPS:2019ylw}, which were included in the fit in the present study. The new data on $T$, $E$, $P$, $H$, and $G$ represent a vital addition to the $\gamma p\to\eta p$ data base as data on $P$ were very scarce before and data on $H$ and $G$ not available at all. Our fit results for those data are shown in Figs.~\ref{fig:TetapCBELSA2020} to \ref{fig:PGHetapCBELSA2020} in Appendix~\ref{app:furtherfitresults}. It is interesting to note that the BnGa group observes a very similar development for the $\eta N$ branching ratio of the $N(1650)1/2^-$ when fitting the data, as discussed in detail in Ref.~\cite{CBELSATAPS:2019ylw}.

Both $S_{11}$ states show an increased coupling to the $KY$ channels, referable to the newly included $\gamma p\to K\Sigma$ reaction. We see no indications for additional poles as, e.g, the $N(1895)1/2^-$ in this partial wave.

\begin{figure}
\begin{center}
\includegraphics[width=1.\linewidth]{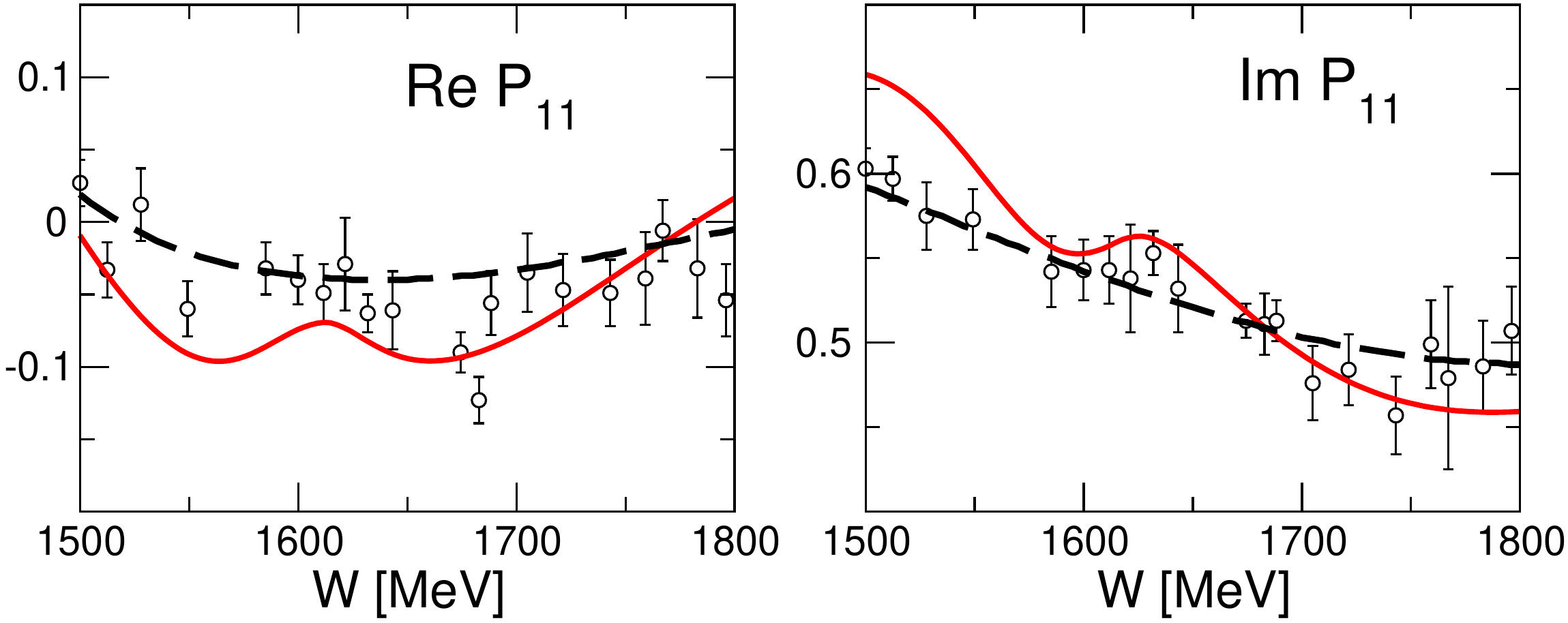} 
\end{center}
\caption{$\pi N\to\pi N$ $P_{11}$ partial wave in the region of the $N(1710)1/2^+$: Solid (red) lines: J\"uBo2022; data points: single-energy solution WI08 GWU/SAID~\cite{Workman:2012hx}; dashed (black) lines: energy-dependent solution WI08~\cite{Workman:2012hx}.
Remarkably, the J\"uBo2022 fit is to that (structureless) energy-dependent solution and blind to the single-energy solution.
}
\label{fig:P11piN}
\end{figure}

In the $P_{11}$ partial wave, we observe a significant change in the analytic structure compared to the J\"uBo2017 solution. While the Roper resonance is still dynamically generated as in previous studies~{\cite{Krehl:1999km}}, the $N(1710)1/2^+$, previously an $s$-channel pole, is now dynamically generated and moved to a much lower pole position close to the $K\Lambda$ threshold. We also observe a major increase of the residues into the $\eta N$ and $K\Lambda$ channels. In addition, the $N(1710)1/2^+$ resonance follows the pronounced structure of the SAID single energy solution (data points in Fig.~\ref{fig:P11piN}) although the fit (solid red lines) is to the energy-dependent solution (dashed lines) which is rather smooth in the shown energy region. This remarkable property was present in the J\"uBo2012 and J\"uBo2015 solutions~\cite{Ronchen:2012eg, Ronchen:2015vfa} but disappeared in the J\"uBo2017 solution~\cite{Ronchen:2018ury}, until the present study. We interpret this as a strong indication of existence for this resonance.

In contrast to the dynamical $N(1710)1/2^+$, the second $s$-channel pole in $P_{11}$ (apart from the nucleon), moved far into the complex plane  ($W_0=1513-i405$~MeV). Therefore, we do not include this singularity in the resonance tables. The dynamically generated state at $W_0\simeq 1750-i160$~MeV observed in previous J\"uBo solutions is not seen anymore.

Another change of the analytic structure took place in the $P_{33}$ partial wave. While the $\Delta(1232)$ is very stable with respect to previous J\"uBo studies~\cite{Ronchen:2018ury,Ronchen:2015vfa} and close to the PDG values, the formerly dynamically generated $\Delta(1600)3/2^+$ is now induced by a bare $s$-channel pole. In contrast, the very broad $\Delta(1920)3/2^+$, previously a $s$-channel pole, is dynamically generated in the present study.
The situation in the $P_{11}$ and $P_{33}$ partial wave shows the difficulties in interpreting the dynamical or $s$-channel (``genuine'') pole nature of a resonance in a complicated multi-channel environment with strong dressing effects that cannot be uniquely separated from ``bare'' states~\cite{Ronchen:2012eg, Doring:2009bi}.

\begin{figure}
\begin{center}
\includegraphics[width=1.\linewidth]{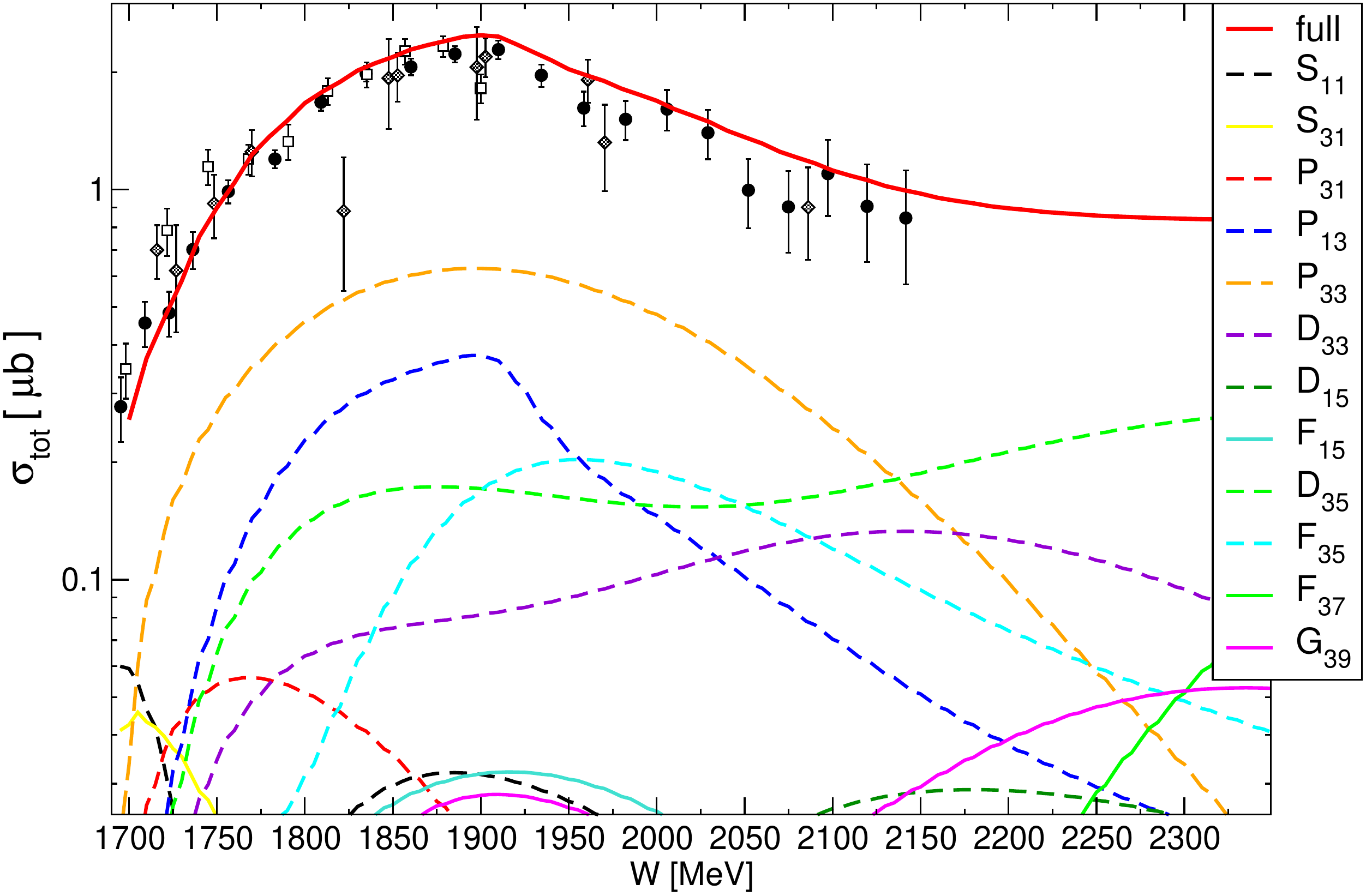} 
\end{center}
\caption{Total cross section of the reaction $\gamma p\to K^+\Sigma^0$ and its partial-wave content on a logarithmic scale (small partial waves not shown). Data: filled (black) circles: SAPHIR 1998~\cite{SAPHIR:1998fev}; empty squares: SAPHIR 1994~\cite{Bockhorst:1994jf}; filled (gray) diamonds: ABBHHM 1969~\cite{Aachen-Berlin-Bonn-Hamburg-Heidelberg-Muenchen:1969pjo}). The data are not included in the fit.}
\label{fig:tot_cs_k+s0}
\end{figure}

\begin{figure}
\begin{center}
\includegraphics[width=1.\linewidth]{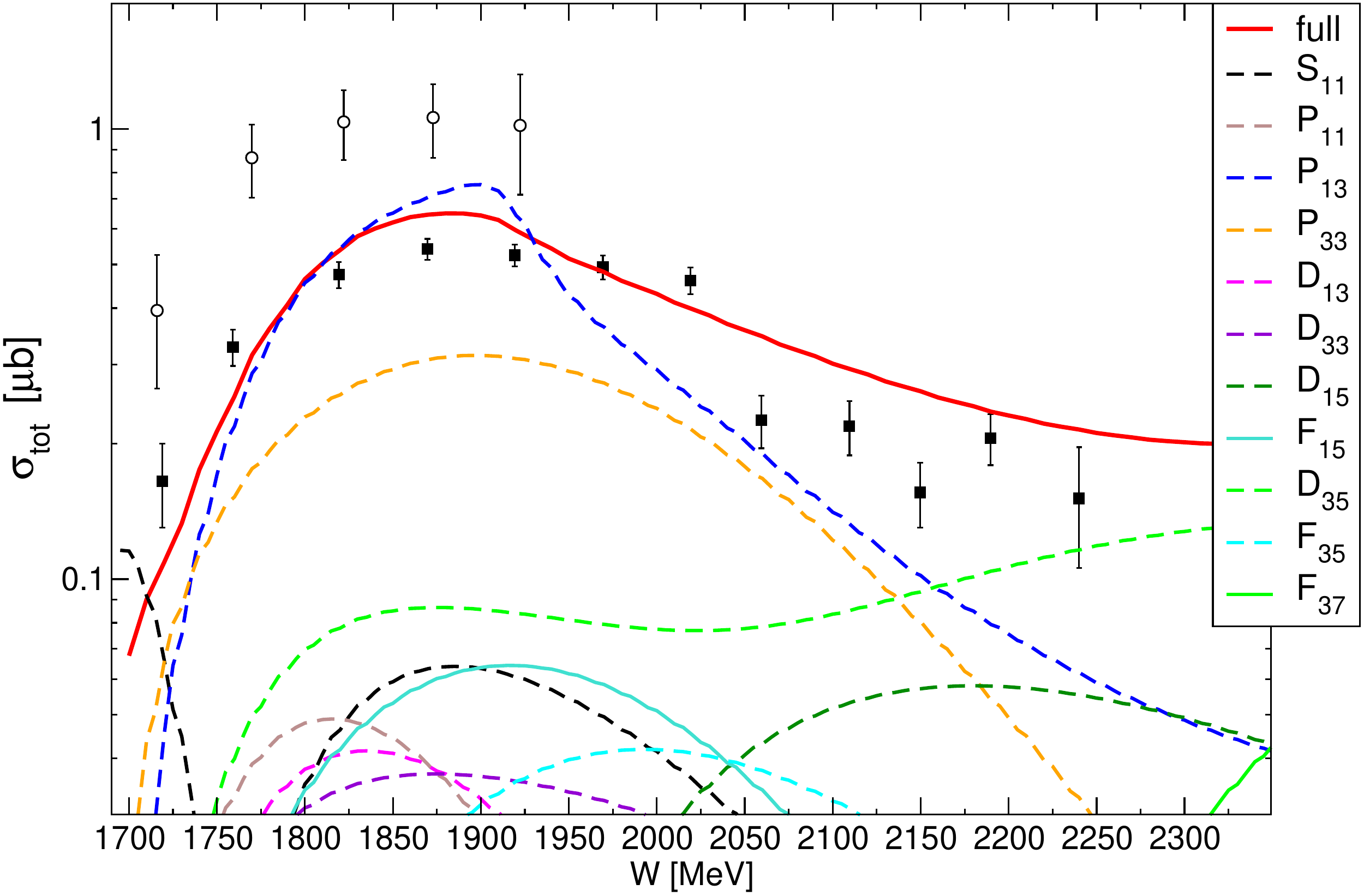} 
\end{center}
\caption{Total cross section of the reaction $\gamma p\to K^0\Sigma^+$ and its partial-wave content on a logarithmic scale (small partial waves not shown). Data: empty circles: SAPHIR 1999~\cite{SAPHIR:1999wfu}; filled (black) squares: CBELSA/TAPS 2012~\cite{CBELSATAPS:2011gly} (data are not included in the fit).}
\label{fig:tot_cs_k0s+}
\end{figure}

The $P_{33}$ wave is one of the dominating partial waves in the $\gamma p\to K\Sigma$ reactions, c.f. Figs.~\ref{fig:tot_cs_k+s0} and \ref{fig:tot_cs_k0s+}, reflected by the large photocoupling of the $\Delta(1920)3/2^+$. Of course, it is difficult to interpret this as a sign that this resonance exists, due to its enormous width of more than 800~MeV. It could well be that this state just appears from the specific way our amplitude is parametrized; all that is certain is that the pertaining partial wave, $P_{33}$, plays a major role in $K\Sigma$ photoproduction.

\begin{figure}
\begin{center}
\includegraphics[width=1.\linewidth]{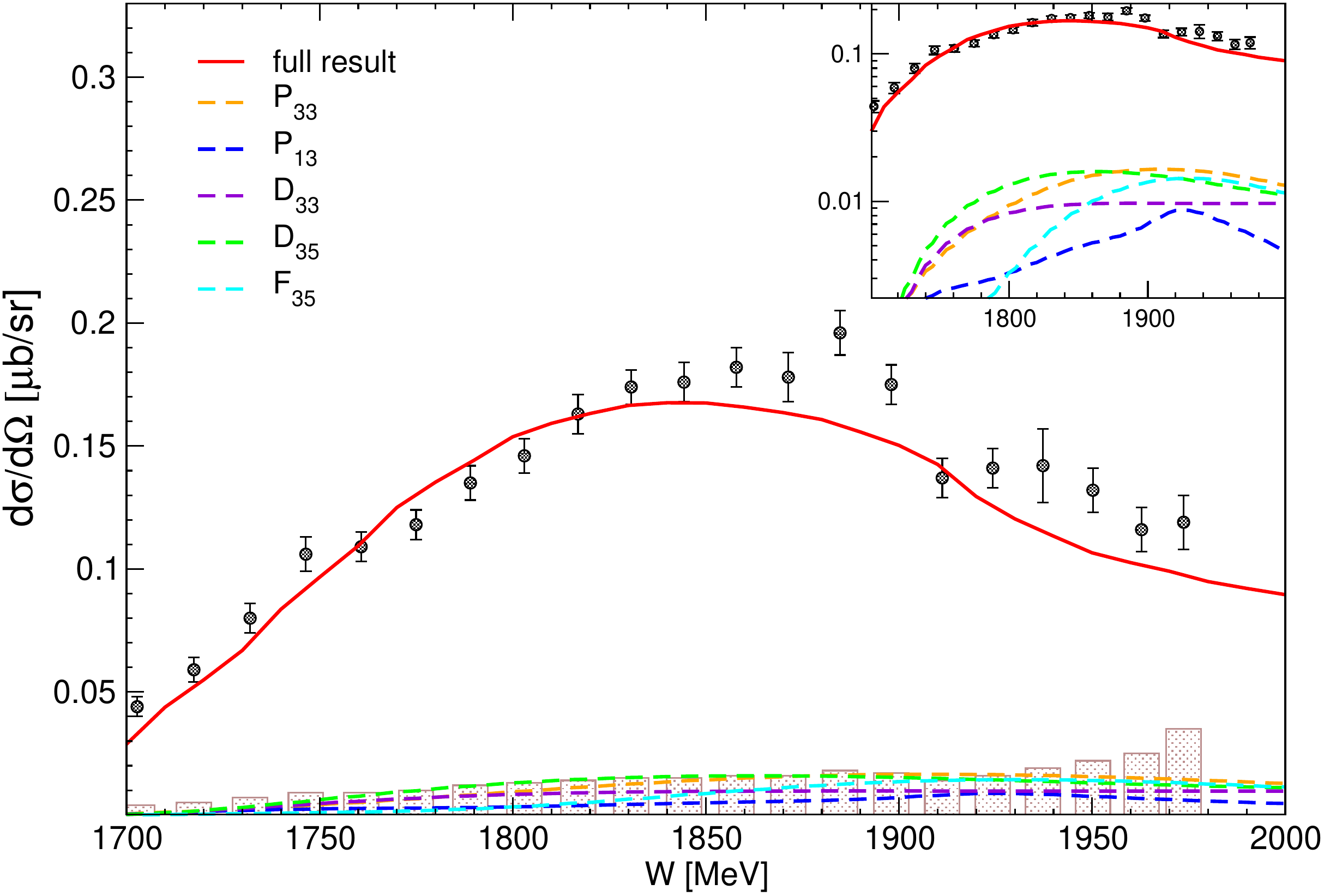} 
\end{center}
\caption{Forward differential cross section ($\cos\theta > 0.9$) reaction $\gamma p\to K^+\Sigma^0$. Only dominant partial waves are shown. The box in the upper right corner shows the same picture on a logarithmic scale.  Data from Ref.~\cite{Jude:2020byj} (not included in the fit). We fit the data shown in Fig.~\ref{fig:dsdok+s0BGOOD} from the same reference. Systematic uncertainties of the data are shown as brown bars. }
\label{fig:forwad_cs_k+s0}
\end{figure}

\begin{figure}
\begin{center}
\includegraphics[width=0.47\linewidth]{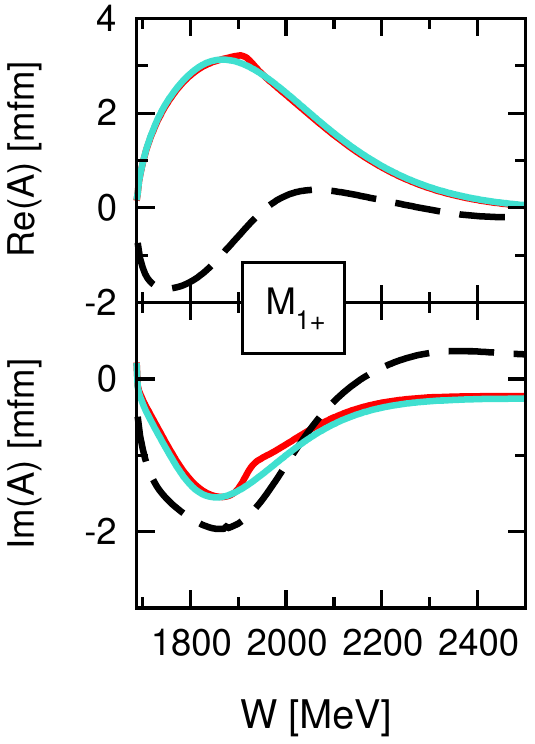} 
\includegraphics[width=0.425\linewidth]{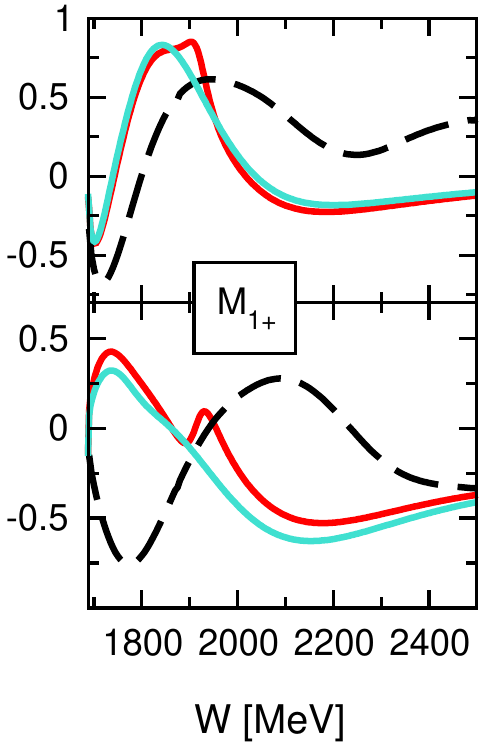} 
\end{center}
\caption{$M_{1+}$ multipole for the reaction $\gamma p\to K^+\Sigma^0$ (left) and $K^0\Sigma^+$ (right): (Red) solid lines: full solution J\"uBo2022. (Turquoise) solid lines: J\"uBo2022 with $N(1900)3/2^+$ switched off. (Black) dashed lines: BnGa2019~\cite{CBELSATAPS:2019ylw}. }
\label{fig:M1+_KS}
\end{figure}

The $P_{13}$ partial wave features two $s$-channel poles: the $N(1720)3/2^+$ and the $N(1900)3/2^+$, both rated with 4 stars by the PDG. It was noted by a number of other analysis groups that the latter state plays an important role in $KY$ photoproduction, see, e.g., Refs.~\cite{Anisovich:2017bsk,Cao:2013psa,Mart:2017mwj,Clymton:2021wof,Wei:2022nqp}. Also in the J\"uBo model the $N(1900)3/2^+$ was included to improve the fit to $K\Lambda$ photoproduction data~\cite{Ronchen:2018ury}.  The pole positions of the two $P_{13}$ states found in the present analysis are close to the PDG values. The width of the $N(1900)3/2^+$ is much reduced compared to previous J\"uBo studies. Although the residue into the $K\Sigma$ channel is considerably smaller than in 2017, the pole is clearly visible in the multipole amplitudes for $\gamma p\to K\Sigma$, especially in $M_{1+}$ of $K^0\Sigma^+$, c.f. Fig.~\ref{fig:M1+_KS}, where a distinct peak arises at the pole position of the $N(1900)3/2^+$. In the present analysis, this resonance is responsible for the drop of the total cross section in $\gamma p\to K^+\Sigma^0$ around $W\sim 1900$~MeV (c.f. Fig.~\ref{fig:tot_cs_k+s0}) and it also qualitatively explains the ``cusp-like" structure observed in a recent measurements of the forward differential cross section for $K^+\Sigma^0$ photoproduction by the BGOOD experiment~\cite{Jude:2020byj}, as can be seen in Fig.~\ref{fig:forwad_cs_k+s0}. This cusp in the data might be sharper at extreme forward angles~\cite{Jude:2020byj}, but our data description in forward direction is, overall, quite good except for a few points, see Fig.~\ref{fig:dsdok+s0BGOOD}.

Although this resonance is also responsible for a sharp drop of the $P_{13}$ contribution to the total cross section of the $K^0\Sigma^+$ final state as shown in Fig.~\ref{fig:tot_cs_k0s+}, it does not reproduce the sharp drop of the data around 2~GeV. Note that we do not include data on total cross section in the fit. A hypothesis for this drop was presented in Ref.~\cite{Ramos:2013wua} that explained it as a cusp from coupled $K^*\Lambda$ and $K^*\Sigma$ channels and their interference; a similar mechanism to explain the sharp structure at around $W\approx 1.68$~GeV in $\gamma n \to \eta n$ was discussed in Ref.~\cite{Doring:2009qr}. In any case, these channels alone cannot explain the differential cross sections, either~\cite{Ramos:2013wua}, while the forward peak in our solution is at least qualitatively described as Fig.~\ref{fig:dsdok0s+} shows. Once our approach includes $K^*Y$ channels, it will be possible to test the cusp hypothesis quantitatively with all available data.

\begin{figure}
\begin{center}
\includegraphics[width=1.\linewidth]{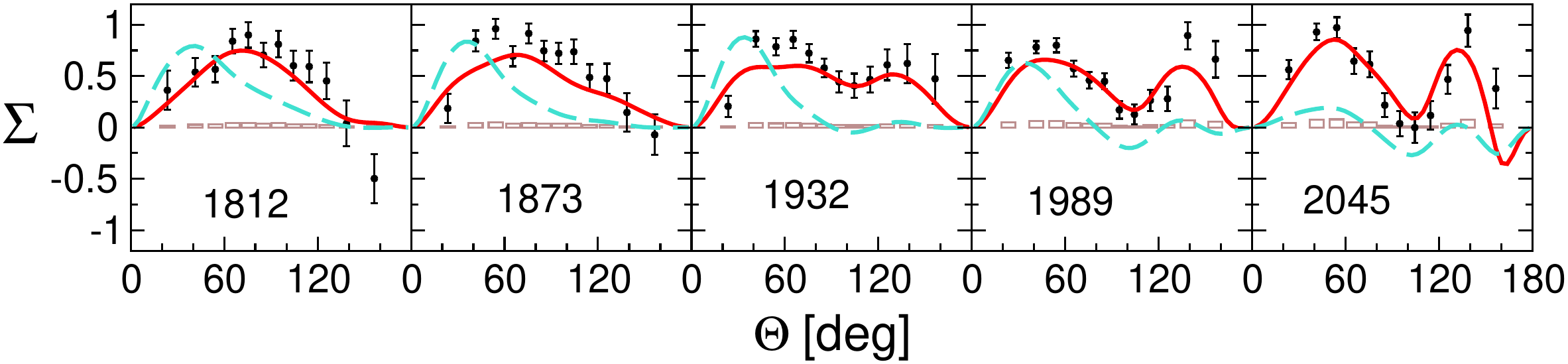} \end{center}
\caption{Fit results for the beam asymetry $\Sigma$ of the reaction $\gamma p\to \eta p$. Solid (red) lines: full results J\"uBo2022. Dashed (turquoise) lines: J\"uBo2022 with $N(1720)3/2^+$ and $N(1900)3/2^+$ switched off.  Only CBELSA/TAPS~\cite{CBELSATAPS:2020cwk} data are shown but other data~\cite{CLAS:2017rxe, GRAAL:2007gsc} are also included in the analysis. }
\label{fig:SetapCBLESA2019}
\end{figure}

On the other hand, in combination with the $N(1720)3/2^+$, which has a larger $\eta N$ residue, the $N(1900)3/2^+$ generates the backward peak in the recent beam asymmetry data for $\gamma p\to \eta p$ by the CBELSA/TAPS Collaboration~\cite{CBELSATAPS:2020cwk}, as shown in Fig.~\ref{fig:SetapCBLESA2019}: by switching off the two resonance states, the peak for large scattering angle, which occurs in the energy region of the $\eta^\prime p$ threshold of $W=1896$~MeV, vanishes. In Ref.~\cite{CBELSATAPS:2020cwk}, this backward peak is associated with the $\eta^\prime p$ and further evidence for the $N(1895)1/2^-$ is claimed based on a fit of the data within the BnGa approach. In the current analysis we do not find evidence for the latter state. Note that while the $\eta^\prime p$ channel is not yet included in J\"uBo, and therefore the $\eta^\prime p$ branch point is missing, the $N(1900)3/2^+$ does not simply simulate this branch point as this state is especially important in $KY$ photoproduction. Furthermore, also the BnGa analysis includes the $N(1900)3/2^+$, despite including the $\eta^\prime p$ explicitly.

As reported in Ref.~\cite{CBELSATAPS:2020cwk}, the BnGa solution introduced the $\eta' N$ channel not only to fit the $\eta$ beam asymmetry data, but also the $\eta$ differential cross sections from Ref.~\cite{A2:2017gwp}. We stress that in the present solution, the data from from both References~\cite{A2:2017gwp, CBELSATAPS:2020cwk} are also well described~\cite{Juelichmodel:online}. More definite statements can be made once the $\eta'$ channel is included in our analysis.

In the $D_{13}$ partial wave, besides the established $N(1520)3/2^-$, we see indications for a dynamically generated $N(1875)3/2^-$ at $W_0=1906(1)-i333(1)$~MeV with a strong coupling to the $\pi\Delta$ channel. While the mass is in agreement with the PDG value of $1900\pm 50$~MeV~\cite{Workman:2022ynf}, our width is much broader than the PDG value of $160\pm 60$~MeV. 

A broad dynamically generated pole is also observed in the $D_{33}$ partial wave at $W_0=2118(10)-i356(73)$~MeV, which could be associated with the $\Delta(1940)3/2^-$ listed by the PDG~\cite{Workman:2022ynf}, although our pole positions differs to some extent from the PDG value of $W_0=1950\pm 100-i(175\pm75)$~MeV. This state couples strongly to the $K\Sigma$ channel and, accordingly, around 2.1~GeV the influence of  the $D_{33}$ on the $\gamma p\to K^+\Sigma^0$ cross section is increasing and it becomes one of the most important partial waves, c.f. Fig.~\ref{fig:tot_cs_k+s0}. Because both states, $N(1875)3/2^-$  and $\Delta(1940)3/2^-$, need further confirmation we do not list them in Tabs.~\ref{tab:poles1} and \ref{tab:poles2}.

In the $D_{15}$ partial wave, the $N(1675)5/2^-$ couples strongly to the $\pi N$ channel and we find resonance parameters similar to the PDG values. The $N(2060)5/2^-$ that was dynamically generated in J\"uBo2017~\cite{Ronchen:2018ury} is not seen in the present study.

An interesting interplay can be observed in the $I=3/2$ and $J=5/2$ partial waves: The $\Delta(1930)5/2^-$ is much broader than in J\"uBo2017 while the width of the $\Delta(1905)5/2^+$ is significantly reduced. A similar switch of large and small width was observed in previous J\"uBo analyses. At that time, however, the important information from the $K\Sigma$ photoproduction data for the $I=3/2$ states was still missing. In the current study, the two partial waves, $D_{35}$ and $F_{35}$, play an important role in the description of those data especially for the $K^+\Sigma^0$ final state, as can be seen in Figs.~\ref{fig:tot_cs_k+s0} to \ref{fig:tot_cs_k0s+}, leading to modified pole positions compared to the J\"uBo2017 solution. 

In the $F_{17}$ partial wave, the pole position of the $N(1990)7/2^+$ changes considerably compared to J\"uBo2017. The mass is now much closer to the PDG value and the width of 72~MeV is much more narrow. This state is rated with only 2 stars by the PDG, meaning the evidence for existence is only fair. In our present analysis, however, the forward peak in the beam asymmetry data for $\gamma p\to\eta p$ is to a large extent caused by the $N(1990)7/2^+$.  We noticed in previous J\"uBo studies that the properties of the $N(1990)7/2^+$ are hard to determine. Here, we note that further studies are needed to confirm its remarkably small imaginary part.

At higher energies, also the $F_{37}$ and $G_{39}$ partial waves gain more influence in the $K\Sigma$ cross sections, although they remain comparably small. While the mass of the $\Delta(1950)7/2^+$ is in the same energy range as in 2017, the pole position of the $\Delta(2400)9/2^-$ changed significantly and is now closer to the PDG value. In the $G_{39}$ partial wave an additional pole at $W_0=1941(12)-i260(24)$~MeV is found that couples strongly to $\pi N$ and $\pi\Delta$. This pole is not listed in Tabs.~\ref{tab:photo} and \ref{tab:poles2} as further evidence for this state is required.  

On the whole, we find that the reaction $\gamma p\to K^+\Sigma^0$ is dominated by isospin $I=3/2$ resonances, with the important exception of the $P_{13}$ partial wave, as can be seen in Fig.~\ref{fig:tot_cs_k+s0}. Accordingly, the current resonance analysis leads to refined values for the $\Delta$ states. The dominance of $I=3/2$ partial waves in $\gamma p\to K^+\Sigma^0$ was also noted by the BnGa group in Ref.~\cite{Anisovich:2013vpa}, while the dominant contributions to $\gamma p\to K^0\Sigma^+$ in that study were nucleon partial waves. Looking at Fig.~\ref{fig:tot_cs_k0s+} we can confirm that $I=1/2$ partial waves play a much bigger role in the $K^0\Sigma^+$  than in the $K^+\Sigma^0$ final state.

Furthermore, we observe that the uncertainties of the $\Delta$ resonances are of a similar size as for the $N^*$ states. This was different in J\"uBo2017~\cite{Ronchen:2018ury}, where the mixed-isopspin $K\Sigma$ photoproduction channels were not yet included and the $\Delta$ states exhibited larger uncertainties than the $N^*$ states in general.


\section{Conclusion}

In the present study, the J\"ulich-Bonn dynamical coupled-channel model was  extended to $K\Sigma$ photoproduction off the proton and recent data sets for other meson photoproduction reactions were included. The approach now describes the pion- and photon-induced production of the $\pi N$, $\eta N$, $K\Lambda$ and $K\Sigma$ channels. The amplitudes were determined in a simultaneous fit of all reactions to almost {72,000} data points. Based on those fit results, the spectrum of $N^*$ and $\Delta$ resonances was extracted in terms of complex pole positions and residues. All 4-star resonances up to $J=9/2$, except for the $N(1895)1/2^+$, and a number of states rated with less than 4 stars are observed. In addition, we see indication for dynamically generated poles that were not seen in previous J\"uBo studies and require further confirmation. 

We find that the $\gamma p\to K^+\Sigma^0$ reaction is dominated by $I=3/2$ partial waves, with the exception of the $P_{13}$ wave. In that partial wave, the $N(1900)3/2^+$ state explains qualitatively the cusp-like structure in the recent BGOOD data~\cite{Jude:2020byj} for $K^+\Sigma^0$. Moreover, in combination with the $N(1720)3/2^+$ it is responsible for the backward peak observed in a recent CBELSA/TAPS measurement of the beam asymmetry $\Sigma$ in $\gamma p\to\eta p$~\cite{CBELSATAPS:2020cwk}. 
The explanation of a sharp decrease of the cross section in $K^0\Sigma^+$ at $W\approx 2$ GeV remains a challenge in the present approach.

Furthermore, the inclusion of the new polarization data $T$, $E$, $P$, $H$, and $G$ in $\gamma p\to\eta p$~\cite{CBELSATAPS:2019ylw} led to a significant change in the $\eta N$ residue of the $N(1650)1/2^-$, which is almost twice as large as in previous studies, reducing the striking difference of the $\eta N$ residue of the two $S_{11}$ states.
In many cases, the inclusion of the mixed isospin $\gamma p\to K\Sigma$ data led to refined values for the $\Delta$ resonances. 

\acknowledgments
We thank F. Afzal, R. Beck, J. Hartmann, D. Ireland, T. Jude, H. Schmieden, I. Strakovsky, U. Thoma, and R. Workman for useful discussion and for providing data.
The authors gratefully acknowledge the computing time granted by the JARA Vergabegremium and provided on the JARA Partition part of the supercomputer JURECA at Forschungszentrum J\"ulich.
This work is supported in part by DFG and NSFC through funds provided to the Sino-German CRC 110 ``Symmetry and the Emergence of Structure in QCD" (NSFC Grant No. 11621131001, DFG Grant No. TRR110), and by VolkswagenStiftung (grant No. 93562). M.D. is supported
by 
the U.S. Department of Energy, Office of Science, Office
of Nuclear Physics under grant No. DE-SC0016582 and contract DE-AC05-06OR23177. This material is based upon work supported by the National Science Foundation under Grant No. PHY 2012289.
The work of UGM was also supported by the CAS President's International Fellowship Initiative (PIFI) (Grant No. 2018DM0034).

\appendix

\section{$K\Sigma$ photoproduction multipoles}
\label{app:multipoles}

\begin{figure*}
\begin{center}
\includegraphics[width=1.\linewidth]{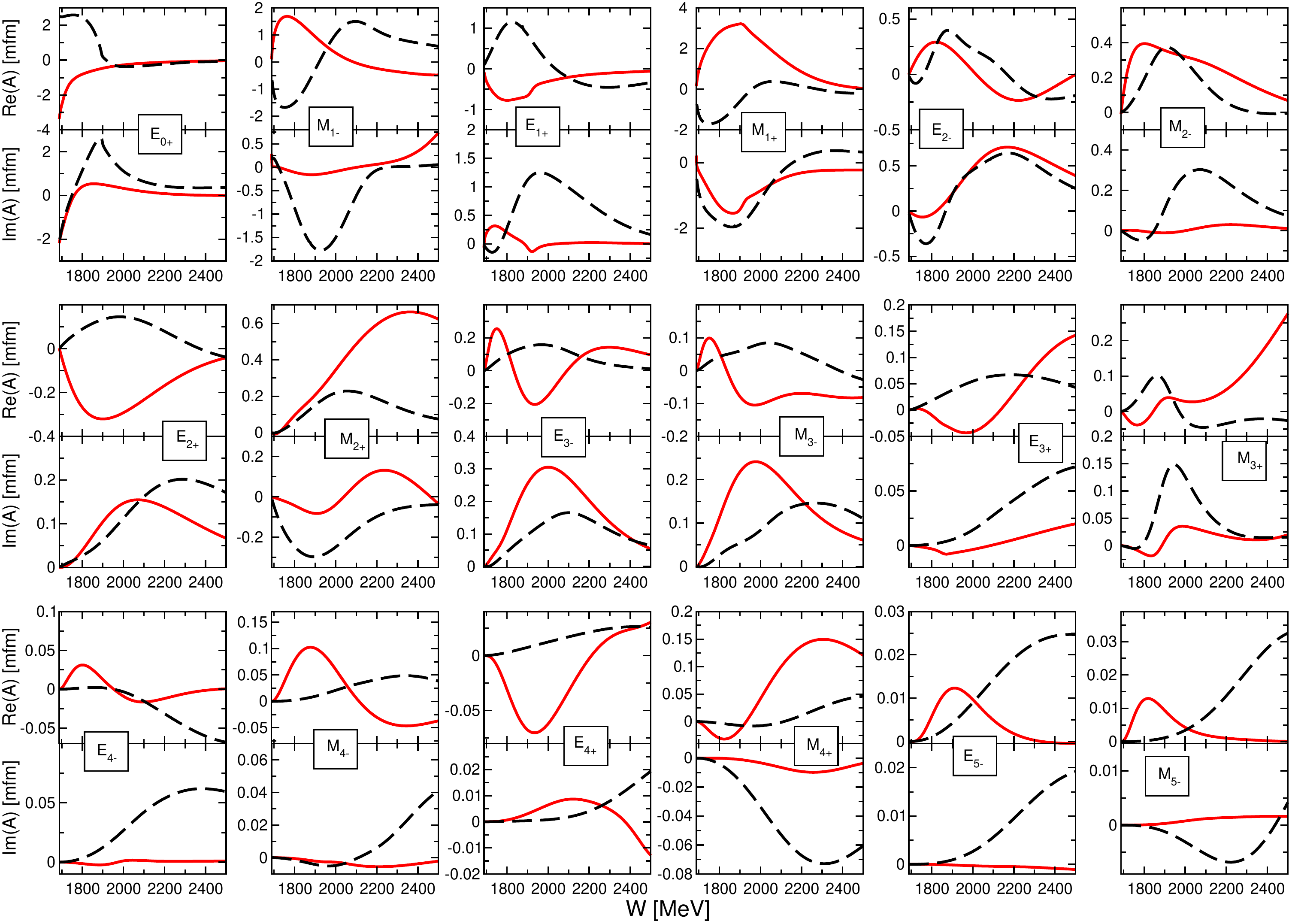} 
\end{center}
\caption{Electric and magnetic multipoles for the reaction $\gamma p\to K^+\Sigma^0$: (Red) solid lines: J\"uBo2022 (this solution). (Black) dashed lines: BnGa2019~\cite{CBELSATAPS:2019ylw}.}
\label{fig:mltpkps0}
\end{figure*}

\begin{figure*}
\begin{center}
\includegraphics[width=1.\linewidth]{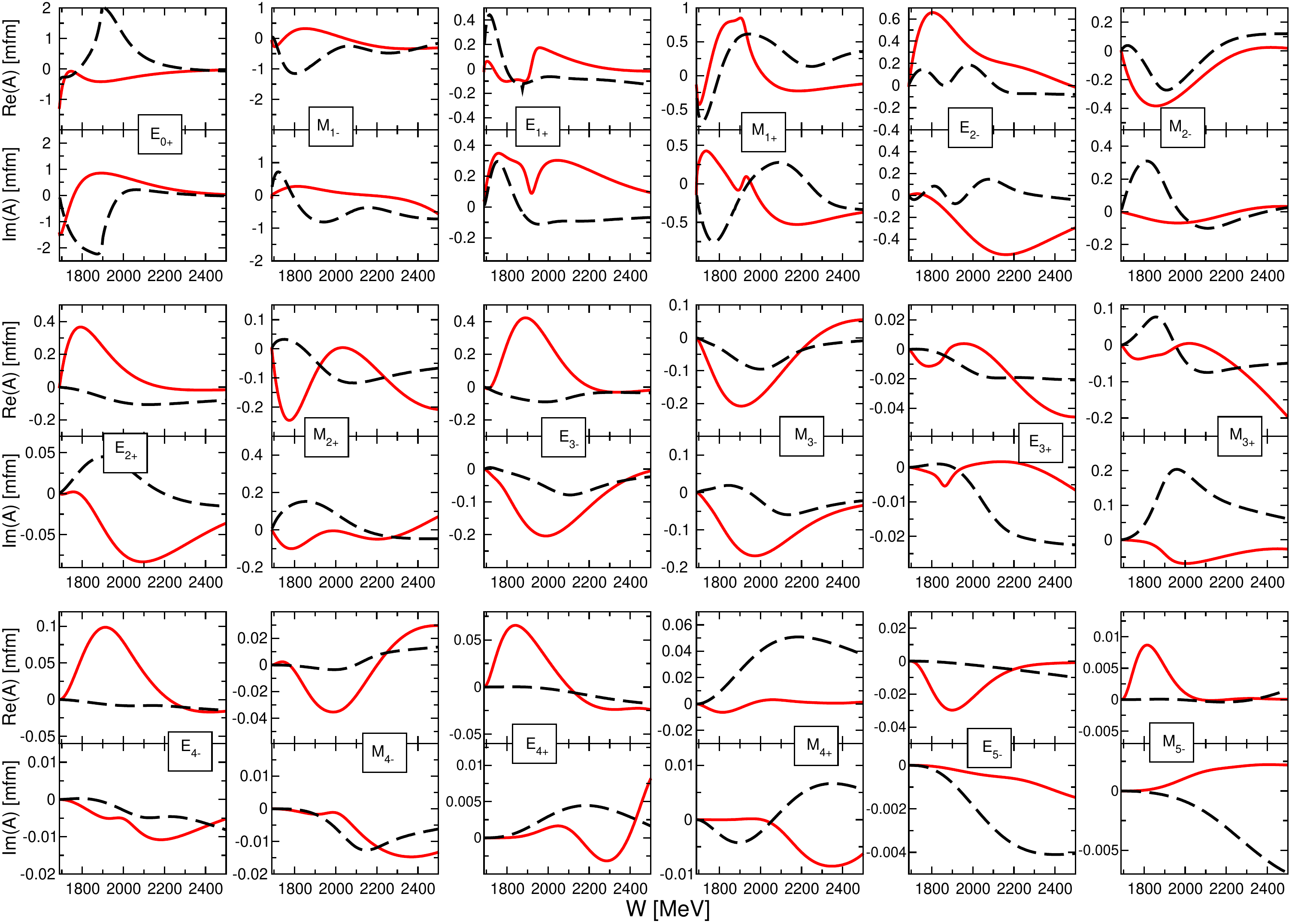} 
\end{center}
\caption{Electric and magnetic multipoles for the reaction $\gamma p\to K^0\Sigma^+$: (Red) solid lines: J\"uBo2022 (this solution). (Black) dashed lines: BnGa2019~\cite{CBELSATAPS:2019ylw}.}
\label{fig:mltpk0sp}
\end{figure*}

In Figs.~\ref{fig:mltpkps0} and \ref{fig:mltpk0sp} we show the multipoles for the reactions $\gamma p\to K^+\Sigma^0$ and $\gamma p\to K^0\Sigma^+$, respectively. The current solution J\"uBo2022 is shown together with the Bonn-Gatchina BnGa2019 solution~\cite{CBELSATAPS:2019ylw}.

\section{Further fit results}
\label{app:furtherfitresults}

\begin{figure}
\begin{center}
\includegraphics[width=1.\linewidth]{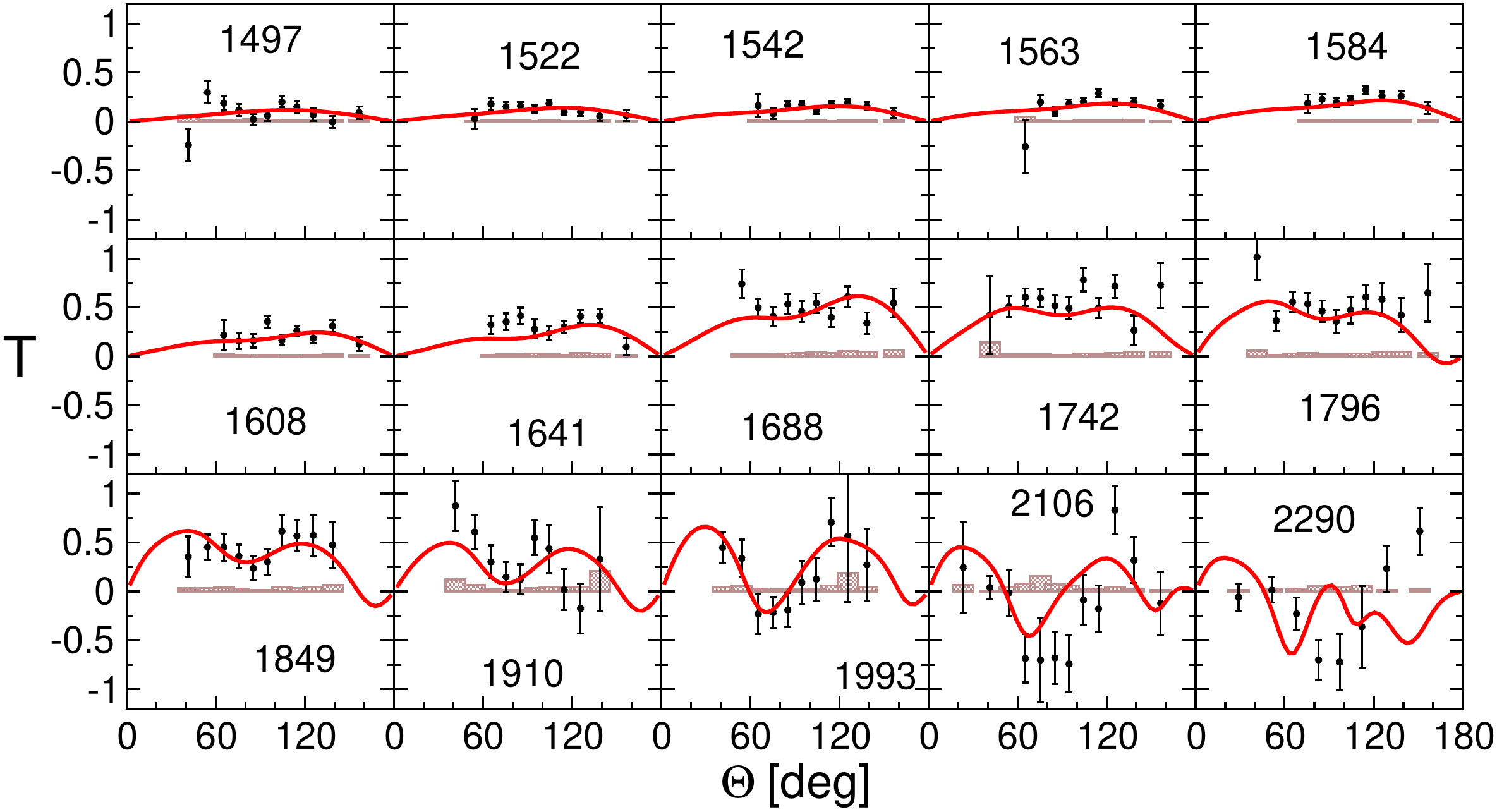} 
\end{center}
\caption{Fit results of $T$ for the reaction $\gamma p\to\eta p$. Data: CBELSA/TAPS 2020~\cite{CBELSATAPS:2019ylw}.}
\label{fig:TetapCBELSA2020}
\end{figure}

\begin{figure}
\begin{center}
\includegraphics[width=1.\linewidth]{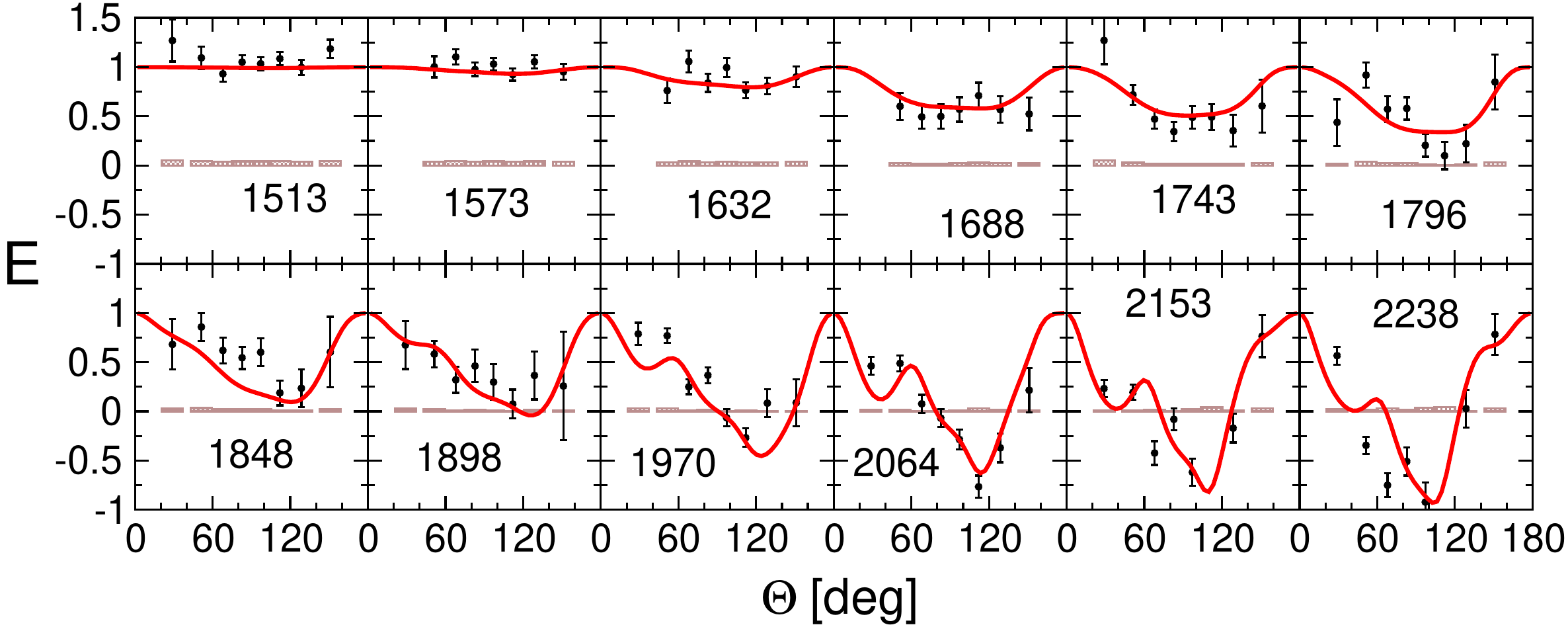} 
\end{center}
\caption{Fit results of $E$ for the reaction $\gamma p\to\eta p$. Data: CBELSA/TAPS 2020~\cite{CBELSATAPS:2019ylw}.}
\label{fig:EetapCBELSA2020}
\end{figure}

\begin{figure}
\begin{center}
\includegraphics[width=1.\linewidth]{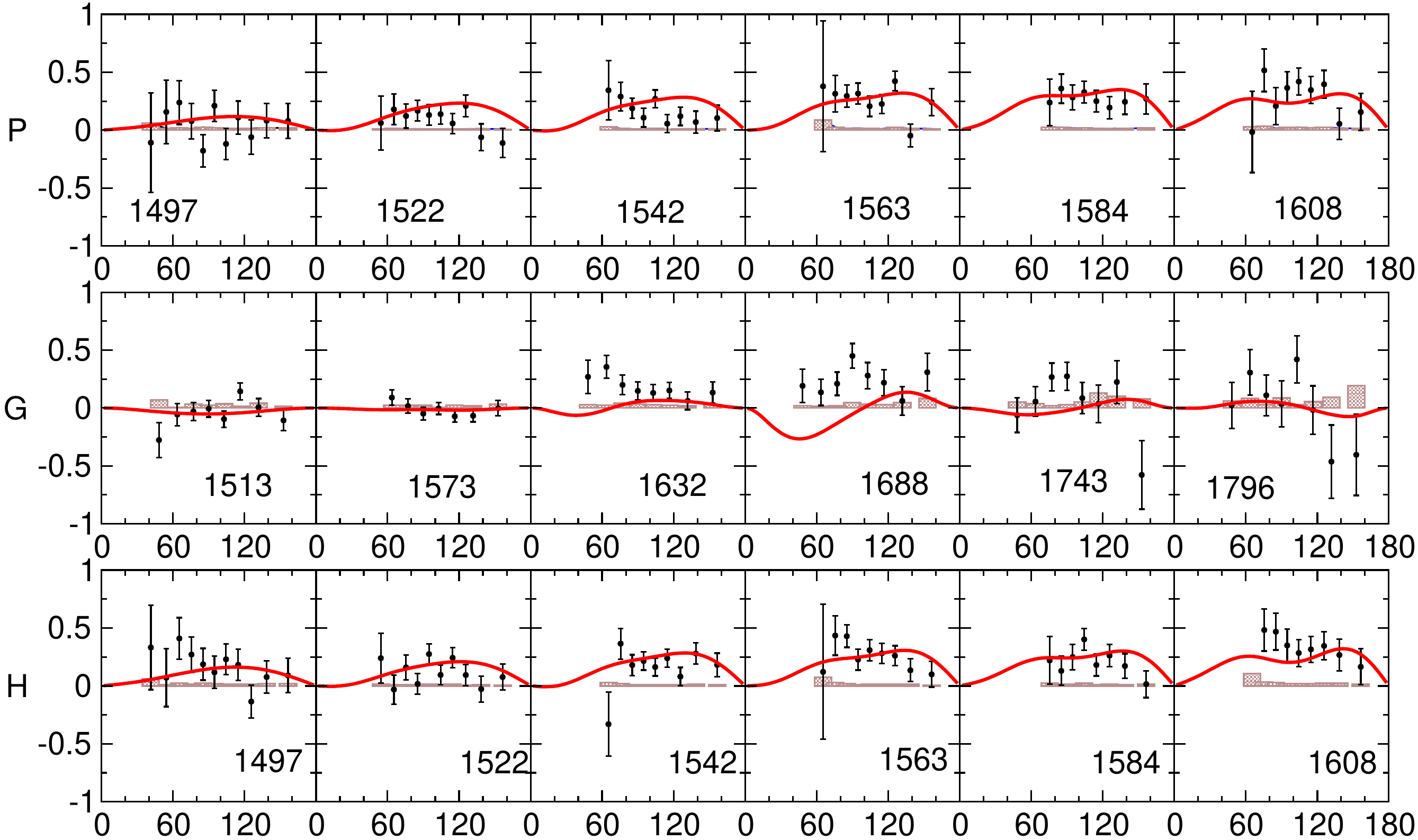} 
\end{center}
\caption{Fit results of $P$, $G$, and $H$ for the reaction $\gamma p\to\eta p$. Data: CBELSA/TAPS 2020~\cite{CBELSATAPS:2019ylw}.}
\label{fig:PGHetapCBELSA2020}
\end{figure}

In Figs.~\ref{fig:TetapCBELSA2020} to \ref{fig:PGHetapCBELSA2020} we show fit results for the polarization observables $T$, $E$, $P$, $G$, and $H$ in $\gamma p\to\eta p$ by the CBELSA/TAPS collaboration~\cite{CBELSATAPS:2019ylw}. Although this paper is about $K\Sigma$ photoproduction, the shown data were published between the previous iteration of the J\"uBo analysis and the present one and are now included in the fits. Also newly included are the MAMI A2 data for the differential cross section in $\gamma p\to \pi^0 p$ from Ref~\cite{A2:2015mhs} and in $\gamma p\to\eta p$ from Ref.~\cite{A2:2017gwp}, as well as the CBELSA/TAPS data on $E$ in $\gamma p\to\pi^0p$ from Ref.~\cite{CBELSATAPS:2019hhr} and on $\Sigma$ in $\gamma p\to\eta p$ from Ref.~\cite{CBELSATAPS:2020cwk} and the BGOOD data on $d\sigma/d\Omega$ and $P$ at forward angles for $K^+\Lambda$ photoproduction~\cite{Alef:2020yul}. The corresponding fit results can be found online~\cite{Juelichmodel:online}.

\section{Weights applied in the fit to the $\gamma p\to K^+\Sigma^0$, $K^0\Sigma^+$ data}
\label{app:weights}

The data weights in the $\chi^2$ calculation are shown in Table~\ref{tab:weightskps0}.

\begin{table*}[]
\caption{Weights applied in the fit to the $\gamma p\to K\Sigma$ data and energy ranges. }
\begin{center}
\renewcommand{\arraystretch}{1.4}
\begin {tabular}{l|lcc|ccc} 
\hline\hline
&Observable & Energy range [MeV] &\hspace{0.25cm} Weight \hspace{0.25cm} & \hspace{0.2cm} Observable \hspace{0.2cm} & Energy range [MeV] &\hspace{0.25cm} Weight \hspace{0.25cm}\\ \hline
$\gamma p\to K^+\Sigma^0\quad$&$d\sigma/d\Omega$ & $1695<W<1750$ & 4 & $\Sigma$ & $1755<W<2300$ & 8  \\
&				&$1750< W<1950$ & 32 & $\Sigma$~\cite{Paterson:2016vmc}& $1737<W<2170$  & 20 \\
&				&$1950<W<2355$ & 4 & &&\\ 
&$d\sigma/d\Omega$ of Ref.~\cite{Jude:2020byj} & $1688<W<1974$	&  40 \\			\cline{2-7}
&$P$ & $1706<W<1850$ & 40  &  $T$& $1737<W<1900$ &175\\
&	& $1850<W<2360$ & 20 & & $1900<W<2170$ & 35 \\ \cline{2-7}
&$C_{x^\prime}$ & $1787<W<2454$ &  70 &  $O_x$ & $1737<W<$2000 & 68  \\
&&& & &   $2000<W<$2170 & 136\\ \cline{2-7}
&  $C_{z^\prime}$ & $1787<W<2454$  & 72  &$O_z$ &$1737<W<2170$  & 65  \\ \hline
$\gamma p\to K^0\Sigma^+\quad$ &$d\sigma/d\Omega$ & $1690<W<2258$ & 50 &  $P$ & $1730<W<2460$ & 57 \\
&				&$1921< W<2020$, $\theta<66\deg$ & 250 & \\ \hline \hline
\end {tabular}
\end{center}
\label{tab:weightskps0}
\end{table*}

\bibliography{BIB}

\begin{thebibliography}{126}%
\makeatletter
\providecommand \@ifxundefined [1]{%
 \@ifx{#1\undefined}
}%
\providecommand \@ifnum [1]{%
 \ifnum #1\expandafter \@firstoftwo
 \else \expandafter \@secondoftwo
 \fi
}%
\providecommand \@ifx [1]{%
 \ifx #1\expandafter \@firstoftwo
 \else \expandafter \@secondoftwo
 \fi
}%
\providecommand \natexlab [1]{#1}%
\providecommand \enquote  [1]{``#1''}%
\providecommand \bibnamefont  [1]{#1}%
\providecommand \bibfnamefont [1]{#1}%
\providecommand \citenamefont [1]{#1}%
\providecommand \href@noop [0]{\@secondoftwo}%
\providecommand \href [0]{\begingroup \@sanitize@url \@href}%
\providecommand \@href[1]{\@@startlink{#1}\@@href}%
\providecommand \@@href[1]{\endgroup#1\@@endlink}%
\providecommand \@sanitize@url [0]{\catcode `\\12\catcode `\$12\catcode
  `\&12\catcode `\#12\catcode `\^12\catcode `\_12\catcode `\%12\relax}%
\providecommand \@@startlink[1]{}%
\providecommand \@@endlink[0]{}%
\providecommand \url  [0]{\begingroup\@sanitize@url \@url }%
\providecommand \@url [1]{\endgroup\@href {#1}{\urlprefix }}%
\providecommand \urlprefix  [0]{URL }%
\providecommand \Eprint [0]{\href }%
\providecommand \doibase [0]{http://dx.doi.org/}%
\providecommand \selectlanguage [0]{\@gobble}%
\providecommand \bibinfo  [0]{\@secondoftwo}%
\providecommand \bibfield  [0]{\@secondoftwo}%
\providecommand \translation [1]{[#1]}%
\providecommand \BibitemOpen [0]{}%
\providecommand \bibitemStop [0]{}%
\providecommand \bibitemNoStop [0]{.\EOS\space}%
\providecommand \EOS [0]{\spacefactor3000\relax}%
\providecommand \BibitemShut  [1]{\csname bibitem#1\endcsname}%
\let\auto@bib@innerbib\@empty
\bibitem [{\citenamefont {Capstick}\ and\ \citenamefont
  {Roberts}(1994)}]{Capstick:1993kb}%
  \BibitemOpen
  \bibfield  {author} {\bibinfo {author} {\bibfnamefont {Simon}\ \bibnamefont
  {Capstick}}\ and\ \bibinfo {author} {\bibfnamefont {Winston}\ \bibnamefont
  {Roberts}},\ }\bibfield  {title} {\enquote {\bibinfo {title} {{Quasi two-body
  decays of nonstrange baryons}},}\ }\href {\doibase 10.1103/PhysRevD.49.4570}
  {\bibfield  {journal} {\bibinfo  {journal} {Phys. Rev. D}\ }\textbf {\bibinfo
  {volume} {49}},\ \bibinfo {pages} {4570--4586} (\bibinfo {year} {1994})},\
  \Eprint {http://arxiv.org/abs/nucl-th/9310030} {arXiv:nucl-th/9310030}
  \BibitemShut {NoStop}%
\bibitem [{\citenamefont {Ronniger}\ and\ \citenamefont
  {Metsch}(2011)}]{Ronniger:2011td}%
  \BibitemOpen
  \bibfield  {author} {\bibinfo {author} {\bibfnamefont {M.}~\bibnamefont
  {Ronniger}}\ and\ \bibinfo {author} {\bibfnamefont {B.~C.}\ \bibnamefont
  {Metsch}},\ }\bibfield  {title} {\enquote {\bibinfo {title} {{Effects of a
  spin-flavour dependent interaction on the baryon mass spectrum}},}\ }\href
  {\doibase 10.1140/epja/i2011-11162-8} {\bibfield  {journal} {\bibinfo
  {journal} {Eur. Phys. J. A}\ }\textbf {\bibinfo {volume} {47}},\ \bibinfo
  {pages} {162} (\bibinfo {year} {2011})},\ \Eprint
  {http://arxiv.org/abs/1111.3835} {arXiv:1111.3835 [hep-ph]} \BibitemShut
  {NoStop}%
\bibitem [{\citenamefont {Qin}\ \emph {et~al.}(2019)\citenamefont {Qin},
  \citenamefont {Roberts},\ and\ \citenamefont {Schmidt}}]{Qin:2019hgk}%
  \BibitemOpen
  \bibfield  {author} {\bibinfo {author} {\bibfnamefont {Si-xue}\ \bibnamefont
  {Qin}}, \bibinfo {author} {\bibfnamefont {Craig~D}\ \bibnamefont {Roberts}},
  \ and\ \bibinfo {author} {\bibfnamefont {Sebastian~M}\ \bibnamefont
  {Schmidt}},\ }\bibfield  {title} {\enquote {\bibinfo {title} {{Spectrum of
  light- and heavy-baryons}},}\ }\href {\doibase 10.1007/s00601-019-1488-x}
  {\bibfield  {journal} {\bibinfo  {journal} {Few Body Syst.}\ }\textbf
  {\bibinfo {volume} {60}},\ \bibinfo {pages} {26} (\bibinfo {year} {2019})},\
  \Eprint {http://arxiv.org/abs/1902.00026} {arXiv:1902.00026 [nucl-th]}
  \BibitemShut {NoStop}%
\bibitem [{\citenamefont {Eichmann}\ \emph {et~al.}(2016)\citenamefont
  {Eichmann}, \citenamefont {Fischer},\ and\ \citenamefont
  {Sanchis-Alepuz}}]{Eichmann:2016hgl}%
  \BibitemOpen
  \bibfield  {author} {\bibinfo {author} {\bibfnamefont {Gernot}\ \bibnamefont
  {Eichmann}}, \bibinfo {author} {\bibfnamefont {Christian~S.}\ \bibnamefont
  {Fischer}}, \ and\ \bibinfo {author} {\bibfnamefont {Helios}\ \bibnamefont
  {Sanchis-Alepuz}},\ }\bibfield  {title} {\enquote {\bibinfo {title} {{Light
  baryons and their excitations}},}\ }\href {\doibase
  10.1103/PhysRevD.94.094033} {\bibfield  {journal} {\bibinfo  {journal} {Phys.
  Rev. D}\ }\textbf {\bibinfo {volume} {94}},\ \bibinfo {pages} {094033}
  (\bibinfo {year} {2016})},\ \Eprint {http://arxiv.org/abs/1607.05748}
  {arXiv:1607.05748 [hep-ph]} \BibitemShut {NoStop}%
\bibitem [{\citenamefont {Edwards}\ \emph {et~al.}(2013)\citenamefont
  {Edwards}, \citenamefont {Mathur}, \citenamefont {Richards},\ and\
  \citenamefont {Wallace}}]{Edwards:2012fx}%
  \BibitemOpen
  \bibfield  {author} {\bibinfo {author} {\bibfnamefont {Robert~G.}\
  \bibnamefont {Edwards}}, \bibinfo {author} {\bibfnamefont {Nilmani}\
  \bibnamefont {Mathur}}, \bibinfo {author} {\bibfnamefont {David~G.}\
  \bibnamefont {Richards}}, \ and\ \bibinfo {author} {\bibfnamefont
  {Stephen~J.}\ \bibnamefont {Wallace}} (\bibinfo {collaboration} {Hadron
  Spectrum}),\ }\bibfield  {title} {\enquote {\bibinfo {title} {{Flavor
  structure of the excited baryon spectra from lattice QCD}},}\ }\href
  {\doibase 10.1103/PhysRevD.87.054506} {\bibfield  {journal} {\bibinfo
  {journal} {Phys. Rev. D}\ }\textbf {\bibinfo {volume} {87}},\ \bibinfo
  {pages} {054506} (\bibinfo {year} {2013})},\ \Eprint
  {http://arxiv.org/abs/1212.5236} {arXiv:1212.5236 [hep-ph]} \BibitemShut
  {NoStop}%
\bibitem [{\citenamefont {Engel}\ \emph {et~al.}(2013)\citenamefont {Engel},
  \citenamefont {Lang}, \citenamefont {Mohler},\ and\ \citenamefont
  {Sch\"afer}}]{Engel:2013ig}%
  \BibitemOpen
  \bibfield  {author} {\bibinfo {author} {\bibfnamefont {Georg~P.}\
  \bibnamefont {Engel}}, \bibinfo {author} {\bibfnamefont {C.~B.}\ \bibnamefont
  {Lang}}, \bibinfo {author} {\bibfnamefont {Daniel}\ \bibnamefont {Mohler}}, \
  and\ \bibinfo {author} {\bibfnamefont {Andreas}\ \bibnamefont {Sch\"afer}}
  (\bibinfo {collaboration} {BGR}),\ }\bibfield  {title} {\enquote {\bibinfo
  {title} {{QCD with Two Light Dynamical Chirally Improved Quarks: Baryons}},}\
  }\href {\doibase 10.1103/PhysRevD.87.074504} {\bibfield  {journal} {\bibinfo
  {journal} {Phys. Rev. D}\ }\textbf {\bibinfo {volume} {87}},\ \bibinfo
  {pages} {074504} (\bibinfo {year} {2013})},\ \Eprint
  {http://arxiv.org/abs/1301.4318} {arXiv:1301.4318 [hep-lat]} \BibitemShut
  {NoStop}%
\bibitem [{\citenamefont {Lang}\ \emph {et~al.}(2017)\citenamefont {Lang},
  \citenamefont {Leskovec}, \citenamefont {Padmanath},\ and\ \citenamefont
  {Prelovsek}}]{Lang:2016hnn}%
  \BibitemOpen
  \bibfield  {author} {\bibinfo {author} {\bibfnamefont {C.~B.}\ \bibnamefont
  {Lang}}, \bibinfo {author} {\bibfnamefont {L.}~\bibnamefont {Leskovec}},
  \bibinfo {author} {\bibfnamefont {M.}~\bibnamefont {Padmanath}}, \ and\
  \bibinfo {author} {\bibfnamefont {S.}~\bibnamefont {Prelovsek}},\ }\bibfield
  {title} {\enquote {\bibinfo {title} {{Pion-nucleon scattering in the Roper
  channel from lattice QCD}},}\ }\href {\doibase 10.1103/PhysRevD.95.014510}
  {\bibfield  {journal} {\bibinfo  {journal} {Phys. Rev. D}\ }\textbf {\bibinfo
  {volume} {95}},\ \bibinfo {pages} {014510} (\bibinfo {year} {2017})},\
  \Eprint {http://arxiv.org/abs/1610.01422} {arXiv:1610.01422 [hep-lat]}
  \BibitemShut {NoStop}%
\bibitem [{\citenamefont {Andersen}\ \emph {et~al.}(2018)\citenamefont
  {Andersen}, \citenamefont {Bulava}, \citenamefont {H\"orz},\ and\
  \citenamefont {Morningstar}}]{Andersen:2017una}%
  \BibitemOpen
  \bibfield  {author} {\bibinfo {author} {\bibfnamefont {Christian~Walther}\
  \bibnamefont {Andersen}}, \bibinfo {author} {\bibfnamefont {John}\
  \bibnamefont {Bulava}}, \bibinfo {author} {\bibfnamefont {Ben}\ \bibnamefont
  {H\"orz}}, \ and\ \bibinfo {author} {\bibfnamefont {Colin}\ \bibnamefont
  {Morningstar}},\ }\bibfield  {title} {\enquote {\bibinfo {title} {{Elastic
  $I=3/2$ p-wave nucleon-pion scattering amplitude and the $\Delta$(1232)
  resonance from N$_f$=2+1 lattice QCD}},}\ }\href {\doibase
  10.1103/PhysRevD.97.014506} {\bibfield  {journal} {\bibinfo  {journal} {Phys.
  Rev. D}\ }\textbf {\bibinfo {volume} {97}},\ \bibinfo {pages} {014506}
  (\bibinfo {year} {2018})},\ \Eprint {http://arxiv.org/abs/1710.01557}
  {arXiv:1710.01557 [hep-lat]} \BibitemShut {NoStop}%
\bibitem [{\citenamefont {Silvi}\ \emph {et~al.}(2021)\citenamefont {Silvi}
  \emph {et~al.}}]{Silvi:2021uya}%
  \BibitemOpen
  \bibfield  {author} {\bibinfo {author} {\bibfnamefont {Giorgio}\ \bibnamefont
  {Silvi}} \emph {et~al.},\ }\bibfield  {title} {\enquote {\bibinfo {title}
  {{$P$-wave nucleon-pion scattering amplitude in the $\Delta$(1232) channel
  from lattice QCD}},}\ }\href {\doibase 10.1103/PhysRevD.103.094508}
  {\bibfield  {journal} {\bibinfo  {journal} {Phys. Rev. D}\ }\textbf {\bibinfo
  {volume} {103}},\ \bibinfo {pages} {094508} (\bibinfo {year} {2021})},\
  \Eprint {http://arxiv.org/abs/2101.00689} {arXiv:2101.00689 [hep-lat]}
  \BibitemShut {NoStop}%
\bibitem [{\citenamefont {H{\"o}hler}(1983)}]{Hoehler1}%
  \BibitemOpen
  \bibfield  {author} {\bibinfo {author} {\bibfnamefont {G.}~\bibnamefont
  {H{\"o}hler}},\ }\href@noop {} {\emph {\bibinfo {title} {Pion Nucleon
  Scattering}}},\ edited by\ \bibinfo {editor} {\bibfnamefont {H.}~\bibnamefont
  {Schopper}},\ Landolt-B{\"o}rnstein\ (\bibinfo  {publisher} {Springer},\
  \bibinfo {address} {New York, NY},\ \bibinfo {year} {1983})\BibitemShut
  {NoStop}%
\bibitem [{\citenamefont {Cutkosky}\ \emph {et~al.}(1979)\citenamefont
  {Cutkosky}, \citenamefont {Forsyth}, \citenamefont {Hendrick},\ and\
  \citenamefont {Kelly}}]{Cutkosky:1979fy}%
  \BibitemOpen
  \bibfield  {author} {\bibinfo {author} {\bibfnamefont {R.~E.}\ \bibnamefont
  {Cutkosky}}, \bibinfo {author} {\bibfnamefont {C.~P.}\ \bibnamefont
  {Forsyth}}, \bibinfo {author} {\bibfnamefont {R.~E.}\ \bibnamefont
  {Hendrick}}, \ and\ \bibinfo {author} {\bibfnamefont {R.~L.}\ \bibnamefont
  {Kelly}},\ }\bibfield  {title} {\enquote {\bibinfo {title} {{Pion - Nucleon
  Partial Wave Amplitudes}},}\ }\href {\doibase 10.1103/PhysRevD.20.2839}
  {\bibfield  {journal} {\bibinfo  {journal} {Phys. Rev. D}\ }\textbf {\bibinfo
  {volume} {20}},\ \bibinfo {pages} {2839} (\bibinfo {year}
  {1979})}\BibitemShut {NoStop}%
\bibitem [{\citenamefont {Arndt}\ \emph {et~al.}(2006)\citenamefont {Arndt},
  \citenamefont {Briscoe}, \citenamefont {Strakovsky},\ and\ \citenamefont
  {Workman}}]{Arndt:2006bf}%
  \BibitemOpen
  \bibfield  {author} {\bibinfo {author} {\bibfnamefont {R.~A.}\ \bibnamefont
  {Arndt}}, \bibinfo {author} {\bibfnamefont {W.~J.}\ \bibnamefont {Briscoe}},
  \bibinfo {author} {\bibfnamefont {I.~I.}\ \bibnamefont {Strakovsky}}, \ and\
  \bibinfo {author} {\bibfnamefont {R.~L.}\ \bibnamefont {Workman}},\
  }\bibfield  {title} {\enquote {\bibinfo {title} {{Extended partial-wave
  analysis of $\pi N$ scattering data}},}\ }\href {\doibase
  10.1103/PhysRevC.74.045205} {\bibfield  {journal} {\bibinfo  {journal} {Phys.
  Rev. C}\ }\textbf {\bibinfo {volume} {74}},\ \bibinfo {pages} {045205}
  (\bibinfo {year} {2006})},\ \Eprint {http://arxiv.org/abs/nucl-th/0605082}
  {arXiv:nucl-th/0605082} \BibitemShut {NoStop}%
\bibitem [{\citenamefont {Mai}\ \emph {et~al.}(2022)\citenamefont {Mai},
  \citenamefont {Mei\ss{}ner},\ and\ \citenamefont {Urbach}}]{Mai:2022eur}%
  \BibitemOpen
  \bibfield  {author} {\bibinfo {author} {\bibfnamefont {Maxim}\ \bibnamefont
  {Mai}}, \bibinfo {author} {\bibfnamefont {U.-G.}\ \bibnamefont
  {Mei\ss{}ner}}, \ and\ \bibinfo {author} {\bibfnamefont {Carsten}\
  \bibnamefont {Urbach}},\ }\bibfield  {title} {\enquote {\bibinfo {title}
  {{Towards a theory of hadron resonances}},}\ }\href@noop {} {\  (\bibinfo
  {year} {2022})},\ \Eprint {http://arxiv.org/abs/2206.01477} {arXiv:2206.01477
  [hep-ph]} \BibitemShut {NoStop}%
\bibitem [{\citenamefont {Workman}(2022)}]{Workman:2022ynf}%
  \BibitemOpen
  \bibfield  {author} {\bibinfo {author} {\bibfnamefont {R.~L.}\ \bibnamefont
  {Workman}} (\bibinfo {collaboration} {Particle Data Group}),\ }\bibfield
  {title} {\enquote {\bibinfo {title} {{Review of Particle Physics}},}\
  }\href@noop {} {\bibfield  {journal} {\bibinfo  {journal} {PTEP}\ }\textbf
  {\bibinfo {volume} {2022}},\ \bibinfo {pages} {083C01} (\bibinfo {year}
  {2022})}\BibitemShut {NoStop}%
\bibitem [{\citenamefont {Rönchen}\ \emph {et~al.}(2013)\citenamefont
  {Rönchen}, \citenamefont {Döring}, \citenamefont {Huang}, \citenamefont
  {Haberzettl}, \citenamefont {Haidenbauer}, \citenamefont {Hanhart},
  \citenamefont {Krewald}, \citenamefont {Mei{\ss}ner},\ and\ \citenamefont
  {Nakayama}}]{Ronchen:2012eg}%
  \BibitemOpen
  \bibfield  {author} {\bibinfo {author} {\bibfnamefont {D.}~\bibnamefont
  {Rönchen}}, \bibinfo {author} {\bibfnamefont {M.}~\bibnamefont {Döring}},
  \bibinfo {author} {\bibfnamefont {F.}~\bibnamefont {Huang}}, \bibinfo
  {author} {\bibfnamefont {H.}~\bibnamefont {Haberzettl}}, \bibinfo {author}
  {\bibfnamefont {J.}~\bibnamefont {Haidenbauer}}, \bibinfo {author}
  {\bibfnamefont {C.}~\bibnamefont {Hanhart}}, \bibinfo {author} {\bibfnamefont
  {S.}~\bibnamefont {Krewald}}, \bibinfo {author} {\bibfnamefont {U.-G.}\
  \bibnamefont {Mei{\ss}ner}}, \ and\ \bibinfo {author} {\bibfnamefont
  {K.}~\bibnamefont {Nakayama}},\ }\bibfield  {title} {\enquote {\bibinfo
  {title} {{Coupled-channel dynamics in the reactions piN --\ensuremath{>} piN,
  etaN, KLambda, KSigma}},}\ }\href {\doibase 10.1140/epja/i2013-13044-5}
  {\bibfield  {journal} {\bibinfo  {journal} {Eur. Phys. J. A}\ }\textbf
  {\bibinfo {volume} {49}},\ \bibinfo {pages} {44} (\bibinfo {year} {2013})},\
  \Eprint {http://arxiv.org/abs/1211.6998} {arXiv:1211.6998 [nucl-th]}
  \BibitemShut {NoStop}%
\bibitem [{\citenamefont {Anisovich}\ \emph {et~al.}(2013)\citenamefont
  {Anisovich}, \citenamefont {Klempt}, \citenamefont {Nikonov}, \citenamefont
  {Sarantsev},\ and\ \citenamefont {Thoma}}]{Anisovich:2013vpa}%
  \BibitemOpen
  \bibfield  {author} {\bibinfo {author} {\bibfnamefont {A.~V.}\ \bibnamefont
  {Anisovich}}, \bibinfo {author} {\bibfnamefont {E.}~\bibnamefont {Klempt}},
  \bibinfo {author} {\bibfnamefont {V.~A.}\ \bibnamefont {Nikonov}}, \bibinfo
  {author} {\bibfnamefont {A.~V.}\ \bibnamefont {Sarantsev}}, \ and\ \bibinfo
  {author} {\bibfnamefont {U.}~\bibnamefont {Thoma}},\ }\bibfield  {title}
  {\enquote {\bibinfo {title} {{Sign ambiguity in the K$\Sigma$ channel}},}\
  }\href {\doibase 10.1140/epja/i2013-13158-8} {\bibfield  {journal} {\bibinfo
  {journal} {Eur. Phys. J. A}\ }\textbf {\bibinfo {volume} {49}},\ \bibinfo
  {pages} {158} (\bibinfo {year} {2013})},\ \Eprint
  {http://arxiv.org/abs/1310.3610} {arXiv:1310.3610 [nucl-ex]} \BibitemShut
  {NoStop}%
\bibitem [{\citenamefont {Ireland}\ \emph {et~al.}(2020)\citenamefont
  {Ireland}, \citenamefont {Pasyuk},\ and\ \citenamefont
  {Strakovsky}}]{Ireland:2019uwn}%
  \BibitemOpen
  \bibfield  {author} {\bibinfo {author} {\bibfnamefont {David~G.}\
  \bibnamefont {Ireland}}, \bibinfo {author} {\bibfnamefont {Eugene}\
  \bibnamefont {Pasyuk}}, \ and\ \bibinfo {author} {\bibfnamefont {Igor}\
  \bibnamefont {Strakovsky}},\ }\bibfield  {title} {\enquote {\bibinfo {title}
  {{Photoproduction Reactions and Non-Strange Baryon Spectroscopy}},}\ }\href
  {\doibase 10.1016/j.ppnp.2019.103752} {\bibfield  {journal} {\bibinfo
  {journal} {Prog. Part. Nucl. Phys.}\ }\textbf {\bibinfo {volume} {111}},\
  \bibinfo {pages} {103752} (\bibinfo {year} {2020})},\ \Eprint
  {http://arxiv.org/abs/1906.04228} {arXiv:1906.04228 [nucl-ex]} \BibitemShut
  {NoStop}%
\bibitem [{\citenamefont {Thiel}\ \emph {et~al.}(2022)\citenamefont {Thiel},
  \citenamefont {Afzal},\ and\ \citenamefont {Wunderlich}}]{Thiel:2022xtb}%
  \BibitemOpen
  \bibfield  {author} {\bibinfo {author} {\bibfnamefont {Annika}\ \bibnamefont
  {Thiel}}, \bibinfo {author} {\bibfnamefont {Farah}\ \bibnamefont {Afzal}}, \
  and\ \bibinfo {author} {\bibfnamefont {Yannick}\ \bibnamefont {Wunderlich}},\
  }\bibfield  {title} {\enquote {\bibinfo {title} {{Light Baryon
  Spectroscopy}},}\ }\href {\doibase 10.1016/j.ppnp.2022.103949} {\bibfield
  {journal} {\bibinfo  {journal} {Prog. Part. Nucl. Phys.}\ }\textbf {\bibinfo
  {volume} {125}},\ \bibinfo {pages} {103949} (\bibinfo {year} {2022})},\
  \Eprint {http://arxiv.org/abs/2202.05055} {arXiv:2202.05055 [nucl-ex]}
  \BibitemShut {NoStop}%
\bibitem [{\citenamefont {Jude}\ \emph {et~al.}(2019)\citenamefont {Jude} \emph
  {et~al.}}]{Jude:2019qqd}%
  \BibitemOpen
  \bibfield  {author} {\bibinfo {author} {\bibfnamefont {T.~C.}\ \bibnamefont
  {Jude}} \emph {et~al.},\ }\bibfield  {title} {\enquote {\bibinfo {title}
  {{Strangeness Photoproduction at the BGO-OD Experiment}},}\ }\href {\doibase
  10.1134/S1063779619050113} {\bibfield  {journal} {\bibinfo  {journal} {Phys.
  Part. Nucl.}\ }\textbf {\bibinfo {volume} {50}},\ \bibinfo {pages} {493--500}
  (\bibinfo {year} {2019})},\ \bibinfo {note} {[Erratum: Phys.Part.Nucl. 51,
  122 (2020)]}\BibitemShut {NoStop}%
\bibitem [{\citenamefont {Barker}\ \emph {et~al.}(1975)\citenamefont {Barker},
  \citenamefont {Donnachie},\ and\ \citenamefont {Storrow}}]{Barker:1975bp}%
  \BibitemOpen
  \bibfield  {author} {\bibinfo {author} {\bibfnamefont {I.~S.}\ \bibnamefont
  {Barker}}, \bibinfo {author} {\bibfnamefont {A.}~\bibnamefont {Donnachie}}, \
  and\ \bibinfo {author} {\bibfnamefont {J.~K.}\ \bibnamefont {Storrow}},\
  }\bibfield  {title} {\enquote {\bibinfo {title} {{Complete Experiments in
  Pseudoscalar Photoproduction}},}\ }\href {\doibase
  10.1016/0550-3213(75)90049-8} {\bibfield  {journal} {\bibinfo  {journal}
  {Nucl. Phys. B}\ }\textbf {\bibinfo {volume} {95}},\ \bibinfo {pages}
  {347--356} (\bibinfo {year} {1975})}\BibitemShut {NoStop}%
\bibitem [{\citenamefont {Chiang}\ and\ \citenamefont
  {Tabakin}(1997)}]{Chiang:1996em}%
  \BibitemOpen
  \bibfield  {author} {\bibinfo {author} {\bibfnamefont {Wen-Tai}\ \bibnamefont
  {Chiang}}\ and\ \bibinfo {author} {\bibfnamefont {Frank}\ \bibnamefont
  {Tabakin}},\ }\bibfield  {title} {\enquote {\bibinfo {title} {{Completeness
  rules for spin observables in pseudoscalar meson photoproduction}},}\ }\href
  {\doibase 10.1103/PhysRevC.55.2054} {\bibfield  {journal} {\bibinfo
  {journal} {Phys. Rev. C}\ }\textbf {\bibinfo {volume} {55}},\ \bibinfo
  {pages} {2054--2066} (\bibinfo {year} {1997})},\ \Eprint
  {http://arxiv.org/abs/nucl-th/9611053} {arXiv:nucl-th/9611053} \BibitemShut
  {NoStop}%
\bibitem [{\citenamefont {Keaton}\ and\ \citenamefont
  {Workman}(1996)}]{Keaton:1996pe}%
  \BibitemOpen
  \bibfield  {author} {\bibinfo {author} {\bibfnamefont {Greg}\ \bibnamefont
  {Keaton}}\ and\ \bibinfo {author} {\bibfnamefont {Ron}\ \bibnamefont
  {Workman}},\ }\bibfield  {title} {\enquote {\bibinfo {title} {{Ambiguities in
  the partial wave analysis of pseudoscalar meson photoproduction}},}\ }\href
  {\doibase 10.1103/PhysRevC.54.1437} {\bibfield  {journal} {\bibinfo
  {journal} {Phys. Rev. C}\ }\textbf {\bibinfo {volume} {54}},\ \bibinfo
  {pages} {1437--1440} (\bibinfo {year} {1996})},\ \Eprint
  {http://arxiv.org/abs/nucl-th/9606052} {arXiv:nucl-th/9606052} \BibitemShut
  {NoStop}%
\bibitem [{\citenamefont {Sandorfi}\ \emph {et~al.}(2011)\citenamefont
  {Sandorfi}, \citenamefont {Hoblit}, \citenamefont {Kamano},\ and\
  \citenamefont {Lee}}]{Sandorfi:2010uv}%
  \BibitemOpen
  \bibfield  {author} {\bibinfo {author} {\bibfnamefont {A.~M.}\ \bibnamefont
  {Sandorfi}}, \bibinfo {author} {\bibfnamefont {S.}~\bibnamefont {Hoblit}},
  \bibinfo {author} {\bibfnamefont {H.}~\bibnamefont {Kamano}}, \ and\ \bibinfo
  {author} {\bibfnamefont {T.~S.~H.}\ \bibnamefont {Lee}},\ }\bibfield  {title}
  {\enquote {\bibinfo {title} {{Determining pseudoscalar meson photo-production
  amplitudes from complete experiments}},}\ }\href {\doibase
  10.1088/0954-3899/38/5/053001} {\bibfield  {journal} {\bibinfo  {journal} {J.
  Phys. G}\ }\textbf {\bibinfo {volume} {38}},\ \bibinfo {pages} {053001}
  (\bibinfo {year} {2011})},\ \Eprint {http://arxiv.org/abs/1010.4555}
  {arXiv:1010.4555 [nucl-th]} \BibitemShut {NoStop}%
\bibitem [{\citenamefont {Ireland}(2010)}]{Ireland:2010bi}%
  \BibitemOpen
  \bibfield  {author} {\bibinfo {author} {\bibfnamefont {D.~G.}\ \bibnamefont
  {Ireland}},\ }\bibfield  {title} {\enquote {\bibinfo {title} {{Information
  Content of Polarization Measurements}},}\ }\href {\doibase
  10.1103/PhysRevC.82.025204} {\bibfield  {journal} {\bibinfo  {journal} {Phys.
  Rev. C}\ }\textbf {\bibinfo {volume} {82}},\ \bibinfo {pages} {025204}
  (\bibinfo {year} {2010})},\ \Eprint {http://arxiv.org/abs/1004.5250}
  {arXiv:1004.5250 [hep-ph]} \BibitemShut {NoStop}%
\bibitem [{\citenamefont {Nys}\ \emph {et~al.}(2015)\citenamefont {Nys},
  \citenamefont {Vrancx},\ and\ \citenamefont {Ryckebusch}}]{Nys:2015kqa}%
  \BibitemOpen
  \bibfield  {author} {\bibinfo {author} {\bibfnamefont {Jannes}\ \bibnamefont
  {Nys}}, \bibinfo {author} {\bibfnamefont {Tom}\ \bibnamefont {Vrancx}}, \
  and\ \bibinfo {author} {\bibfnamefont {Jan}\ \bibnamefont {Ryckebusch}},\
  }\bibfield  {title} {\enquote {\bibinfo {title} {{Amplitude extraction in
  pseudoscalar-meson photoproduction: towards a situation of complete
  information}},}\ }\href {\doibase 10.1088/0954-3899/42/3/034016} {\bibfield
  {journal} {\bibinfo  {journal} {J. Phys. G}\ }\textbf {\bibinfo {volume}
  {42}},\ \bibinfo {pages} {034016} (\bibinfo {year} {2015})},\ \Eprint
  {http://arxiv.org/abs/1502.01259} {arXiv:1502.01259 [nucl-th]} \BibitemShut
  {NoStop}%
\bibitem [{\citenamefont {Tiator}\ \emph {et~al.}(2017)\citenamefont {Tiator},
  \citenamefont {Workman}, \citenamefont {Wunderlich},\ and\ \citenamefont
  {Haberzettl}}]{Tiator:2017cde}%
  \BibitemOpen
  \bibfield  {author} {\bibinfo {author} {\bibfnamefont {L.}~\bibnamefont
  {Tiator}}, \bibinfo {author} {\bibfnamefont {R.~L.}\ \bibnamefont {Workman}},
  \bibinfo {author} {\bibfnamefont {Y.}~\bibnamefont {Wunderlich}}, \ and\
  \bibinfo {author} {\bibfnamefont {H.}~\bibnamefont {Haberzettl}},\ }\bibfield
   {title} {\enquote {\bibinfo {title} {{Amplitude reconstruction from complete
  electroproduction experiments and truncated partial-wave expansions}},}\
  }\href {\doibase 10.1103/PhysRevC.96.025210} {\bibfield  {journal} {\bibinfo
  {journal} {Phys. Rev. C}\ }\textbf {\bibinfo {volume} {96}},\ \bibinfo
  {pages} {025210} (\bibinfo {year} {2017})},\ \Eprint
  {http://arxiv.org/abs/1702.08375} {arXiv:1702.08375 [nucl-th]} \BibitemShut
  {NoStop}%
\bibitem [{\citenamefont {Wunderlich}\ \emph {et~al.}(2020)\citenamefont
  {Wunderlich}, \citenamefont {Kroenert}, \citenamefont {Afzal},\ and\
  \citenamefont {Thiel}}]{Wunderlich:2020umg}%
  \BibitemOpen
  \bibfield  {author} {\bibinfo {author} {\bibfnamefont {Y.}~\bibnamefont
  {Wunderlich}}, \bibinfo {author} {\bibfnamefont {P.}~\bibnamefont
  {Kroenert}}, \bibinfo {author} {\bibfnamefont {F.}~\bibnamefont {Afzal}}, \
  and\ \bibinfo {author} {\bibfnamefont {A.}~\bibnamefont {Thiel}},\ }\bibfield
   {title} {\enquote {\bibinfo {title} {{Moravcsik's theorem on complete sets
  of polarization observables reexamined}},}\ }\href {\doibase
  10.1103/PhysRevC.102.034605} {\bibfield  {journal} {\bibinfo  {journal}
  {Phys. Rev. C}\ }\textbf {\bibinfo {volume} {102}},\ \bibinfo {pages}
  {034605} (\bibinfo {year} {2020})},\ \Eprint
  {http://arxiv.org/abs/2004.14483} {arXiv:2004.14483 [nucl-th]} \BibitemShut
  {NoStop}%
\bibitem [{\citenamefont {Workman}\ \emph {et~al.}(2017)\citenamefont
  {Workman}, \citenamefont {Tiator}, \citenamefont {Wunderlich}, \citenamefont
  {D\"oring},\ and\ \citenamefont {Haberzettl}}]{Workman:2016irf}%
  \BibitemOpen
  \bibfield  {author} {\bibinfo {author} {\bibfnamefont {R.~L.}\ \bibnamefont
  {Workman}}, \bibinfo {author} {\bibfnamefont {L.}~\bibnamefont {Tiator}},
  \bibinfo {author} {\bibfnamefont {Y.}~\bibnamefont {Wunderlich}}, \bibinfo
  {author} {\bibfnamefont {M.}~\bibnamefont {D\"oring}}, \ and\ \bibinfo
  {author} {\bibfnamefont {H.}~\bibnamefont {Haberzettl}},\ }\bibfield  {title}
  {\enquote {\bibinfo {title} {{Amplitude reconstruction from complete
  photoproduction experiments and truncated partial-wave expansions}},}\ }\href
  {\doibase 10.1103/PhysRevC.95.015206} {\bibfield  {journal} {\bibinfo
  {journal} {Phys. Rev. C}\ }\textbf {\bibinfo {volume} {95}},\ \bibinfo
  {pages} {015206} (\bibinfo {year} {2017})},\ \Eprint
  {http://arxiv.org/abs/1611.04434} {arXiv:1611.04434 [nucl-th]} \BibitemShut
  {NoStop}%
\bibitem [{\citenamefont {Wunderlich}(2021)}]{Wunderlich:2021xhp}%
  \BibitemOpen
  \bibfield  {author} {\bibinfo {author} {\bibfnamefont {Y.}~\bibnamefont
  {Wunderlich}},\ }\bibfield  {title} {\enquote {\bibinfo {title} {{New
  graphical criterion for the selection of complete sets of polarization
  observables and its application to single-meson photoproduction as well as
  electroproduction}},}\ }\href {\doibase 10.1103/PhysRevC.104.045203}
  {\bibfield  {journal} {\bibinfo  {journal} {Phys. Rev. C}\ }\textbf {\bibinfo
  {volume} {104}},\ \bibinfo {pages} {045203} (\bibinfo {year} {2021})},\
  \Eprint {http://arxiv.org/abs/2106.00486} {arXiv:2106.00486 [nucl-th]}
  \BibitemShut {NoStop}%
\bibitem [{\citenamefont {Steininger}\ and\ \citenamefont
  {Mei\ss{}ner}(1997)}]{Steininger:1996xw}%
  \BibitemOpen
  \bibfield  {author} {\bibinfo {author} {\bibfnamefont {Sven}\ \bibnamefont
  {Steininger}}\ and\ \bibinfo {author} {\bibfnamefont {U.-G.}\ \bibnamefont
  {Mei\ss{}ner}},\ }\bibfield  {title} {\enquote {\bibinfo {title} {{Threshold
  kaon photoproduction and electroproduction in SU(3) baryon chiral
  perturbation theory}},}\ }\href {\doibase 10.1016/S0370-2693(96)01490-6}
  {\bibfield  {journal} {\bibinfo  {journal} {Phys. Lett. B}\ }\textbf
  {\bibinfo {volume} {391}},\ \bibinfo {pages} {446--450} (\bibinfo {year}
  {1997})},\ \Eprint {http://arxiv.org/abs/nucl-th/9609051}
  {arXiv:nucl-th/9609051} \BibitemShut {NoStop}%
\bibitem [{\citenamefont {Mai}\ \emph {et~al.}(2009)\citenamefont {Mai},
  \citenamefont {Bruns}, \citenamefont {Kubis},\ and\ \citenamefont
  {Mei\ss{}ner}}]{Mai:2009ce}%
  \BibitemOpen
  \bibfield  {author} {\bibinfo {author} {\bibfnamefont {Maxim}\ \bibnamefont
  {Mai}}, \bibinfo {author} {\bibfnamefont {Peter~C.}\ \bibnamefont {Bruns}},
  \bibinfo {author} {\bibfnamefont {Bastian}\ \bibnamefont {Kubis}}, \ and\
  \bibinfo {author} {\bibfnamefont {U.-G.}\ \bibnamefont {Mei\ss{}ner}},\
  }\bibfield  {title} {\enquote {\bibinfo {title} {{Aspects of meson-baryon
  scattering in three and two-flavor chiral perturbation theory}},}\ }\href
  {\doibase 10.1103/PhysRevD.80.094006} {\bibfield  {journal} {\bibinfo
  {journal} {Phys. Rev. D}\ }\textbf {\bibinfo {volume} {80}},\ \bibinfo
  {pages} {094006} (\bibinfo {year} {2009})},\ \Eprint
  {http://arxiv.org/abs/0905.2810} {arXiv:0905.2810 [hep-ph]} \BibitemShut
  {NoStop}%
\bibitem [{\citenamefont {Kaiser}\ \emph {et~al.}(1997)\citenamefont {Kaiser},
  \citenamefont {Waas},\ and\ \citenamefont {Weise}}]{Kaiser:1996js}%
  \BibitemOpen
  \bibfield  {author} {\bibinfo {author} {\bibfnamefont {Norbert}\ \bibnamefont
  {Kaiser}}, \bibinfo {author} {\bibfnamefont {T.}~\bibnamefont {Waas}}, \ and\
  \bibinfo {author} {\bibfnamefont {W.}~\bibnamefont {Weise}},\ }\bibfield
  {title} {\enquote {\bibinfo {title} {{SU(3) chiral dynamics with coupled
  channels: Eta and kaon photoproduction}},}\ }\href {\doibase
  10.1016/S0375-9474(96)00321-1} {\bibfield  {journal} {\bibinfo  {journal}
  {Nucl. Phys. A}\ }\textbf {\bibinfo {volume} {612}},\ \bibinfo {pages}
  {297--320} (\bibinfo {year} {1997})},\ \Eprint
  {http://arxiv.org/abs/hep-ph/9607459} {arXiv:hep-ph/9607459} \BibitemShut
  {NoStop}%
\bibitem [{\citenamefont {Borasoy}\ \emph {et~al.}(2007)\citenamefont
  {Borasoy}, \citenamefont {Bruns}, \citenamefont {Mei\ss{}ner},\ and\
  \citenamefont {Ni\ss{}ler}}]{Borasoy:2007ku}%
  \BibitemOpen
  \bibfield  {author} {\bibinfo {author} {\bibfnamefont {B.}~\bibnamefont
  {Borasoy}}, \bibinfo {author} {\bibfnamefont {P.~C.}\ \bibnamefont {Bruns}},
  \bibinfo {author} {\bibfnamefont {U.-G.}\ \bibnamefont {Mei\ss{}ner}}, \ and\
  \bibinfo {author} {\bibfnamefont {R.}~\bibnamefont {Ni\ss{}ler}},\ }\bibfield
   {title} {\enquote {\bibinfo {title} {{A Gauge invariant chiral unitary
  framework for kaon photo- and electroproduction on the proton}},}\ }\href
  {\doibase 10.1140/epja/i2007-10492-4} {\bibfield  {journal} {\bibinfo
  {journal} {Eur. Phys. J. A}\ }\textbf {\bibinfo {volume} {34}},\ \bibinfo
  {pages} {161--183} (\bibinfo {year} {2007})},\ \Eprint
  {http://arxiv.org/abs/0709.3181} {arXiv:0709.3181 [nucl-th]} \BibitemShut
  {NoStop}%
\bibitem [{\citenamefont {Golli}\ and\ \citenamefont
  {\v{S}irca}(2016)}]{Golli:2016dlj}%
  \BibitemOpen
  \bibfield  {author} {\bibinfo {author} {\bibfnamefont {Bojan}\ \bibnamefont
  {Golli}}\ and\ \bibinfo {author} {\bibfnamefont {Simon}\ \bibnamefont
  {\v{S}irca}},\ }\bibfield  {title} {\enquote {\bibinfo {title} {{Eta and kaon
  production in a chiral quark model}},}\ }\href {\doibase
  10.1140/epja/i2016-16279-6} {\bibfield  {journal} {\bibinfo  {journal} {Eur.
  Phys. J. A}\ }\textbf {\bibinfo {volume} {52}},\ \bibinfo {pages} {279}
  (\bibinfo {year} {2016})},\ \Eprint {http://arxiv.org/abs/1604.01937}
  {arXiv:1604.01937 [hep-ph]} \BibitemShut {NoStop}%
\bibitem [{\citenamefont {Maxwell}(2015)}]{Maxwell:2015psa}%
  \BibitemOpen
  \bibfield  {author} {\bibinfo {author} {\bibfnamefont {Oren~V.}\ \bibnamefont
  {Maxwell}},\ }\bibfield  {title} {\enquote {\bibinfo {title} {{New fit to the
  reaction $\gamma p \to K^+ \Sigma^0$}},}\ }\href {\doibase
  10.1103/PhysRevC.92.044614} {\bibfield  {journal} {\bibinfo  {journal} {Phys.
  Rev. C}\ }\textbf {\bibinfo {volume} {92}},\ \bibinfo {pages} {044614}
  (\bibinfo {year} {2015})}\BibitemShut {NoStop}%
\bibitem [{\citenamefont {Wei}\ \emph {et~al.}(2022)\citenamefont {Wei},
  \citenamefont {Wang}, \citenamefont {Huang},\ and\ \citenamefont
  {Nakayama}}]{Wei:2022nqp}%
  \BibitemOpen
  \bibfield  {author} {\bibinfo {author} {\bibfnamefont {Neng-Chang}\
  \bibnamefont {Wei}}, \bibinfo {author} {\bibfnamefont {Ai-Chao}\ \bibnamefont
  {Wang}}, \bibinfo {author} {\bibfnamefont {Fei}\ \bibnamefont {Huang}}, \
  and\ \bibinfo {author} {\bibfnamefont {Kanzo}\ \bibnamefont {Nakayama}},\
  }\bibfield  {title} {\enquote {\bibinfo {title} {{Nucleon and
  \ensuremath{\Delta} resonances in
  \ensuremath{\gamma}n\textrightarrow{}K+\ensuremath{\Sigma}-
  photoproduction}},}\ }\href {\doibase 10.1103/PhysRevD.105.094017} {\bibfield
   {journal} {\bibinfo  {journal} {Phys. Rev. D}\ }\textbf {\bibinfo {volume}
  {105}},\ \bibinfo {pages} {094017} (\bibinfo {year} {2022})},\ \Eprint
  {http://arxiv.org/abs/2204.13922} {arXiv:2204.13922 [hep-ph]} \BibitemShut
  {NoStop}%
\bibitem [{\citenamefont {Mart}\ and\ \citenamefont
  {Sakinah}(2017)}]{Mart:2017mwj}%
  \BibitemOpen
  \bibfield  {author} {\bibinfo {author} {\bibfnamefont {T.}~\bibnamefont
  {Mart}}\ and\ \bibinfo {author} {\bibfnamefont {S.}~\bibnamefont {Sakinah}},\
  }\bibfield  {title} {\enquote {\bibinfo {title} {{Multipoles model for $K^+
  \Lambda$ photoproduction on the nucleon reexamined}},}\ }\href {\doibase
  10.1103/PhysRevC.95.045205} {\bibfield  {journal} {\bibinfo  {journal} {Phys.
  Rev. C}\ }\textbf {\bibinfo {volume} {95}},\ \bibinfo {pages} {045205}
  (\bibinfo {year} {2017})}\BibitemShut {NoStop}%
\bibitem [{\citenamefont {Mart}\ and\ \citenamefont
  {Kholili}(2019)}]{Mart:2019fau}%
  \BibitemOpen
  \bibfield  {author} {\bibinfo {author} {\bibfnamefont {T.}~\bibnamefont
  {Mart}}\ and\ \bibinfo {author} {\bibfnamefont {M.~J.}\ \bibnamefont
  {Kholili}},\ }\bibfield  {title} {\enquote {\bibinfo {title} {{Partial wave
  analysis for $K\Sigma$ photoproduction on the nucleon valid from threshold up
  to $W$ = 2.8 GeV}},}\ }\href {\doibase 10.1088/1361-6471/ab34c6} {\bibfield
  {journal} {\bibinfo  {journal} {J. Phys. G}\ }\textbf {\bibinfo {volume}
  {46}},\ \bibinfo {pages} {105112} (\bibinfo {year} {2019})}\BibitemShut
  {NoStop}%
\bibitem [{\citenamefont {Clymton}\ and\ \citenamefont
  {Mart}(2021)}]{Clymton:2021wof}%
  \BibitemOpen
  \bibfield  {author} {\bibinfo {author} {\bibfnamefont {S.}~\bibnamefont
  {Clymton}}\ and\ \bibinfo {author} {\bibfnamefont {T.}~\bibnamefont {Mart}},\
  }\bibfield  {title} {\enquote {\bibinfo {title} {{Extracting the pole and
  Breit-Wigner properties of nucleon and \ensuremath{\Delta} resonances from
  the \ensuremath{\gamma}N\textrightarrow{}K\ensuremath{\Sigma}
  photoproduction}},}\ }\href {\doibase 10.1103/PhysRevD.104.056015} {\bibfield
   {journal} {\bibinfo  {journal} {Phys. Rev. D}\ }\textbf {\bibinfo {volume}
  {104}},\ \bibinfo {pages} {056015} (\bibinfo {year} {2021})},\ \Eprint
  {http://arxiv.org/abs/2104.10333} {arXiv:2104.10333 [hep-ph]} \BibitemShut
  {NoStop}%
\bibitem [{\citenamefont {Luthfiyah}\ and\ \citenamefont
  {Mart}(2021)}]{Luthfiyah:2021yqe}%
  \BibitemOpen
  \bibfield  {author} {\bibinfo {author} {\bibfnamefont {N.~H.}\ \bibnamefont
  {Luthfiyah}}\ and\ \bibinfo {author} {\bibfnamefont {T.}~\bibnamefont
  {Mart}},\ }\bibfield  {title} {\enquote {\bibinfo {title} {{Role of the
  high-spin nucleon and delta resonances in the $K\Lambda$ and $K\Sigma$
  photoproduction off the nucleon}},}\ }\href {\doibase
  10.1103/PhysRevD.104.076022} {\bibfield  {journal} {\bibinfo  {journal}
  {Phys. Rev. D}\ }\textbf {\bibinfo {volume} {104}},\ \bibinfo {pages}
  {076022} (\bibinfo {year} {2021})},\ \Eprint
  {http://arxiv.org/abs/2110.01789} {arXiv:2110.01789 [hep-ph]} \BibitemShut
  {NoStop}%
\bibitem [{\citenamefont {Egorov}\ and\ \citenamefont
  {Postnikov}(2021)}]{Egorov:2021scy}%
  \BibitemOpen
  \bibfield  {author} {\bibinfo {author} {\bibfnamefont {M.~V.}\ \bibnamefont
  {Egorov}}\ and\ \bibinfo {author} {\bibfnamefont {V.~I.}\ \bibnamefont
  {Postnikov}},\ }\bibfield  {title} {\enquote {\bibinfo {title} {{Kaon
  Electroproduction on the Proton}},}\ }\href {\doibase
  10.1134/S1063776121060121} {\bibfield  {journal} {\bibinfo  {journal} {J.
  Exp. Theor. Phys.}\ }\textbf {\bibinfo {volume} {133}},\ \bibinfo {pages}
  {32--43} (\bibinfo {year} {2021})}\BibitemShut {NoStop}%
\bibitem [{\citenamefont {Lee}\ \emph {et~al.}(2001)\citenamefont {Lee},
  \citenamefont {Mart}, \citenamefont {Bennhold},\ and\ \citenamefont
  {Wright}}]{Lee:1999kd}%
  \BibitemOpen
  \bibfield  {author} {\bibinfo {author} {\bibfnamefont {F.~X.}\ \bibnamefont
  {Lee}}, \bibinfo {author} {\bibfnamefont {T.}~\bibnamefont {Mart}}, \bibinfo
  {author} {\bibfnamefont {C.}~\bibnamefont {Bennhold}}, \ and\ \bibinfo
  {author} {\bibfnamefont {L.~E.}\ \bibnamefont {Wright}},\ }\bibfield  {title}
  {\enquote {\bibinfo {title} {{Quasifree kaon photoproduction on nuclei}},}\
  }\href {\doibase 10.1016/S0375-9474(01)01098-3} {\bibfield  {journal}
  {\bibinfo  {journal} {Nucl. Phys. A}\ }\textbf {\bibinfo {volume} {695}},\
  \bibinfo {pages} {237--272} (\bibinfo {year} {2001})},\ \Eprint
  {http://arxiv.org/abs/nucl-th/9907119} {arXiv:nucl-th/9907119} \BibitemShut
  {NoStop}%
\bibitem [{\citenamefont {Tiator}(2018)}]{Tiator:2018pjq}%
  \BibitemOpen
  \bibfield  {author} {\bibinfo {author} {\bibfnamefont {Lothar}\ \bibnamefont
  {Tiator}},\ }\bibfield  {title} {\enquote {\bibinfo {title} {{The MAID Legacy
  and Future}},}\ }\href {\doibase 10.1007/s00601-018-1343-5} {\bibfield
  {journal} {\bibinfo  {journal} {Few Body Syst.}\ }\textbf {\bibinfo {volume}
  {59}},\ \bibinfo {pages} {21} (\bibinfo {year} {2018})},\ \Eprint
  {http://arxiv.org/abs/1801.04777} {arXiv:1801.04777 [nucl-th]} \BibitemShut
  {NoStop}%
\bibitem [{\citenamefont {Corthals}\ \emph {et~al.}(2007)\citenamefont
  {Corthals}, \citenamefont {Ireland}, \citenamefont {Van~Cauteren},\ and\
  \citenamefont {Ryckebusch}}]{Corthals:2006nz}%
  \BibitemOpen
  \bibfield  {author} {\bibinfo {author} {\bibfnamefont {T.}~\bibnamefont
  {Corthals}}, \bibinfo {author} {\bibfnamefont {D.~G.}\ \bibnamefont
  {Ireland}}, \bibinfo {author} {\bibfnamefont {T.}~\bibnamefont
  {Van~Cauteren}}, \ and\ \bibinfo {author} {\bibfnamefont {J.}~\bibnamefont
  {Ryckebusch}},\ }\bibfield  {title} {\enquote {\bibinfo {title}
  {{Regge-plus-resonance treatment of the $p(\gamma,K^+)\Sigma^0$ and
  $p(\gamma,K^0)\Sigma^+$ reactions at forward kaon angles}},}\ }\href
  {\doibase 10.1103/PhysRevC.75.045204} {\bibfield  {journal} {\bibinfo
  {journal} {Phys. Rev. C}\ }\textbf {\bibinfo {volume} {75}},\ \bibinfo
  {pages} {045204} (\bibinfo {year} {2007})},\ \Eprint
  {http://arxiv.org/abs/nucl-th/0612085} {arXiv:nucl-th/0612085} \BibitemShut
  {NoStop}%
\bibitem [{\citenamefont {Anisovich}\ \emph {et~al.}(2012)\citenamefont
  {Anisovich}, \citenamefont {Beck}, \citenamefont {Klempt}, \citenamefont
  {Nikonov}, \citenamefont {Sarantsev},\ and\ \citenamefont
  {Thoma}}]{Anisovich:2011fc}%
  \BibitemOpen
  \bibfield  {author} {\bibinfo {author} {\bibfnamefont {A.~V.}\ \bibnamefont
  {Anisovich}}, \bibinfo {author} {\bibfnamefont {R.}~\bibnamefont {Beck}},
  \bibinfo {author} {\bibfnamefont {E.}~\bibnamefont {Klempt}}, \bibinfo
  {author} {\bibfnamefont {V.~A.}\ \bibnamefont {Nikonov}}, \bibinfo {author}
  {\bibfnamefont {A.~V.}\ \bibnamefont {Sarantsev}}, \ and\ \bibinfo {author}
  {\bibfnamefont {U.}~\bibnamefont {Thoma}},\ }\bibfield  {title} {\enquote
  {\bibinfo {title} {{Properties of baryon resonances from a multichannel
  partial wave analysis}},}\ }\href {\doibase 10.1140/epja/i2012-12015-8}
  {\bibfield  {journal} {\bibinfo  {journal} {Eur. Phys. J. A}\ }\textbf
  {\bibinfo {volume} {48}},\ \bibinfo {pages} {15} (\bibinfo {year} {2012})},\
  \Eprint {http://arxiv.org/abs/1112.4937} {arXiv:1112.4937 [hep-ph]}
  \BibitemShut {NoStop}%
\bibitem [{\citenamefont {M\"uller}\ \emph {et~al.}(2020)\citenamefont
  {M\"uller} \emph {et~al.}}]{CBELSATAPS:2019ylw}%
  \BibitemOpen
  \bibfield  {author} {\bibinfo {author} {\bibfnamefont {J.}~\bibnamefont
  {M\"uller}} \emph {et~al.} (\bibinfo {collaboration} {CBELSA/TAPS}),\
  }\bibfield  {title} {\enquote {\bibinfo {title} {{New data on $\vec{\gamma}
  \vec{p}\rightarrow \eta p$ with polarized photons and protons and their
  implications for $N^* \to N\eta$ decays}},}\ }\href {\doibase
  10.1016/j.physletb.2020.135323} {\bibfield  {journal} {\bibinfo  {journal}
  {Phys. Lett. B}\ }\textbf {\bibinfo {volume} {803}},\ \bibinfo {pages}
  {135323} (\bibinfo {year} {2020})},\ \Eprint
  {http://arxiv.org/abs/1909.08464} {arXiv:1909.08464 [nucl-ex]} \BibitemShut
  {NoStop}%
\bibitem [{\citenamefont {Cao}\ \emph {et~al.}(2013)\citenamefont {Cao},
  \citenamefont {Shklyar},\ and\ \citenamefont {Lenske}}]{Cao:2013psa}%
  \BibitemOpen
  \bibfield  {author} {\bibinfo {author} {\bibfnamefont {Xu}~\bibnamefont
  {Cao}}, \bibinfo {author} {\bibfnamefont {V.}~\bibnamefont {Shklyar}}, \ and\
  \bibinfo {author} {\bibfnamefont {H.}~\bibnamefont {Lenske}},\ }\bibfield
  {title} {\enquote {\bibinfo {title} {{Coupled-channel analysis of
  K\ensuremath{\Sigma} production on the nucleon up to 2.0 GeV}},}\ }\href
  {\doibase 10.1103/PhysRevC.88.055204} {\bibfield  {journal} {\bibinfo
  {journal} {Phys. Rev. C}\ }\textbf {\bibinfo {volume} {88}},\ \bibinfo
  {pages} {055204} (\bibinfo {year} {2013})},\ \Eprint
  {http://arxiv.org/abs/1303.2604} {arXiv:1303.2604 [nucl-th]} \BibitemShut
  {NoStop}%
\bibitem [{\citenamefont {Kamano}\ \emph {et~al.}(2013)\citenamefont {Kamano},
  \citenamefont {Nakamura}, \citenamefont {Lee},\ and\ \citenamefont
  {Sato}}]{Kamano:2013iva}%
  \BibitemOpen
  \bibfield  {author} {\bibinfo {author} {\bibfnamefont {H.}~\bibnamefont
  {Kamano}}, \bibinfo {author} {\bibfnamefont {S.~X.}\ \bibnamefont
  {Nakamura}}, \bibinfo {author} {\bibfnamefont {T.~S.~H.}\ \bibnamefont
  {Lee}}, \ and\ \bibinfo {author} {\bibfnamefont {T.}~\bibnamefont {Sato}},\
  }\bibfield  {title} {\enquote {\bibinfo {title} {{Nucleon resonances within a
  dynamical coupled-channels model of $\pi N$ and $\gamma N$ reactions}},}\
  }\href {\doibase 10.1103/PhysRevC.88.035209} {\bibfield  {journal} {\bibinfo
  {journal} {Phys. Rev. C}\ }\textbf {\bibinfo {volume} {88}},\ \bibinfo
  {pages} {035209} (\bibinfo {year} {2013})},\ \Eprint
  {http://arxiv.org/abs/1305.4351} {arXiv:1305.4351 [nucl-th]} \BibitemShut
  {NoStop}%
\bibitem [{\citenamefont {Kamano}\ \emph {et~al.}(2016)\citenamefont {Kamano},
  \citenamefont {Nakamura}, \citenamefont {Lee},\ and\ \citenamefont
  {Sato}}]{Kamano:2016bgm}%
  \BibitemOpen
  \bibfield  {author} {\bibinfo {author} {\bibfnamefont {H.}~\bibnamefont
  {Kamano}}, \bibinfo {author} {\bibfnamefont {S.~X.}\ \bibnamefont
  {Nakamura}}, \bibinfo {author} {\bibfnamefont {T.~S.~H.}\ \bibnamefont
  {Lee}}, \ and\ \bibinfo {author} {\bibfnamefont {T.}~\bibnamefont {Sato}},\
  }\bibfield  {title} {\enquote {\bibinfo {title} {{Isospin decomposition of
  $\gamma N \to N^*$ transitions within a dynamical coupled-channels model}},}\
  }\href {\doibase 10.1103/PhysRevC.94.015201} {\bibfield  {journal} {\bibinfo
  {journal} {Phys. Rev. C}\ }\textbf {\bibinfo {volume} {94}},\ \bibinfo
  {pages} {015201} (\bibinfo {year} {2016})},\ \Eprint
  {http://arxiv.org/abs/1605.00363} {arXiv:1605.00363 [nucl-th]} \BibitemShut
  {NoStop}%
\bibitem [{\citenamefont {Kamano}\ \emph {et~al.}(2019)\citenamefont {Kamano},
  \citenamefont {Lee}, \citenamefont {Nakamura},\ and\ \citenamefont
  {Sato}}]{Kamano:2019gtm}%
  \BibitemOpen
  \bibfield  {author} {\bibinfo {author} {\bibfnamefont {H.}~\bibnamefont
  {Kamano}}, \bibinfo {author} {\bibfnamefont {T.~S.~H.}\ \bibnamefont {Lee}},
  \bibinfo {author} {\bibfnamefont {S.~X.}\ \bibnamefont {Nakamura}}, \ and\
  \bibinfo {author} {\bibfnamefont {T.}~\bibnamefont {Sato}},\ }\bibfield
  {title} {\enquote {\bibinfo {title} {{The ANL-Osaka Partial-Wave Amplitudes
  of $\pi N$ and $\gamma N$ Reactions}},}\ }\href@noop {} {\  (\bibinfo {year}
  {2019})},\ \Eprint {http://arxiv.org/abs/1909.11935} {arXiv:1909.11935
  [nucl-th]} \BibitemShut {NoStop}%
\bibitem [{\citenamefont {{Argonne/ANL-Osaka website}}()}]{anlweb}%
  \BibitemOpen
  \bibfield  {author} {\bibinfo {author} {\bibnamefont {{Argonne/ANL-Osaka
  website}}},\ }\href@noop {} {}\bibinfo {howpublished}
  {\url{https://www.phy.anl.gov/theory/research/anl-osaka- pwa}}\BibitemShut
  {NoStop}%
\bibitem [{\citenamefont {Sch{\"u}tz}\ \emph {et~al.}(1994)\citenamefont
  {Sch{\"u}tz}, \citenamefont {Durso}, \citenamefont {Holinde},\ and\
  \citenamefont {Speth}}]{Schutz:1994ue}%
  \BibitemOpen
  \bibfield  {author} {\bibinfo {author} {\bibfnamefont {C.}~\bibnamefont
  {Sch{\"u}tz}}, \bibinfo {author} {\bibfnamefont {J.~W.}\ \bibnamefont
  {Durso}}, \bibinfo {author} {\bibfnamefont {K.}~\bibnamefont {Holinde}}, \
  and\ \bibinfo {author} {\bibfnamefont {J.}~\bibnamefont {Speth}},\ }\bibfield
   {title} {\enquote {\bibinfo {title} {{Role of correlated two pion exchange
  in $\pi N$ scattering}},}\ }\href {\doibase 10.1103/PhysRevC.49.2671}
  {\bibfield  {journal} {\bibinfo  {journal} {Phys. Rev. C}\ }\textbf {\bibinfo
  {volume} {49}},\ \bibinfo {pages} {2671--2687} (\bibinfo {year}
  {1994})}\BibitemShut {NoStop}%
\bibitem [{\citenamefont {Döring}\ \emph
  {et~al.}(2009{\natexlab{a}})\citenamefont {Döring}, \citenamefont {Hanhart},
  \citenamefont {Huang}, \citenamefont {Krewald},\ and\ \citenamefont
  {Mei\ss{}ner}}]{Doring:2009bi}%
  \BibitemOpen
  \bibfield  {author} {\bibinfo {author} {\bibfnamefont {M.}~\bibnamefont
  {Döring}}, \bibinfo {author} {\bibfnamefont {C.}~\bibnamefont {Hanhart}},
  \bibinfo {author} {\bibfnamefont {F.}~\bibnamefont {Huang}}, \bibinfo
  {author} {\bibfnamefont {S.}~\bibnamefont {Krewald}}, \ and\ \bibinfo
  {author} {\bibfnamefont {U.-G.}\ \bibnamefont {Mei\ss{}ner}},\ }\bibfield
  {title} {\enquote {\bibinfo {title} {{The Role of the background in the
  extraction of resonance contributions from meson-baryon scattering}},}\
  }\href {\doibase 10.1016/j.physletb.2009.09.052} {\bibfield  {journal}
  {\bibinfo  {journal} {Phys. Lett. B}\ }\textbf {\bibinfo {volume} {681}},\
  \bibinfo {pages} {26--31} (\bibinfo {year} {2009}{\natexlab{a}})},\ \Eprint
  {http://arxiv.org/abs/0903.1781} {arXiv:0903.1781 [nucl-th]} \BibitemShut
  {NoStop}%
\bibitem [{\citenamefont {Döring}\ \emph
  {et~al.}(2009{\natexlab{b}})\citenamefont {Döring}, \citenamefont {Hanhart},
  \citenamefont {Huang}, \citenamefont {Krewald},\ and\ \citenamefont
  {Mei\ss{}ner}}]{Doring:2009yv}%
  \BibitemOpen
  \bibfield  {author} {\bibinfo {author} {\bibfnamefont {M.}~\bibnamefont
  {Döring}}, \bibinfo {author} {\bibfnamefont {C.}~\bibnamefont {Hanhart}},
  \bibinfo {author} {\bibfnamefont {F.}~\bibnamefont {Huang}}, \bibinfo
  {author} {\bibfnamefont {S.}~\bibnamefont {Krewald}}, \ and\ \bibinfo
  {author} {\bibfnamefont {U.-G.}\ \bibnamefont {Mei\ss{}ner}},\ }\bibfield
  {title} {\enquote {\bibinfo {title} {{Analytic properties of the scattering
  amplitude and resonances parameters in a meson exchange model}},}\ }\href
  {\doibase 10.1016/j.nuclphysa.2009.08.010} {\bibfield  {journal} {\bibinfo
  {journal} {Nucl. Phys. A}\ }\textbf {\bibinfo {volume} {829}},\ \bibinfo
  {pages} {170--209} (\bibinfo {year} {2009}{\natexlab{b}})},\ \Eprint
  {http://arxiv.org/abs/0903.4337} {arXiv:0903.4337 [nucl-th]} \BibitemShut
  {NoStop}%
\bibitem [{\citenamefont {Döring}\ \emph {et~al.}(2011)\citenamefont
  {Döring}, \citenamefont {Hanhart}, \citenamefont {Huang}, \citenamefont
  {Krewald}, \citenamefont {Mei\ss{}ner},\ and\ \citenamefont
  {Rönchen}}]{Doring:2010ap}%
  \BibitemOpen
  \bibfield  {author} {\bibinfo {author} {\bibfnamefont {M.}~\bibnamefont
  {Döring}}, \bibinfo {author} {\bibfnamefont {C.}~\bibnamefont {Hanhart}},
  \bibinfo {author} {\bibfnamefont {F.}~\bibnamefont {Huang}}, \bibinfo
  {author} {\bibfnamefont {S.}~\bibnamefont {Krewald}}, \bibinfo {author}
  {\bibfnamefont {U.-G.}\ \bibnamefont {Mei\ss{}ner}}, \ and\ \bibinfo {author}
  {\bibfnamefont {D.}~\bibnamefont {Rönchen}},\ }\bibfield  {title} {\enquote
  {\bibinfo {title} {{The reaction $\pi^+ p \to K^+\Sigma^+$ in a unitary
  coupled-channels model}},}\ }\href {\doibase 10.1016/j.nuclphysa.2010.12.010}
  {\bibfield  {journal} {\bibinfo  {journal} {Nucl. Phys. A}\ }\textbf
  {\bibinfo {volume} {851}},\ \bibinfo {pages} {58--98} (\bibinfo {year}
  {2011})},\ \Eprint {http://arxiv.org/abs/1009.3781} {arXiv:1009.3781
  [nucl-th]} \BibitemShut {NoStop}%
\bibitem [{\citenamefont {Rönchen}\ \emph {et~al.}(2014)\citenamefont
  {Rönchen}, \citenamefont {Döring}, \citenamefont {Huang}, \citenamefont
  {Haberzettl}, \citenamefont {Haidenbauer}, \citenamefont {Hanhart},
  \citenamefont {Krewald}, \citenamefont {Mei\ss{}ner},\ and\ \citenamefont
  {Nakayama}}]{Ronchen:2014cna}%
  \BibitemOpen
  \bibfield  {author} {\bibinfo {author} {\bibfnamefont {D.}~\bibnamefont
  {Rönchen}}, \bibinfo {author} {\bibfnamefont {M.}~\bibnamefont {Döring}},
  \bibinfo {author} {\bibfnamefont {F.}~\bibnamefont {Huang}}, \bibinfo
  {author} {\bibfnamefont {H.}~\bibnamefont {Haberzettl}}, \bibinfo {author}
  {\bibfnamefont {J.}~\bibnamefont {Haidenbauer}}, \bibinfo {author}
  {\bibfnamefont {C.}~\bibnamefont {Hanhart}}, \bibinfo {author} {\bibfnamefont
  {S.}~\bibnamefont {Krewald}}, \bibinfo {author} {\bibfnamefont {U.-G.}\
  \bibnamefont {Mei\ss{}ner}}, \ and\ \bibinfo {author} {\bibfnamefont
  {K.}~\bibnamefont {Nakayama}},\ }\bibfield  {title} {\enquote {\bibinfo
  {title} {{Photocouplings at the Pole from Pion Photoproduction}},}\ }\href
  {\doibase 10.1140/epja/i2014-14101-3} {\bibfield  {journal} {\bibinfo
  {journal} {Eur. Phys. J. A}\ }\textbf {\bibinfo {volume} {50}},\ \bibinfo
  {pages} {101} (\bibinfo {year} {2014})},\ \bibinfo {note} {[Erratum:
  Eur.Phys.J.A 51, 63 (2015)]},\ \Eprint {http://arxiv.org/abs/1401.0634}
  {arXiv:1401.0634 [nucl-th]} \BibitemShut {NoStop}%
\bibitem [{\citenamefont {R\"onchen}\ \emph {et~al.}(2015)\citenamefont
  {R\"onchen}, \citenamefont {D\"oring}, \citenamefont {Haberzettl},
  \citenamefont {Haidenbauer}, \citenamefont {Mei\ss{}ner},\ and\ \citenamefont
  {Nakayama}}]{Ronchen:2015vfa}%
  \BibitemOpen
  \bibfield  {author} {\bibinfo {author} {\bibfnamefont {D.}~\bibnamefont
  {R\"onchen}}, \bibinfo {author} {\bibfnamefont {M.}~\bibnamefont {D\"oring}},
  \bibinfo {author} {\bibfnamefont {H.}~\bibnamefont {Haberzettl}}, \bibinfo
  {author} {\bibfnamefont {J.}~\bibnamefont {Haidenbauer}}, \bibinfo {author}
  {\bibfnamefont {U.-G.}\ \bibnamefont {Mei\ss{}ner}}, \ and\ \bibinfo {author}
  {\bibfnamefont {K.}~\bibnamefont {Nakayama}},\ }\bibfield  {title} {\enquote
  {\bibinfo {title} {{Eta photoproduction in a combined analysis of pion- and
  photon-induced reactions}},}\ }\href {\doibase 10.1140/epja/i2015-15070-7}
  {\bibfield  {journal} {\bibinfo  {journal} {Eur. Phys. J. A}\ }\textbf
  {\bibinfo {volume} {51}},\ \bibinfo {pages} {70} (\bibinfo {year} {2015})},\
  \Eprint {http://arxiv.org/abs/1504.01643} {arXiv:1504.01643 [nucl-th]}
  \BibitemShut {NoStop}%
\bibitem [{\citenamefont {R\"onchen}\ \emph {et~al.}(2018)\citenamefont
  {R\"onchen}, \citenamefont {D\"oring},\ and\ \citenamefont
  {Mei\ss{}ner}}]{Ronchen:2018ury}%
  \BibitemOpen
  \bibfield  {author} {\bibinfo {author} {\bibfnamefont {D.}~\bibnamefont
  {R\"onchen}}, \bibinfo {author} {\bibfnamefont {M.}~\bibnamefont {D\"oring}},
  \ and\ \bibinfo {author} {\bibfnamefont {U.~G}\ \bibnamefont {Mei\ss{}ner}},\
  }\bibfield  {title} {\enquote {\bibinfo {title} {{The impact of
  $K^{+}\Lambda$ photoproduction on the resonance spectrum}},}\ }\href
  {\doibase 10.1140/epja/i2018-12541-3} {\bibfield  {journal} {\bibinfo
  {journal} {Eur. Phys. J. A}\ }\textbf {\bibinfo {volume} {54}},\ \bibinfo
  {pages} {110} (\bibinfo {year} {2018})},\ \Eprint
  {http://arxiv.org/abs/1801.10458} {arXiv:1801.10458 [nucl-th]} \BibitemShut
  {NoStop}%
\bibitem [{\citenamefont {Ablikim}\ \emph {et~al.}(2019)\citenamefont {Ablikim}
  \emph {et~al.}}]{BESIII:2018cnd}%
  \BibitemOpen
  \bibfield  {author} {\bibinfo {author} {\bibfnamefont {M.}~\bibnamefont
  {Ablikim}} \emph {et~al.} (\bibinfo {collaboration} {BESIII}),\ }\bibfield
  {title} {\enquote {\bibinfo {title} {{Polarization and Entanglement in
  Baryon-Antibaryon Pair Production in Electron-Positron Annihilation}},}\
  }\href {\doibase 10.1038/s41567-019-0494-8} {\bibfield  {journal} {\bibinfo
  {journal} {Nature Phys.}\ }\textbf {\bibinfo {volume} {15}},\ \bibinfo
  {pages} {631--634} (\bibinfo {year} {2019})},\ \Eprint
  {http://arxiv.org/abs/1808.08917} {arXiv:1808.08917 [hep-ex]} \BibitemShut
  {NoStop}%
\bibitem [{\citenamefont {Ireland}\ \emph {et~al.}(2019)\citenamefont
  {Ireland}, \citenamefont {D\"oring}, \citenamefont {Glazier}, \citenamefont
  {Haidenbauer}, \citenamefont {Mai}, \citenamefont {Murray-Smith},\ and\
  \citenamefont {R\"onchen}}]{Ireland:2019uja}%
  \BibitemOpen
  \bibfield  {author} {\bibinfo {author} {\bibfnamefont {D.~G.}\ \bibnamefont
  {Ireland}}, \bibinfo {author} {\bibfnamefont {M.}~\bibnamefont {D\"oring}},
  \bibinfo {author} {\bibfnamefont {D.~I.}\ \bibnamefont {Glazier}}, \bibinfo
  {author} {\bibfnamefont {J.}~\bibnamefont {Haidenbauer}}, \bibinfo {author}
  {\bibfnamefont {M.}~\bibnamefont {Mai}}, \bibinfo {author} {\bibfnamefont
  {R.}~\bibnamefont {Murray-Smith}}, \ and\ \bibinfo {author} {\bibfnamefont
  {D.}~\bibnamefont {R\"onchen}},\ }\bibfield  {title} {\enquote {\bibinfo
  {title} {{Kaon Photoproduction and the $\Lambda$ Decay Parameter
  $\alpha_-$}},}\ }\href {\doibase 10.1103/PhysRevLett.123.182301} {\bibfield
  {journal} {\bibinfo  {journal} {Phys. Rev. Lett.}\ }\textbf {\bibinfo
  {volume} {123}},\ \bibinfo {pages} {182301} (\bibinfo {year} {2019})},\
  \Eprint {http://arxiv.org/abs/1904.07616} {arXiv:1904.07616 [nucl-ex]}
  \BibitemShut {NoStop}%
\bibitem [{\citenamefont {Ablikim}\ \emph {et~al.}(2022)\citenamefont {Ablikim}
  \emph {et~al.}}]{BESIII:2022qax}%
  \BibitemOpen
  \bibfield  {author} {\bibinfo {author} {\bibfnamefont {M.}~\bibnamefont
  {Ablikim}} \emph {et~al.} (\bibinfo {collaboration} {BESIII}),\ }\bibfield
  {title} {\enquote {\bibinfo {title} {{Precision measurements of decay
  parameters and $CP$ asymmetry in $\Lambda$ decays}},}\ }\href@noop {} {\
  (\bibinfo {year} {2022})},\ \Eprint {http://arxiv.org/abs/2204.11058}
  {arXiv:2204.11058 [hep-ex]} \BibitemShut {NoStop}%
\bibitem [{\citenamefont {Mai}\ \emph {et~al.}(2021{\natexlab{a}})\citenamefont
  {Mai}, \citenamefont {D\"oring}, \citenamefont {Granados}, \citenamefont
  {Haberzettl}, \citenamefont {Mei\ss{}ner}, \citenamefont {R\"onchen},
  \citenamefont {Strakovsky},\ and\ \citenamefont {Workman}}]{Mai:2021vsw}%
  \BibitemOpen
  \bibfield  {author} {\bibinfo {author} {\bibfnamefont {Maxim}\ \bibnamefont
  {Mai}}, \bibinfo {author} {\bibfnamefont {Michael}\ \bibnamefont {D\"oring}},
  \bibinfo {author} {\bibfnamefont {Carlos}\ \bibnamefont {Granados}}, \bibinfo
  {author} {\bibfnamefont {Helmut}\ \bibnamefont {Haberzettl}}, \bibinfo
  {author} {\bibfnamefont {U.-G.}\ \bibnamefont {Mei\ss{}ner}}, \bibinfo
  {author} {\bibfnamefont {Deborah}\ \bibnamefont {R\"onchen}}, \bibinfo
  {author} {\bibfnamefont {Igor}\ \bibnamefont {Strakovsky}}, \ and\ \bibinfo
  {author} {\bibfnamefont {Ron}\ \bibnamefont {Workman}} (\bibinfo
  {collaboration} {J\"ulich-Bonn-Washington}),\ }\bibfield  {title} {\enquote
  {\bibinfo {title} {{J\"ulich-Bonn-Washington model for pion electroproduction
  multipoles}},}\ }\href {\doibase 10.1103/PhysRevC.103.065204} {\bibfield
  {journal} {\bibinfo  {journal} {Phys. Rev. C}\ }\textbf {\bibinfo {volume}
  {103}},\ \bibinfo {pages} {065204} (\bibinfo {year} {2021}{\natexlab{a}})},\
  \Eprint {http://arxiv.org/abs/2104.07312} {arXiv:2104.07312 [nucl-th]}
  \BibitemShut {NoStop}%
\bibitem [{\citenamefont {Mai}\ \emph {et~al.}(2021{\natexlab{b}})\citenamefont
  {Mai}, \citenamefont {D\"oring}, \citenamefont {Granados}, \citenamefont
  {Haberzettl}, \citenamefont {Hergenrather}, \citenamefont {Mei\ss{}ner},
  \citenamefont {R\"onchen}, \citenamefont {Strakovsky},\ and\ \citenamefont
  {Workman}}]{Mai:2021aui}%
  \BibitemOpen
  \bibfield  {author} {\bibinfo {author} {\bibfnamefont {Maxim}\ \bibnamefont
  {Mai}}, \bibinfo {author} {\bibfnamefont {Michael}\ \bibnamefont {D\"oring}},
  \bibinfo {author} {\bibfnamefont {Carlos}\ \bibnamefont {Granados}}, \bibinfo
  {author} {\bibfnamefont {Helmut}\ \bibnamefont {Haberzettl}}, \bibinfo
  {author} {\bibfnamefont {Jackson}\ \bibnamefont {Hergenrather}}, \bibinfo
  {author} {\bibfnamefont {U.-G.}\ \bibnamefont {Mei\ss{}ner}}, \bibinfo
  {author} {\bibfnamefont {Deborah}\ \bibnamefont {R\"onchen}}, \bibinfo
  {author} {\bibfnamefont {Igor}\ \bibnamefont {Strakovsky}}, \ and\ \bibinfo
  {author} {\bibfnamefont {Ron}\ \bibnamefont {Workman}} (\bibinfo
  {collaboration} {J\"ulich-Bonn-Washington}),\ }\bibfield  {title} {\enquote
  {\bibinfo {title} {{Coupled-channel analysis of pion- and
  eta-electroproduction with the J\"ulich-Bonn-Washington model}},}\ }\href
  {\doibase 10.1103/PhysRevC.106.015201} {\  (\bibinfo {year}
  {2021}{\natexlab{b}}),\ 10.1103/PhysRevC.106.015201},\ \Eprint
  {http://arxiv.org/abs/2111.04774} {arXiv:2111.04774 [nucl-th]} \BibitemShut
  {NoStop}%
\bibitem [{\citenamefont {Carman}\ \emph {et~al.}(2022)\citenamefont {Carman}
  \emph {et~al.}}]{CLAS:2022yzd}%
  \BibitemOpen
  \bibfield  {author} {\bibinfo {author} {\bibfnamefont {D.~S.}\ \bibnamefont
  {Carman}} \emph {et~al.} (\bibinfo {collaboration} {CLAS}),\ }\bibfield
  {title} {\enquote {\bibinfo {title} {{Beam-Recoil Transferred Polarization in
  $K^+Y$ Electroproduction in the Nucleon Resonance Region with CLAS12}},}\
  }\href {\doibase 10.1103/PhysRevC.105.065201} {\bibfield  {journal} {\bibinfo
   {journal} {Phys. Rev. C}\ }\textbf {\bibinfo {volume} {105}},\ \bibinfo
  {pages} {065201} (\bibinfo {year} {2022})},\ \Eprint
  {http://arxiv.org/abs/2202.03398} {arXiv:2202.03398 [nucl-ex]} \BibitemShut
  {NoStop}%
\bibitem [{\citenamefont {Carman}(2020)}]{Carman:2019lkk}%
  \BibitemOpen
  \bibfield  {author} {\bibinfo {author} {\bibfnamefont {Daniel~S.}\
  \bibnamefont {Carman}} (\bibinfo {collaboration} {CLAS}),\ }\bibfield
  {title} {\enquote {\bibinfo {title} {{Excited nucleon spectrum and structure
  studies with CLAS and CLAS12}},}\ }\href {\doibase 10.1063/5.0008932}
  {\bibfield  {journal} {\bibinfo  {journal} {AIP Conf. Proc.}\ }\textbf
  {\bibinfo {volume} {2249}},\ \bibinfo {pages} {030004} (\bibinfo {year}
  {2020})},\ \Eprint {http://arxiv.org/abs/1907.02407} {arXiv:1907.02407
  [nucl-ex]} \BibitemShut {NoStop}%
\bibitem [{\citenamefont {Shen}\ \emph {et~al.}(2018)\citenamefont {Shen},
  \citenamefont {R\"onchen}, \citenamefont {Mei\ss{}ner},\ and\ \citenamefont
  {Zou}}]{Shen:2017ayv}%
  \BibitemOpen
  \bibfield  {author} {\bibinfo {author} {\bibfnamefont {Chao-Wei}\
  \bibnamefont {Shen}}, \bibinfo {author} {\bibfnamefont {Deborah}\
  \bibnamefont {R\"onchen}}, \bibinfo {author} {\bibfnamefont {U.-G.}\
  \bibnamefont {Mei\ss{}ner}}, \ and\ \bibinfo {author} {\bibfnamefont
  {Bing-Song}\ \bibnamefont {Zou}},\ }\bibfield  {title} {\enquote {\bibinfo
  {title} {{Exploratory study of possible resonances in heavy meson - heavy
  baryon coupled-channel interactions}},}\ }\href {\doibase
  10.1088/1674-1137/42/2/023106} {\bibfield  {journal} {\bibinfo  {journal}
  {Chin. Phys. C}\ }\textbf {\bibinfo {volume} {42}},\ \bibinfo {pages}
  {023106} (\bibinfo {year} {2018})},\ \Eprint
  {http://arxiv.org/abs/1710.03885} {arXiv:1710.03885 [hep-ph]} \BibitemShut
  {NoStop}%
\bibitem [{\citenamefont {Wang}\ \emph
  {et~al.}(2022{\natexlab{a}})\citenamefont {Wang}, \citenamefont {Shen},
  \citenamefont {R\"onchen}, \citenamefont {Mei\ss{}ner},\ and\ \citenamefont
  {Zou}}]{Wang:2022oof}%
  \BibitemOpen
  \bibfield  {author} {\bibinfo {author} {\bibfnamefont {Zheng-Li}\
  \bibnamefont {Wang}}, \bibinfo {author} {\bibfnamefont {Chao-Wei}\
  \bibnamefont {Shen}}, \bibinfo {author} {\bibfnamefont {Deborah}\
  \bibnamefont {R\"onchen}}, \bibinfo {author} {\bibfnamefont {U.-G.}\
  \bibnamefont {Mei\ss{}ner}}, \ and\ \bibinfo {author} {\bibfnamefont
  {Bing-Song}\ \bibnamefont {Zou}},\ }\bibfield  {title} {\enquote {\bibinfo
  {title} {{Resonances in heavy meson\textendash{}heavy baryon coupled-channel
  interactions}},}\ }\href {\doibase 10.1140/epjc/s10052-022-10462-2}
  {\bibfield  {journal} {\bibinfo  {journal} {Eur. Phys. J. C}\ }\textbf
  {\bibinfo {volume} {82}},\ \bibinfo {pages} {497} (\bibinfo {year}
  {2022}{\natexlab{a}})},\ \Eprint {http://arxiv.org/abs/2204.12122}
  {arXiv:2204.12122 [hep-ph]} \BibitemShut {NoStop}%
\bibitem [{\citenamefont {Wang}\ \emph
  {et~al.}(2022{\natexlab{b}})\citenamefont {Wang}, \citenamefont {R\"onchen},
  \citenamefont {Mei\ss{}ner}, \citenamefont {Lu}, \citenamefont {Shen},\ and\
  \citenamefont {Wu}}]{Wang:2022osj}%
  \BibitemOpen
  \bibfield  {author} {\bibinfo {author} {\bibfnamefont {Yu-Fei}\ \bibnamefont
  {Wang}}, \bibinfo {author} {\bibfnamefont {Deborah}\ \bibnamefont
  {R\"onchen}}, \bibinfo {author} {\bibfnamefont {Ulf-G.}\ \bibnamefont
  {Mei\ss{}ner}}, \bibinfo {author} {\bibfnamefont {Yu}~\bibnamefont {Lu}},
  \bibinfo {author} {\bibfnamefont {Chao-Wei}\ \bibnamefont {Shen}}, \ and\
  \bibinfo {author} {\bibfnamefont {Jia-Jun}\ \bibnamefont {Wu}},\ }\bibfield
  {title} {\enquote {\bibinfo {title} {{The reaction $\pi N \to \omega N$ in a
  dynamical coupled-channel approach}},}\ }\href@noop {} {\  (\bibinfo {year}
  {2022}{\natexlab{b}})},\ \Eprint {http://arxiv.org/abs/2208.03061}
  {arXiv:2208.03061 [nucl-th]} \BibitemShut {NoStop}%
\bibitem [{\citenamefont {Mai}\ \emph {et~al.}(2017)\citenamefont {Mai},
  \citenamefont {Hu}, \citenamefont {D{\"o}ring}, \citenamefont {Pilloni},\
  and\ \citenamefont {Szczepaniak}}]{Mai:2017vot}%
  \BibitemOpen
  \bibfield  {author} {\bibinfo {author} {\bibfnamefont {M.}~\bibnamefont
  {Mai}}, \bibinfo {author} {\bibfnamefont {B.}~\bibnamefont {Hu}}, \bibinfo
  {author} {\bibfnamefont {M.}~\bibnamefont {D{\"o}ring}}, \bibinfo {author}
  {\bibfnamefont {A.}~\bibnamefont {Pilloni}}, \ and\ \bibinfo {author}
  {\bibfnamefont {A.}~\bibnamefont {Szczepaniak}},\ }\bibfield  {title}
  {\enquote {\bibinfo {title} {{Three-body Unitarity with Isobars
  Revisited}},}\ }\href {\doibase 10.1140/epja/i2017-12368-4} {\bibfield
  {journal} {\bibinfo  {journal} {Eur. Phys. J. A}\ }\textbf {\bibinfo {volume}
  {53}},\ \bibinfo {pages} {177} (\bibinfo {year} {2017})},\ \Eprint
  {http://arxiv.org/abs/1706.06118} {arXiv:1706.06118 [nucl-th]} \BibitemShut
  {NoStop}%
\bibitem [{\citenamefont {Huang}\ \emph {et~al.}(2012)\citenamefont {Huang},
  \citenamefont {Döring}, \citenamefont {Haberzettl}, \citenamefont
  {Haidenbauer}, \citenamefont {Hanhart}, \citenamefont {Krewald},
  \citenamefont {Mei\ss{}ner},\ and\ \citenamefont {Nakayama}}]{Huang:2011as}%
  \BibitemOpen
  \bibfield  {author} {\bibinfo {author} {\bibfnamefont {F.}~\bibnamefont
  {Huang}}, \bibinfo {author} {\bibfnamefont {M.}~\bibnamefont {Döring}},
  \bibinfo {author} {\bibfnamefont {H.}~\bibnamefont {Haberzettl}}, \bibinfo
  {author} {\bibfnamefont {J.}~\bibnamefont {Haidenbauer}}, \bibinfo {author}
  {\bibfnamefont {C.}~\bibnamefont {Hanhart}}, \bibinfo {author} {\bibfnamefont
  {S.}~\bibnamefont {Krewald}}, \bibinfo {author} {\bibfnamefont {U.-G.}\
  \bibnamefont {Mei\ss{}ner}}, \ and\ \bibinfo {author} {\bibfnamefont
  {K.}~\bibnamefont {Nakayama}},\ }\bibfield  {title} {\enquote {\bibinfo
  {title} {{Pion photoproduction in a dynamical coupled-channels model}},}\
  }\href {\doibase 10.1103/PhysRevC.85.054003} {\bibfield  {journal} {\bibinfo
  {journal} {Phys. Rev. C}\ }\textbf {\bibinfo {volume} {85}},\ \bibinfo
  {pages} {054003} (\bibinfo {year} {2012})},\ \Eprint
  {http://arxiv.org/abs/1110.3833} {arXiv:1110.3833 [nucl-th]} \BibitemShut
  {NoStop}%
\bibitem [{\citenamefont {Workman}\ \emph {et~al.}(2012)\citenamefont
  {Workman}, \citenamefont {Arndt}, \citenamefont {Briscoe}, \citenamefont
  {Paris},\ and\ \citenamefont {Strakovsky}}]{Workman:2012hx}%
  \BibitemOpen
  \bibfield  {author} {\bibinfo {author} {\bibfnamefont {R.~L.}\ \bibnamefont
  {Workman}}, \bibinfo {author} {\bibfnamefont {R.~A.}\ \bibnamefont {Arndt}},
  \bibinfo {author} {\bibfnamefont {W.~J.}\ \bibnamefont {Briscoe}}, \bibinfo
  {author} {\bibfnamefont {M.~W.}\ \bibnamefont {Paris}}, \ and\ \bibinfo
  {author} {\bibfnamefont {I.~I.}\ \bibnamefont {Strakovsky}},\ }\bibfield
  {title} {\enquote {\bibinfo {title} {{Parameterization dependence of T matrix
  poles and eigenphases from a fit to $\pi$N elastic scattering data}},}\
  }\href {\doibase 10.1103/PhysRevC.86.035202} {\bibfield  {journal} {\bibinfo
  {journal} {Phys. Rev. C}\ }\textbf {\bibinfo {volume} {86}},\ \bibinfo
  {pages} {035202} (\bibinfo {year} {2012})},\ \Eprint
  {http://arxiv.org/abs/1204.2277} {arXiv:1204.2277 [hep-ph]} \BibitemShut
  {NoStop}%
\bibitem [{\citenamefont {{SAID/GWU website}}()}]{SAID}%
  \BibitemOpen
  \bibfield  {author} {\bibinfo {author} {\bibnamefont {{SAID/GWU website}}},\
  }\href@noop {} {}\bibinfo {howpublished}
  {\url{http://gwdac.phys.gwu.edu}}\BibitemShut {NoStop}%
\bibitem [{\citenamefont {{Website of Bonn-Gatchina group with analysis
  results}}()}]{BnGa_web}%
  \BibitemOpen
  \bibfield  {author} {\bibinfo {author} {\bibnamefont {{Website of
  Bonn-Gatchina group with analysis results}}},\ }\href@noop {} {}\bibinfo
  {howpublished} {\url{https://pwa.hiskp.uni-bonn.de}}\BibitemShut {NoStop}%
\bibitem [{\citenamefont {{Figures representing the full fit result of this
  study, including a display of all data}}()}]{Juelichmodel:online}%
  \BibitemOpen
  \bibfield  {author} {\bibinfo {author} {\bibnamefont {{Figures representing
  the full fit result of this study, including a display of all data}}},\
  }\href@noop {} {}\bibinfo {howpublished}
  {\url{http://collaborations.fz-juelich.de/ikp/meson-baryon/juelich_amplitudes.html}}\BibitemShut
  {NoStop}%
\bibitem [{\citenamefont {Zachariou}\ \emph {et~al.}(2022)\citenamefont
  {Zachariou} \emph {et~al.}}]{CLAS:2021hex}%
  \BibitemOpen
  \bibfield  {author} {\bibinfo {author} {\bibfnamefont {N.}~\bibnamefont
  {Zachariou}} \emph {et~al.} (\bibinfo {collaboration} {CLAS}),\ }\bibfield
  {title} {\enquote {\bibinfo {title} {{Beam-spin asymmetry
  $\boldsymbol{\Sigma}$ for $\Sigma^-$ hyperon photoproduction off the
  neutron}},}\ }\href {\doibase 10.1016/j.physletb.2022.136985} {\bibfield
  {journal} {\bibinfo  {journal} {Phys. Lett. B}\ }\textbf {\bibinfo {volume}
  {827}},\ \bibinfo {pages} {136985} (\bibinfo {year} {2022})},\ \Eprint
  {http://arxiv.org/abs/2106.13957} {arXiv:2106.13957 [nucl-ex]} \BibitemShut
  {NoStop}%
\bibitem [{\citenamefont {Kohl}\ \emph {et~al.}(2021)\citenamefont {Kohl} \emph
  {et~al.}}]{BGOOD:2021oxp}%
  \BibitemOpen
  \bibfield  {author} {\bibinfo {author} {\bibfnamefont {K.}~\bibnamefont
  {Kohl}} \emph {et~al.} (\bibinfo {collaboration} {BGOOD}),\ }\bibfield
  {title} {\enquote {\bibinfo {title} {{Measurement of the $\gamma n\rightarrow
  K^0\Sigma^0$ differential cross section over the $K^*$ threshold}},}\
  }\href@noop {} {\  (\bibinfo {year} {2021})},\ \Eprint
  {http://arxiv.org/abs/2108.13319} {arXiv:2108.13319 [nucl-ex]} \BibitemShut
  {NoStop}%
\bibitem [{\citenamefont {Akondi}\ \emph {et~al.}(2019)\citenamefont {Akondi}
  \emph {et~al.}}]{A2:2018doh}%
  \BibitemOpen
  \bibfield  {author} {\bibinfo {author} {\bibfnamefont {C.~S.}\ \bibnamefont
  {Akondi}} \emph {et~al.} (\bibinfo {collaboration} {A2}),\ }\bibfield
  {title} {\enquote {\bibinfo {title} {{Experimental study of the $\gamma
  p\rightarrow K^0\Sigma^+$, $\gamma n\rightarrow K^0\Lambda$, and $\gamma
  n\rightarrow K^0 \Sigma^0$ reactions at the Mainz Microtron}},}\ }\href
  {\doibase 10.1140/epja/i2019-12924-x} {\bibfield  {journal} {\bibinfo
  {journal} {Eur. Phys. J. A}\ }\textbf {\bibinfo {volume} {55}},\ \bibinfo
  {pages} {202} (\bibinfo {year} {2019})},\ \Eprint
  {http://arxiv.org/abs/1811.05547} {arXiv:1811.05547 [nucl-ex]} \BibitemShut
  {NoStop}%
\bibitem [{\citenamefont {Bockhorst}\ \emph {et~al.}(1994)\citenamefont
  {Bockhorst} \emph {et~al.}}]{Bockhorst:1994jf}%
  \BibitemOpen
  \bibfield  {author} {\bibinfo {author} {\bibfnamefont {M.}~\bibnamefont
  {Bockhorst}} \emph {et~al.},\ }\bibfield  {title} {\enquote {\bibinfo {title}
  {{Measurement of $\gamma p \to K^+\Lambda$ and $\gamma p \to K^+ \Sigma^0$ at
  photon energies up to 1.47-GeV}},}\ }\href {\doibase 10.1007/BF01577542}
  {\bibfield  {journal} {\bibinfo  {journal} {Z. Phys. C}\ }\textbf {\bibinfo
  {volume} {63}},\ \bibinfo {pages} {37--47} (\bibinfo {year}
  {1994})}\BibitemShut {NoStop}%
\bibitem [{\citenamefont {Tran}\ \emph {et~al.}(1998)\citenamefont {Tran} \emph
  {et~al.}}]{SAPHIR:1998fev}%
  \BibitemOpen
  \bibfield  {author} {\bibinfo {author} {\bibfnamefont {M.~Q.}\ \bibnamefont
  {Tran}} \emph {et~al.} (\bibinfo {collaboration} {SAPHIR}),\ }\bibfield
  {title} {\enquote {\bibinfo {title} {{Measurement of $\gamma p \to
  K^+\Lambda$ and $\gamma p \to K^+\Sigma^0$ at photon energies up to
  2-GeV}},}\ }\href {\doibase 10.1016/S0370-2693(98)01393-8} {\bibfield
  {journal} {\bibinfo  {journal} {Phys. Lett. B}\ }\textbf {\bibinfo {volume}
  {445}},\ \bibinfo {pages} {20--26} (\bibinfo {year} {1998})}\BibitemShut
  {NoStop}%
\bibitem [{\citenamefont {Glander}\ \emph {et~al.}(2004)\citenamefont {Glander}
  \emph {et~al.}}]{Glander:2003jw}%
  \BibitemOpen
  \bibfield  {author} {\bibinfo {author} {\bibfnamefont {K.~H.}\ \bibnamefont
  {Glander}} \emph {et~al.},\ }\bibfield  {title} {\enquote {\bibinfo {title}
  {{Measurement of $\gamma p \to K^+\Lambda$ and $\gamma p \to K^+\Sigma^0$ at
  photon energies up to 2.6-GeV}},}\ }\href {\doibase
  10.1140/epja/i2003-10119-x} {\bibfield  {journal} {\bibinfo  {journal} {Eur.
  Phys. J. A}\ }\textbf {\bibinfo {volume} {19}},\ \bibinfo {pages} {251--273}
  (\bibinfo {year} {2004})},\ \Eprint {http://arxiv.org/abs/nucl-ex/0308025}
  {arXiv:nucl-ex/0308025} \BibitemShut {NoStop}%
\bibitem [{\citenamefont {Dey}\ \emph {et~al.}(2010)\citenamefont {Dey} \emph
  {et~al.}}]{CLAS:2010aen}%
  \BibitemOpen
  \bibfield  {author} {\bibinfo {author} {\bibfnamefont {B.}~\bibnamefont
  {Dey}} \emph {et~al.} (\bibinfo {collaboration} {CLAS}),\ }\bibfield  {title}
  {\enquote {\bibinfo {title} {{Differential cross sections and recoil
  polarizations for the reaction $\gamma p \to K^{+} \Sigma^{0}$}},}\ }\href
  {\doibase 10.1103/PhysRevC.82.025202} {\bibfield  {journal} {\bibinfo
  {journal} {Phys. Rev. C}\ }\textbf {\bibinfo {volume} {82}},\ \bibinfo
  {pages} {025202} (\bibinfo {year} {2010})},\ \Eprint
  {http://arxiv.org/abs/1006.0374} {arXiv:1006.0374 [nucl-ex]} \BibitemShut
  {NoStop}%
\bibitem [{\citenamefont {Jude}\ \emph {et~al.}(2021)\citenamefont {Jude} \emph
  {et~al.}}]{Jude:2020byj}%
  \BibitemOpen
  \bibfield  {author} {\bibinfo {author} {\bibfnamefont {T.~C.}\ \bibnamefont
  {Jude}} \emph {et~al.},\ }\bibfield  {title} {\enquote {\bibinfo {title}
  {{Observation of a cusp-like structure in the $\gamma p\to K^+\Sigma^0$ cross
  section at forward angles and low momentum transfer}},}\ }\href {\doibase
  10.1016/j.physletb.2021.136559} {\bibfield  {journal} {\bibinfo  {journal}
  {Phys. Lett. B}\ }\textbf {\bibinfo {volume} {820}},\ \bibinfo {pages}
  {136559} (\bibinfo {year} {2021})},\ \Eprint
  {http://arxiv.org/abs/2006.12437} {arXiv:2006.12437 [nucl-ex]} \BibitemShut
  {NoStop}%
\bibitem [{\citenamefont {Alef}\ \emph {et~al.}(2021)\citenamefont {Alef} \emph
  {et~al.}}]{Alef:2020yul}%
  \BibitemOpen
  \bibfield  {author} {\bibinfo {author} {\bibfnamefont {S.}~\bibnamefont
  {Alef}} \emph {et~al.},\ }\bibfield  {title} {\enquote {\bibinfo {title}
  {{$K^+\Lambda$ photoproduction at forward angles and low momentum
  transfer}},}\ }\href {\doibase 10.1140/epja/s10050-021-00392-0} {\bibfield
  {journal} {\bibinfo  {journal} {Eur. Phys. J. A}\ }\textbf {\bibinfo {volume}
  {57}},\ \bibinfo {pages} {80} (\bibinfo {year} {2021})},\ \Eprint
  {http://arxiv.org/abs/2006.12350} {arXiv:2006.12350 [nucl-ex]} \BibitemShut
  {NoStop}%
\bibitem [{\citenamefont {Afzal}\ \emph {et~al.}(2020)\citenamefont {Afzal}
  \emph {et~al.}}]{CBELSATAPS:2020cwk}%
  \BibitemOpen
  \bibfield  {author} {\bibinfo {author} {\bibfnamefont {F.}~\bibnamefont
  {Afzal}} \emph {et~al.} (\bibinfo {collaboration} {CBELSA/TAPS}),\ }\bibfield
   {title} {\enquote {\bibinfo {title} {{Observation of the p\ensuremath{\eta}'
  Cusp in the New Precise Beam Asymmetry \ensuremath{\Sigma} Data for
  \ensuremath{\gamma}p\textrightarrow{}p\ensuremath{\eta}}},}\ }\href {\doibase
  10.1103/PhysRevLett.125.152002} {\bibfield  {journal} {\bibinfo  {journal}
  {Phys. Rev. Lett.}\ }\textbf {\bibinfo {volume} {125}},\ \bibinfo {pages}
  {152002} (\bibinfo {year} {2020})},\ \Eprint
  {http://arxiv.org/abs/2009.06248} {arXiv:2009.06248 [nucl-ex]} \BibitemShut
  {NoStop}%
\bibitem [{\citenamefont {Brody}\ \emph {et~al.}(1960)\citenamefont {Brody},
  \citenamefont {Wetherell},\ and\ \citenamefont {Walker}}]{Brody:1960zz}%
  \BibitemOpen
  \bibfield  {author} {\bibinfo {author} {\bibfnamefont {H.~M.}\ \bibnamefont
  {Brody}}, \bibinfo {author} {\bibfnamefont {A.~M.}\ \bibnamefont
  {Wetherell}}, \ and\ \bibinfo {author} {\bibfnamefont {R.~L.}\ \bibnamefont
  {Walker}},\ }\bibfield  {title} {\enquote {\bibinfo {title} {{Photoproduction
  of $K^+$ Mesons in Hydrogen}},}\ }\href {\doibase 10.1103/PhysRev.119.1710}
  {\bibfield  {journal} {\bibinfo  {journal} {Phys. Rev.}\ }\textbf {\bibinfo
  {volume} {119}},\ \bibinfo {pages} {1710--1716} (\bibinfo {year}
  {1960})}\BibitemShut {NoStop}%
\bibitem [{\citenamefont {Anderson}\ \emph {et~al.}(1962)\citenamefont
  {Anderson}, \citenamefont {Gabathuler}, \citenamefont {Jones}, \citenamefont
  {McDaniel},\ and\ \citenamefont {Sadoff}}]{Anderson:1962za}%
  \BibitemOpen
  \bibfield  {author} {\bibinfo {author} {\bibfnamefont {R.~L.}\ \bibnamefont
  {Anderson}}, \bibinfo {author} {\bibfnamefont {E.}~\bibnamefont
  {Gabathuler}}, \bibinfo {author} {\bibfnamefont {D.}~\bibnamefont {Jones}},
  \bibinfo {author} {\bibfnamefont {B.~D.}\ \bibnamefont {McDaniel}}, \ and\
  \bibinfo {author} {\bibfnamefont {A.~J.}\ \bibnamefont {Sadoff}},\ }\bibfield
   {title} {\enquote {\bibinfo {title} {{Photoproduction of K+ mesons in
  hydrogen}},}\ }\href {\doibase 10.1103/PhysRevLett.9.131} {\bibfield
  {journal} {\bibinfo  {journal} {Phys. Rev. Lett.}\ }\textbf {\bibinfo
  {volume} {9}},\ \bibinfo {pages} {131--133} (\bibinfo {year}
  {1962})}\BibitemShut {NoStop}%
\bibitem [{\citenamefont {Fujii}\ \emph {et~al.}(1970)\citenamefont {Fujii}
  \emph {et~al.}}]{Fujii:1970gn}%
  \BibitemOpen
  \bibfield  {author} {\bibinfo {author} {\bibfnamefont {T.}~\bibnamefont
  {Fujii}} \emph {et~al.},\ }\bibfield  {title} {\enquote {\bibinfo {title}
  {{Photoproduction of k+ mesons and polarization of lambda0 hyperons in the
  1-gev range}},}\ }\href {\doibase 10.1103/PhysRevD.2.439} {\bibfield
  {journal} {\bibinfo  {journal} {Phys. Rev. D}\ }\textbf {\bibinfo {volume}
  {2}},\ \bibinfo {pages} {439--448} (\bibinfo {year} {1970})}\BibitemShut
  {NoStop}%
\bibitem [{\citenamefont {Bleckmann}\ \emph {et~al.}(1970)\citenamefont
  {Bleckmann}, \citenamefont {Herda}, \citenamefont {Opara}, \citenamefont
  {Schulz}, \citenamefont {Schwille},\ and\ \citenamefont
  {Urbahn}}]{Bleckmann:1970kb}%
  \BibitemOpen
  \bibfield  {author} {\bibinfo {author} {\bibfnamefont {A.}~\bibnamefont
  {Bleckmann}}, \bibinfo {author} {\bibfnamefont {S.}~\bibnamefont {Herda}},
  \bibinfo {author} {\bibfnamefont {U.}~\bibnamefont {Opara}}, \bibinfo
  {author} {\bibfnamefont {W.}~\bibnamefont {Schulz}}, \bibinfo {author}
  {\bibfnamefont {W.~J.}\ \bibnamefont {Schwille}}, \ and\ \bibinfo {author}
  {\bibfnamefont {H.}~\bibnamefont {Urbahn}},\ }\bibfield  {title} {\enquote
  {\bibinfo {title} {{Photoproduction of k+ lambda and k+ sigma0 from hydrogen
  between 1.3 and 1.45 gev}},}\ }\href {\doibase 10.1007/BF01408507} {\bibfield
   {journal} {\bibinfo  {journal} {Z. Phys.}\ }\textbf {\bibinfo {volume}
  {239}},\ \bibinfo {pages} {1--15} (\bibinfo {year} {1970})}\BibitemShut
  {NoStop}%
\bibitem [{\citenamefont {Feller}\ \emph {et~al.}(1972)\citenamefont {Feller},
  \citenamefont {Menze}, \citenamefont {Opara}, \citenamefont {Schulz},\ and\
  \citenamefont {Schwille}}]{Feller:1972ph}%
  \BibitemOpen
  \bibfield  {author} {\bibinfo {author} {\bibfnamefont {P.}~\bibnamefont
  {Feller}}, \bibinfo {author} {\bibfnamefont {D.}~\bibnamefont {Menze}},
  \bibinfo {author} {\bibfnamefont {U.}~\bibnamefont {Opara}}, \bibinfo
  {author} {\bibfnamefont {W.}~\bibnamefont {Schulz}}, \ and\ \bibinfo {author}
  {\bibfnamefont {W.~J.}\ \bibnamefont {Schwille}},\ }\bibfield  {title}
  {\enquote {\bibinfo {title} {{Photoproduction of k+ lambda0 and k+ sigma0
  from hydrogen at constant momentum transfer t between 1.05 and 2.2 gev}},}\
  }\href {\doibase 10.1016/0550-3213(72)90379-3} {\bibfield  {journal}
  {\bibinfo  {journal} {Nucl. Phys. B}\ }\textbf {\bibinfo {volume} {39}},\
  \bibinfo {pages} {413--420} (\bibinfo {year} {1972})}\BibitemShut {NoStop}%
\bibitem [{\citenamefont {McNabb}\ \emph {et~al.}(2004)\citenamefont {McNabb}
  \emph {et~al.}}]{CLAS:2003zrd}%
  \BibitemOpen
  \bibfield  {author} {\bibinfo {author} {\bibfnamefont {J.~W.~C.}\
  \bibnamefont {McNabb}} \emph {et~al.} (\bibinfo {collaboration} {CLAS}),\
  }\bibfield  {title} {\enquote {\bibinfo {title} {{Hyperon photoproduction in
  the nucleon resonance region}},}\ }\href {\doibase
  10.1103/PhysRevC.69.042201} {\bibfield  {journal} {\bibinfo  {journal} {Phys.
  Rev. C}\ }\textbf {\bibinfo {volume} {69}},\ \bibinfo {pages} {042201}
  (\bibinfo {year} {2004})},\ \Eprint {http://arxiv.org/abs/nucl-ex/0305028}
  {arXiv:nucl-ex/0305028} \BibitemShut {NoStop}%
\bibitem [{\citenamefont {Bradford}\ \emph {et~al.}(2006)\citenamefont
  {Bradford} \emph {et~al.}}]{CLAS:2005lui}%
  \BibitemOpen
  \bibfield  {author} {\bibinfo {author} {\bibfnamefont {R.}~\bibnamefont
  {Bradford}} \emph {et~al.} (\bibinfo {collaboration} {CLAS}),\ }\bibfield
  {title} {\enquote {\bibinfo {title} {{Differential cross sections for gamma +
  p ---\ensuremath{>} K+ + Y for Lambda and Sigma0 hyperons}},}\ }\href
  {\doibase 10.1103/PhysRevC.73.035202} {\bibfield  {journal} {\bibinfo
  {journal} {Phys. Rev. C}\ }\textbf {\bibinfo {volume} {73}},\ \bibinfo
  {pages} {035202} (\bibinfo {year} {2006})},\ \Eprint
  {http://arxiv.org/abs/nucl-ex/0509033} {arXiv:nucl-ex/0509033} \BibitemShut
  {NoStop}%
\bibitem [{\citenamefont {Sumihama}\ \emph {et~al.}(2006)\citenamefont
  {Sumihama} \emph {et~al.}}]{LEPS:2005hji}%
  \BibitemOpen
  \bibfield  {author} {\bibinfo {author} {\bibfnamefont {M.}~\bibnamefont
  {Sumihama}} \emph {et~al.} (\bibinfo {collaboration} {LEPS}),\ }\bibfield
  {title} {\enquote {\bibinfo {title} {{The $\vec\gamma p \to K^+ \Lambda$ and
  $\vec\gamma p \to K^+ \Sigma^0$ reactions at forward angles with photon
  energies from 1.5 GeV to 2.4 GeV}},}\ }\href {\doibase
  10.1103/PhysRevC.73.035214} {\bibfield  {journal} {\bibinfo  {journal} {Phys.
  Rev. C}\ }\textbf {\bibinfo {volume} {73}},\ \bibinfo {pages} {035214}
  (\bibinfo {year} {2006})},\ \Eprint {http://arxiv.org/abs/hep-ex/0512053}
  {arXiv:hep-ex/0512053} \BibitemShut {NoStop}%
\bibitem [{\citenamefont {Kohri}\ \emph {et~al.}(2006)\citenamefont {Kohri}
  \emph {et~al.}}]{Kohri:2006yx}%
  \BibitemOpen
  \bibfield  {author} {\bibinfo {author} {\bibfnamefont {H.}~\bibnamefont
  {Kohri}} \emph {et~al.},\ }\bibfield  {title} {\enquote {\bibinfo {title}
  {{Differential cross section and photon beam asymmetry for the $\vec\gamma n
  \to K^+ \Sigma^-$ reaction at $E_\gamma =$ 1.5 GeV-2.4 GeV}},}\ }\href
  {\doibase 10.1103/PhysRevLett.97.082003} {\bibfield  {journal} {\bibinfo
  {journal} {Phys. Rev. Lett.}\ }\textbf {\bibinfo {volume} {97}},\ \bibinfo
  {pages} {082003} (\bibinfo {year} {2006})},\ \Eprint
  {http://arxiv.org/abs/hep-ex/0602015} {arXiv:hep-ex/0602015} \BibitemShut
  {NoStop}%
\bibitem [{\citenamefont {Jude}\ \emph {et~al.}(2014)\citenamefont {Jude} \emph
  {et~al.}}]{CrystalBallatMAMI:2013iig}%
  \BibitemOpen
  \bibfield  {author} {\bibinfo {author} {\bibfnamefont {T.~C.}\ \bibnamefont
  {Jude}} \emph {et~al.} (\bibinfo {collaboration} {Crystal Ball at MAMI}),\
  }\bibfield  {title} {\enquote {\bibinfo {title} {{$K^+\Lambda$ and
  $K^+\Sigma^0$ photoproduction with fine center-of-mass energy resolution}},}\
  }\href {\doibase 10.1016/j.physletb.2014.06.015} {\bibfield  {journal}
  {\bibinfo  {journal} {Phys. Lett. B}\ }\textbf {\bibinfo {volume} {735}},\
  \bibinfo {pages} {112--118} (\bibinfo {year} {2014})},\ \Eprint
  {http://arxiv.org/abs/1308.5659} {arXiv:1308.5659 [nucl-ex]} \BibitemShut
  {NoStop}%
\bibitem [{\citenamefont {Lleres}\ \emph {et~al.}(2007)\citenamefont {Lleres}
  \emph {et~al.}}]{Lleres:2007tx}%
  \BibitemOpen
  \bibfield  {author} {\bibinfo {author} {\bibfnamefont {A.}~\bibnamefont
  {Lleres}} \emph {et~al.},\ }\bibfield  {title} {\enquote {\bibinfo {title}
  {{Polarization observable measurements for $\gamma p \to K^+\Lambda$ and
  $\gamma p \to K^+\Sigma^0$ for energies up to 1.5-GeV}},}\ }\href {\doibase
  10.1140/epja/i2006-10167-8} {\bibfield  {journal} {\bibinfo  {journal} {Eur.
  Phys. J. A}\ }\textbf {\bibinfo {volume} {31}},\ \bibinfo {pages} {79--93}
  (\bibinfo {year} {2007})}\BibitemShut {NoStop}%
\bibitem [{\citenamefont {Zegers}\ \emph {et~al.}(2003)\citenamefont {Zegers}
  \emph {et~al.}}]{Zegers:2003ux}%
  \BibitemOpen
  \bibfield  {author} {\bibinfo {author} {\bibfnamefont {R.~G.~T.}\
  \bibnamefont {Zegers}} \emph {et~al.} (\bibinfo {collaboration} {LEPS}),\
  }\bibfield  {title} {\enquote {\bibinfo {title} {{Beam polarization
  asymmetries for the $p(\gamma, K^+) \Lambda$ and $p(\gamma, K^+) \Sigma^0$
  reactions at $E_\gamma =$ 1.5 Gev - 2.4 GeV}},}\ }\href {\doibase
  10.1103/PhysRevLett.91.092001} {\bibfield  {journal} {\bibinfo  {journal}
  {Phys. Rev. Lett.}\ }\textbf {\bibinfo {volume} {91}},\ \bibinfo {pages}
  {092001} (\bibinfo {year} {2003})},\ \Eprint
  {http://arxiv.org/abs/nucl-ex/0302005} {arXiv:nucl-ex/0302005} \BibitemShut
  {NoStop}%
\bibitem [{\citenamefont {Paterson}\ \emph {et~al.}(2016)\citenamefont
  {Paterson} \emph {et~al.}}]{Paterson:2016vmc}%
  \BibitemOpen
  \bibfield  {author} {\bibinfo {author} {\bibfnamefont {C.~A.}\ \bibnamefont
  {Paterson}} \emph {et~al.} (\bibinfo {collaboration} {CLAS}),\ }\bibfield
  {title} {\enquote {\bibinfo {title} {{Photoproduction of $\Lambda$ and
  $\Sigma^0$ hyperons using linearly polarized photons}},}\ }\href {\doibase
  10.1103/PhysRevC.93.065201} {\bibfield  {journal} {\bibinfo  {journal} {Phys.
  Rev. C}\ }\textbf {\bibinfo {volume} {93}},\ \bibinfo {pages} {065201}
  (\bibinfo {year} {2016})},\ \Eprint {http://arxiv.org/abs/1603.06492}
  {arXiv:1603.06492 [nucl-ex]} \BibitemShut {NoStop}%
\bibitem [{\citenamefont {Bradford}\ \emph {et~al.}(2007)\citenamefont
  {Bradford} \emph {et~al.}}]{Bradford:2006ba}%
  \BibitemOpen
  \bibfield  {author} {\bibinfo {author} {\bibfnamefont {R.~K.}\ \bibnamefont
  {Bradford}} \emph {et~al.} (\bibinfo {collaboration} {CLAS}),\ }\bibfield
  {title} {\enquote {\bibinfo {title} {{First measurement of beam-recoil
  observables C(x) and C(z) in hyperon photoproduction}},}\ }\href {\doibase
  10.1103/PhysRevC.75.035205} {\bibfield  {journal} {\bibinfo  {journal} {Phys.
  Rev. C}\ }\textbf {\bibinfo {volume} {75}},\ \bibinfo {pages} {035205}
  (\bibinfo {year} {2007})},\ \Eprint {http://arxiv.org/abs/nucl-ex/0611034}
  {arXiv:nucl-ex/0611034} \BibitemShut {NoStop}%
\bibitem [{\citenamefont {Carnahan}(2003)}]{Carnahan:2003mk}%
  \BibitemOpen
  \bibfield  {author} {\bibinfo {author} {\bibfnamefont {Bryan}\ \bibnamefont
  {Carnahan}},\ }\emph {\bibinfo {title} {{Strangeness photoproduction in the
  $\gamma p \to K^0 \Sigma^+$ reaction}}},\ \href {\doibase 10.2172/824935}
  {Ph.D. thesis},\ \bibinfo  {school} {Catholic U. America} (\bibinfo {year}
  {2003})\BibitemShut {NoStop}%
\bibitem [{\citenamefont {Castelijns}\ \emph {et~al.}(2008)\citenamefont
  {Castelijns} \emph {et~al.}}]{CBELSATAPS:2007oqn}%
  \BibitemOpen
  \bibfield  {author} {\bibinfo {author} {\bibfnamefont {R.}~\bibnamefont
  {Castelijns}} \emph {et~al.} (\bibinfo {collaboration} {CBELSA/TAPS}),\
  }\bibfield  {title} {\enquote {\bibinfo {title} {{Nucleon resonance decay by
  the $K^0\Sigma^+$ channel}},}\ }\href {\doibase 10.1140/epja/i2007-10529-8}
  {\bibfield  {journal} {\bibinfo  {journal} {Eur. Phys. J. A}\ }\textbf
  {\bibinfo {volume} {35}},\ \bibinfo {pages} {39--45} (\bibinfo {year}
  {2008})},\ \Eprint {http://arxiv.org/abs/nucl-ex/0702033}
  {arXiv:nucl-ex/0702033} \BibitemShut {NoStop}%
\bibitem [{\citenamefont {Ewald}\ \emph {et~al.}(2012)\citenamefont {Ewald}
  \emph {et~al.}}]{CBELSATAPS:2011gly}%
  \BibitemOpen
  \bibfield  {author} {\bibinfo {author} {\bibfnamefont {R.}~\bibnamefont
  {Ewald}} \emph {et~al.} (\bibinfo {collaboration} {CBELSA/TAPS}),\ }\bibfield
   {title} {\enquote {\bibinfo {title} {{Anomaly in the $K^0_S \Sigma^+$
  photoproduction cross section off the proton at the $K^*$ threshold}},}\
  }\href {\doibase 10.1016/j.physletb.2012.05.066} {\bibfield  {journal}
  {\bibinfo  {journal} {Phys. Lett. B}\ }\textbf {\bibinfo {volume} {713}},\
  \bibinfo {pages} {180--185} (\bibinfo {year} {2012})},\ \Eprint
  {http://arxiv.org/abs/1112.0811} {arXiv:1112.0811 [nucl-ex]} \BibitemShut
  {NoStop}%
\bibitem [{\citenamefont {Aguar-Bartolome}\ \emph {et~al.}(2013)\citenamefont
  {Aguar-Bartolome} \emph {et~al.}}]{A2:2013cqk}%
  \BibitemOpen
  \bibfield  {author} {\bibinfo {author} {\bibfnamefont {P.}~\bibnamefont
  {Aguar-Bartolome}} \emph {et~al.} (\bibinfo {collaboration} {A2}),\
  }\bibfield  {title} {\enquote {\bibinfo {title} {{Measurement of the $\gamma
  p \to K^{0} \Sigma^{+}$ reaction with the Crystal Ball/TAPS detectors at the
  Mainz Microtron}},}\ }\href {\doibase 10.1103/PhysRevC.88.044601} {\bibfield
  {journal} {\bibinfo  {journal} {Phys. Rev. C}\ }\textbf {\bibinfo {volume}
  {88}},\ \bibinfo {pages} {044601} (\bibinfo {year} {2013})},\ \Eprint
  {http://arxiv.org/abs/1306.1243} {arXiv:1306.1243 [hep-ex]} \BibitemShut
  {NoStop}%
\bibitem [{\citenamefont {Goers}\ \emph {et~al.}(1999)\citenamefont {Goers}
  \emph {et~al.}}]{SAPHIR:1999wfu}%
  \BibitemOpen
  \bibfield  {author} {\bibinfo {author} {\bibfnamefont {S}~\bibnamefont
  {Goers}} \emph {et~al.} (\bibinfo {collaboration} {SAPHIR}),\ }\bibfield
  {title} {\enquote {\bibinfo {title} {{Measurement of $\gamma p \to
  K^0\Sigma^+$ at photon energies up to 1.55-GeV}},}\ }\href {\doibase
  10.1016/S0370-2693(99)01031-X} {\bibfield  {journal} {\bibinfo  {journal}
  {Phys. Lett. B}\ }\textbf {\bibinfo {volume} {464}},\ \bibinfo {pages}
  {331--338} (\bibinfo {year} {1999})}\BibitemShut {NoStop}%
\bibitem [{\citenamefont {Nepali}\ \emph {et~al.}(2013)\citenamefont {Nepali}
  \emph {et~al.}}]{CLAS:2013owj}%
  \BibitemOpen
  \bibfield  {author} {\bibinfo {author} {\bibfnamefont {C.~S.}\ \bibnamefont
  {Nepali}} \emph {et~al.} (\bibinfo {collaboration} {CLAS}),\ }\bibfield
  {title} {\enquote {\bibinfo {title} {{Transverse polarization of
  $\Sigma^+$(1189) in photoproduction on a hydrogen target in CLAS}},}\ }\href
  {\doibase 10.1103/PhysRevC.87.045206} {\bibfield  {journal} {\bibinfo
  {journal} {Phys. Rev. C}\ }\textbf {\bibinfo {volume} {87}},\ \bibinfo
  {pages} {045206} (\bibinfo {year} {2013})},\ \Eprint
  {http://arxiv.org/abs/1302.0322} {arXiv:1302.0322 [nucl-ex]} \BibitemShut
  {NoStop}%
\bibitem [{\citenamefont {Th{\"o}rnig}(2021)}]{Thornig2021-hh}%
  \BibitemOpen
  \bibfield  {author} {\bibinfo {author} {\bibfnamefont {Philipp}\ \bibnamefont
  {Th{\"o}rnig}},\ }\bibfield  {title} {\enquote {\bibinfo {title} {{JURECA}:
  Data centric and booster modules implementing the modular supercomputing
  architecture at j{\"u}lich supercomputing centre},}\ }\href@noop {}
  {\bibfield  {journal} {\bibinfo  {journal} {J. large-scale res. facil.
  JLSRF}\ }\textbf {\bibinfo {volume} {7}} (\bibinfo {year}
  {2021})}\BibitemShut {NoStop}%
\bibitem [{\citenamefont {Baru}\ \emph {et~al.}(2011)\citenamefont {Baru},
  \citenamefont {Hanhart}, \citenamefont {Hoferichter}, \citenamefont {Kubis},
  \citenamefont {Nogga},\ and\ \citenamefont {Phillips}}]{Baru:2011bw}%
  \BibitemOpen
  \bibfield  {author} {\bibinfo {author} {\bibfnamefont {V.}~\bibnamefont
  {Baru}}, \bibinfo {author} {\bibfnamefont {C.}~\bibnamefont {Hanhart}},
  \bibinfo {author} {\bibfnamefont {M.}~\bibnamefont {Hoferichter}}, \bibinfo
  {author} {\bibfnamefont {B.}~\bibnamefont {Kubis}}, \bibinfo {author}
  {\bibfnamefont {A.}~\bibnamefont {Nogga}}, \ and\ \bibinfo {author}
  {\bibfnamefont {D.~R.}\ \bibnamefont {Phillips}},\ }\bibfield  {title}
  {\enquote {\bibinfo {title} {{Precision calculation of threshold $\pi^-d$
  scattering, $\pi$N scattering lengths, and the GMO sum rule}},}\ }\href
  {\doibase 10.1016/j.nuclphysa.2011.09.015} {\bibfield  {journal} {\bibinfo
  {journal} {Nucl. Phys. A}\ }\textbf {\bibinfo {volume} {872}},\ \bibinfo
  {pages} {69--116} (\bibinfo {year} {2011})},\ \Eprint
  {http://arxiv.org/abs/1107.5509} {arXiv:1107.5509 [nucl-th]} \BibitemShut
  {NoStop}%
\bibitem [{\citenamefont {Landay}\ \emph {et~al.}(2017)\citenamefont {Landay},
  \citenamefont {D\"oring}, \citenamefont {Fern\'andez-Ram\'\i{}rez},
  \citenamefont {Hu},\ and\ \citenamefont {Molina}}]{Landay:2016cjw}%
  \BibitemOpen
  \bibfield  {author} {\bibinfo {author} {\bibfnamefont {J.}~\bibnamefont
  {Landay}}, \bibinfo {author} {\bibfnamefont {M.}~\bibnamefont {D\"oring}},
  \bibinfo {author} {\bibfnamefont {C.}~\bibnamefont
  {Fern\'andez-Ram\'\i{}rez}}, \bibinfo {author} {\bibfnamefont
  {B.}~\bibnamefont {Hu}}, \ and\ \bibinfo {author} {\bibfnamefont
  {R.}~\bibnamefont {Molina}},\ }\bibfield  {title} {\enquote {\bibinfo {title}
  {{Model Selection for Pion Photoproduction}},}\ }\href {\doibase
  10.1103/PhysRevC.95.015203} {\bibfield  {journal} {\bibinfo  {journal} {Phys.
  Rev. C}\ }\textbf {\bibinfo {volume} {95}},\ \bibinfo {pages} {015203}
  (\bibinfo {year} {2017})},\ \Eprint {http://arxiv.org/abs/1610.07547}
  {arXiv:1610.07547 [nucl-th]} \BibitemShut {NoStop}%
\bibitem [{\citenamefont {Landay}\ \emph {et~al.}(2019)\citenamefont {Landay},
  \citenamefont {Mai}, \citenamefont {D\"oring}, \citenamefont {Haberzettl},\
  and\ \citenamefont {Nakayama}}]{Landay:2018wgf}%
  \BibitemOpen
  \bibfield  {author} {\bibinfo {author} {\bibfnamefont {J.}~\bibnamefont
  {Landay}}, \bibinfo {author} {\bibfnamefont {M.}~\bibnamefont {Mai}},
  \bibinfo {author} {\bibfnamefont {M.}~\bibnamefont {D\"oring}}, \bibinfo
  {author} {\bibfnamefont {H.}~\bibnamefont {Haberzettl}}, \ and\ \bibinfo
  {author} {\bibfnamefont {K.}~\bibnamefont {Nakayama}},\ }\bibfield  {title}
  {\enquote {\bibinfo {title} {{Towards the Minimal Spectrum of Excited
  Baryons}},}\ }\href {\doibase 10.1103/PhysRevD.99.016001} {\bibfield
  {journal} {\bibinfo  {journal} {Phys. Rev. D}\ }\textbf {\bibinfo {volume}
  {99}},\ \bibinfo {pages} {016001} (\bibinfo {year} {2019})},\ \Eprint
  {http://arxiv.org/abs/1810.00075} {arXiv:1810.00075 [nucl-th]} \BibitemShut
  {NoStop}%
\bibitem [{\citenamefont {D\"oring}\ \emph {et~al.}(2016)\citenamefont
  {D\"oring}, \citenamefont {Revier}, \citenamefont {R\"onchen},\ and\
  \citenamefont {Workman}}]{Doring:2016snk}%
  \BibitemOpen
  \bibfield  {author} {\bibinfo {author} {\bibfnamefont {M.}~\bibnamefont
  {D\"oring}}, \bibinfo {author} {\bibfnamefont {J.}~\bibnamefont {Revier}},
  \bibinfo {author} {\bibfnamefont {D.}~\bibnamefont {R\"onchen}}, \ and\
  \bibinfo {author} {\bibfnamefont {R.~L.}\ \bibnamefont {Workman}},\
  }\bibfield  {title} {\enquote {\bibinfo {title} {{Correlations of $\pi$N
  partial waves for multireaction analyses}},}\ }\href {\doibase
  10.1103/PhysRevC.93.065205} {\bibfield  {journal} {\bibinfo  {journal} {Phys.
  Rev. C}\ }\textbf {\bibinfo {volume} {93}},\ \bibinfo {pages} {065205}
  (\bibinfo {year} {2016})},\ \Eprint {http://arxiv.org/abs/1603.07265}
  {arXiv:1603.07265 [nucl-th]} \BibitemShut {NoStop}%
\bibitem [{\citenamefont {Arndt}\ \emph {et~al.}(1995)\citenamefont {Arndt},
  \citenamefont {Strakovsky}, \citenamefont {Workman},\ and\ \citenamefont
  {Pavan}}]{Arndt:1995bj}%
  \BibitemOpen
  \bibfield  {author} {\bibinfo {author} {\bibfnamefont {Richard~A.}\
  \bibnamefont {Arndt}}, \bibinfo {author} {\bibfnamefont {Igor~I.}\
  \bibnamefont {Strakovsky}}, \bibinfo {author} {\bibfnamefont {Ron~L.}\
  \bibnamefont {Workman}}, \ and\ \bibinfo {author} {\bibfnamefont
  {Marcello~M.}\ \bibnamefont {Pavan}},\ }\bibfield  {title} {\enquote
  {\bibinfo {title} {{Updated analysis of pi N elastic scattering data to
  2.1-GeV: The Baryon spectrum}},}\ }\href {\doibase 10.1103/PhysRevC.52.2120}
  {\bibfield  {journal} {\bibinfo  {journal} {Phys. Rev. C}\ }\textbf {\bibinfo
  {volume} {52}},\ \bibinfo {pages} {2120--2130} (\bibinfo {year} {1995})},\
  \Eprint {http://arxiv.org/abs/nucl-th/9505040} {arXiv:nucl-th/9505040}
  \BibitemShut {NoStop}%
\bibitem [{\citenamefont {Briscoe}\ \emph {et~al.}(2020)\citenamefont
  {Briscoe}, \citenamefont {Kudryavtsev}, \citenamefont {Strakovsky},
  \citenamefont {Tarasov},\ and\ \citenamefont {Workman}}]{Briscoe:2020qat}%
  \BibitemOpen
  \bibfield  {author} {\bibinfo {author} {\bibfnamefont {W.~J.}\ \bibnamefont
  {Briscoe}}, \bibinfo {author} {\bibfnamefont {A.~E.}\ \bibnamefont
  {Kudryavtsev}}, \bibinfo {author} {\bibfnamefont {I.~I.}\ \bibnamefont
  {Strakovsky}}, \bibinfo {author} {\bibfnamefont {V.~E.}\ \bibnamefont
  {Tarasov}}, \ and\ \bibinfo {author} {\bibfnamefont {R.~L.}\ \bibnamefont
  {Workman}},\ }\bibfield  {title} {\enquote {\bibinfo {title} {{Threshold $\pi
  ^-$ photoproduction on the neutron}},}\ }\href {\doibase
  10.1140/epja/s10050-020-00221-w} {\bibfield  {journal} {\bibinfo  {journal}
  {Eur. Phys. J. A}\ }\textbf {\bibinfo {volume} {56}},\ \bibinfo {pages} {218}
  (\bibinfo {year} {2020})},\ \Eprint {http://arxiv.org/abs/2004.01742}
  {arXiv:2004.01742 [nucl-th]} \BibitemShut {NoStop}%
\bibitem [{\citenamefont {Strauch}\ \emph {et~al.}(2015)\citenamefont {Strauch}
  \emph {et~al.}}]{CLAS:2015ykk}%
  \BibitemOpen
  \bibfield  {author} {\bibinfo {author} {\bibfnamefont {S.}~\bibnamefont
  {Strauch}} \emph {et~al.} (\bibinfo {collaboration} {CLAS}),\ }\bibfield
  {title} {\enquote {\bibinfo {title} {{First Measurement of the Polarization
  Observable E in the $\vec p(\vec \gamma,\pi^+)n$ Reaction up to 2.25 GeV}},}\
  }\href {\doibase 10.1016/j.physletb.2015.08.053} {\bibfield  {journal}
  {\bibinfo  {journal} {Phys. Lett. B}\ }\textbf {\bibinfo {volume} {750}},\
  \bibinfo {pages} {53--58} (\bibinfo {year} {2015})},\ \Eprint
  {http://arxiv.org/abs/1503.05163} {arXiv:1503.05163 [nucl-ex]} \BibitemShut
  {NoStop}%
\bibitem [{\citenamefont {Lawall}\ \emph {et~al.}(2005)\citenamefont {Lawall}
  \emph {et~al.}}]{Lawall:2005np}%
  \BibitemOpen
  \bibfield  {author} {\bibinfo {author} {\bibfnamefont {R.}~\bibnamefont
  {Lawall}} \emph {et~al.},\ }\bibfield  {title} {\enquote {\bibinfo {title}
  {{Measurement of the reaction $\gamma p \to K^0 \Sigma^+$ at photon energies
  up to 2.6 GeV}},}\ }\href {\doibase 10.1140/epja/i2005-10002-x} {\bibfield
  {journal} {\bibinfo  {journal} {Eur. Phys. J. A}\ }\textbf {\bibinfo {volume}
  {24}},\ \bibinfo {pages} {275--286} (\bibinfo {year} {2005})},\ \Eprint
  {http://arxiv.org/abs/nucl-ex/0504014} {arXiv:nucl-ex/0504014} \BibitemShut
  {NoStop}%
\bibitem [{\citenamefont {Sadasivan}\ \emph {et~al.}(2022)\citenamefont
  {Sadasivan}, \citenamefont {Alexandru}, \citenamefont {Akdag}, \citenamefont
  {Amorim}, \citenamefont {Brett}, \citenamefont {Culver}, \citenamefont
  {D{\"o}ring}, \citenamefont {Lee},\ and\ \citenamefont
  {Mai}}]{Sadasivan:2021emk}%
  \BibitemOpen
  \bibfield  {author} {\bibinfo {author} {\bibfnamefont {Daniel}\ \bibnamefont
  {Sadasivan}}, \bibinfo {author} {\bibfnamefont {Andrei}\ \bibnamefont
  {Alexandru}}, \bibinfo {author} {\bibfnamefont {Hakan}\ \bibnamefont
  {Akdag}}, \bibinfo {author} {\bibfnamefont {Felipe}\ \bibnamefont {Amorim}},
  \bibinfo {author} {\bibfnamefont {Ruair\'\i{}}\ \bibnamefont {Brett}},
  \bibinfo {author} {\bibfnamefont {Chris}\ \bibnamefont {Culver}}, \bibinfo
  {author} {\bibfnamefont {Michael}\ \bibnamefont {D{\"o}ring}}, \bibinfo
  {author} {\bibfnamefont {Frank~X.}\ \bibnamefont {Lee}}, \ and\ \bibinfo
  {author} {\bibfnamefont {Maxim}\ \bibnamefont {Mai}},\ }\bibfield  {title}
  {\enquote {\bibinfo {title} {{Pole position of the a1(1260) resonance in a
  three-body unitary framework}},}\ }\href {\doibase
  10.1103/PhysRevD.105.054020} {\bibfield  {journal} {\bibinfo  {journal}
  {Phys. Rev. D}\ }\textbf {\bibinfo {volume} {105}},\ \bibinfo {pages}
  {054020} (\bibinfo {year} {2022})},\ \Eprint
  {http://arxiv.org/abs/2112.03355} {arXiv:2112.03355 [hep-ph]} \BibitemShut
  {NoStop}%
\bibitem [{\citenamefont {Ceci}\ \emph {et~al.}(2011)\citenamefont {Ceci},
  \citenamefont {Döring}, \citenamefont {Hanhart}, \citenamefont {Krewald},
  \citenamefont {Mei\ss{}ner},\ and\ \citenamefont {Svarc}}]{Ceci:2011ae}%
  \BibitemOpen
  \bibfield  {author} {\bibinfo {author} {\bibfnamefont {S.}~\bibnamefont
  {Ceci}}, \bibinfo {author} {\bibfnamefont {M.}~\bibnamefont {Döring}},
  \bibinfo {author} {\bibfnamefont {C.}~\bibnamefont {Hanhart}}, \bibinfo
  {author} {\bibfnamefont {S.}~\bibnamefont {Krewald}}, \bibinfo {author}
  {\bibfnamefont {U.-G.}\ \bibnamefont {Mei\ss{}ner}}, \ and\ \bibinfo {author}
  {\bibfnamefont {A.}~\bibnamefont {Svarc}},\ }\bibfield  {title} {\enquote
  {\bibinfo {title} {{Relevance of complex branch points for partial wave
  analysis}},}\ }\href {\doibase 10.1103/PhysRevC.84.015205} {\bibfield
  {journal} {\bibinfo  {journal} {Phys. Rev. C}\ }\textbf {\bibinfo {volume}
  {84}},\ \bibinfo {pages} {015205} (\bibinfo {year} {2011})},\ \Eprint
  {http://arxiv.org/abs/1104.3490} {arXiv:1104.3490 [nucl-th]} \BibitemShut
  {NoStop}%
\bibitem [{\citenamefont {Sokhoyan}\ \emph {et~al.}(2015)\citenamefont
  {Sokhoyan} \emph {et~al.}}]{Sokhoyan:2015fra}%
  \BibitemOpen
  \bibfield  {author} {\bibinfo {author} {\bibfnamefont {V.}~\bibnamefont
  {Sokhoyan}} \emph {et~al.} (\bibinfo {collaboration} {CBELSA/TAPS}),\
  }\bibfield  {title} {\enquote {\bibinfo {title} {{High-statistics study of
  the reaction $\gamma p\to p\;2\pi^0$}},}\ }\href {\doibase
  10.1140/epja/i2015-15187-7} {\bibfield  {journal} {\bibinfo  {journal} {Eur.
  Phys. J. A}\ }\textbf {\bibinfo {volume} {51}},\ \bibinfo {pages} {95}
  (\bibinfo {year} {2015})},\ \bibinfo {note} {[Erratum: Eur.Phys.J.A 51, 187
  (2015)]},\ \Eprint {http://arxiv.org/abs/1507.02488} {arXiv:1507.02488
  [nucl-ex]} \BibitemShut {NoStop}%
\bibitem [{\citenamefont {Krehl}\ \emph {et~al.}(2000)\citenamefont {Krehl},
  \citenamefont {Hanhart}, \citenamefont {Krewald},\ and\ \citenamefont
  {Speth}}]{Krehl:1999km}%
  \BibitemOpen
  \bibfield  {author} {\bibinfo {author} {\bibfnamefont {O.}~\bibnamefont
  {Krehl}}, \bibinfo {author} {\bibfnamefont {C.}~\bibnamefont {Hanhart}},
  \bibinfo {author} {\bibfnamefont {S.}~\bibnamefont {Krewald}}, \ and\
  \bibinfo {author} {\bibfnamefont {J.}~\bibnamefont {Speth}},\ }\bibfield
  {title} {\enquote {\bibinfo {title} {{What is the structure of the Roper
  resonance?}}}\ }\href {\doibase 10.1103/PhysRevC.62.025207} {\bibfield
  {journal} {\bibinfo  {journal} {Phys. Rev. C}\ }\textbf {\bibinfo {volume}
  {62}},\ \bibinfo {pages} {025207} (\bibinfo {year} {2000})},\ \Eprint
  {http://arxiv.org/abs/nucl-th/9911080} {arXiv:nucl-th/9911080} \BibitemShut
  {NoStop}%
\bibitem [{\citenamefont {Erbe}\ \emph {et~al.}(1969)\citenamefont {Erbe} \emph
  {et~al.}}]{Aachen-Berlin-Bonn-Hamburg-Heidelberg-Muenchen:1969pjo}%
  \BibitemOpen
  \bibfield  {author} {\bibinfo {author} {\bibfnamefont {R.}~\bibnamefont
  {Erbe}} \emph {et~al.} (\bibinfo {collaboration}
  {Aachen-Berlin-Bonn-Hamburg-Heidelberg-Muenchen}),\ }\bibfield  {title}
  {\enquote {\bibinfo {title} {{Multipion and strange-particle photoproduction
  on protons at energies up to 5.8-GeV}},}\ }\href {\doibase
  10.1103/PhysRev.188.2060} {\bibfield  {journal} {\bibinfo  {journal} {Phys.
  Rev.}\ }\textbf {\bibinfo {volume} {188}},\ \bibinfo {pages} {2060--2077}
  (\bibinfo {year} {1969})}\BibitemShut {NoStop}%
\bibitem [{\citenamefont {Anisovich}\ \emph {et~al.}(2017)\citenamefont
  {Anisovich} \emph {et~al.}}]{Anisovich:2017bsk}%
  \BibitemOpen
  \bibfield  {author} {\bibinfo {author} {\bibfnamefont {A.~V.}\ \bibnamefont
  {Anisovich}} \emph {et~al.},\ }\bibfield  {title} {\enquote {\bibinfo {title}
  {{Strong evidence for nucleon resonances near 1900 MeV}},}\ }\href {\doibase
  10.1103/PhysRevLett.119.062004} {\bibfield  {journal} {\bibinfo  {journal}
  {Phys. Rev. Lett.}\ }\textbf {\bibinfo {volume} {119}},\ \bibinfo {pages}
  {062004} (\bibinfo {year} {2017})},\ \Eprint
  {http://arxiv.org/abs/1712.07549} {arXiv:1712.07549 [nucl-ex]} \BibitemShut
  {NoStop}%
\bibitem [{\citenamefont {Ramos}\ and\ \citenamefont
  {Oset}(2013)}]{Ramos:2013wua}%
  \BibitemOpen
  \bibfield  {author} {\bibinfo {author} {\bibfnamefont {A.}~\bibnamefont
  {Ramos}}\ and\ \bibinfo {author} {\bibfnamefont {E.}~\bibnamefont {Oset}},\
  }\bibfield  {title} {\enquote {\bibinfo {title} {{The role of vector-baryon
  channels and resonances in the $\gamma p \to K^0 \Sigma^+$ and $\gamma n \to
  K^0 \Sigma^0$ reactions near the $K^* \Lambda$ threshold}},}\ }\href
  {\doibase 10.1016/j.physletb.2013.10.012} {\bibfield  {journal} {\bibinfo
  {journal} {Phys. Lett. B}\ }\textbf {\bibinfo {volume} {727}},\ \bibinfo
  {pages} {287--292} (\bibinfo {year} {2013})},\ \Eprint
  {http://arxiv.org/abs/1304.7975} {arXiv:1304.7975 [nucl-th]} \BibitemShut
  {NoStop}%
\bibitem [{\citenamefont {D{\"o}ring}\ and\ \citenamefont
  {Nakayama}(2010)}]{Doring:2009qr}%
  \BibitemOpen
  \bibfield  {author} {\bibinfo {author} {\bibfnamefont {M.}~\bibnamefont
  {D{\"o}ring}}\ and\ \bibinfo {author} {\bibfnamefont {K.}~\bibnamefont
  {Nakayama}},\ }\bibfield  {title} {\enquote {\bibinfo {title} {{On the cross
  section ratio $\sigma_n/\sigma_p$ in $\eta$ photoproduction}},}\ }\href
  {\doibase 10.1016/j.physletb.2009.12.029} {\bibfield  {journal} {\bibinfo
  {journal} {Phys. Lett. B}\ }\textbf {\bibinfo {volume} {683}},\ \bibinfo
  {pages} {145--149} (\bibinfo {year} {2010})},\ \Eprint
  {http://arxiv.org/abs/0909.3538} {arXiv:0909.3538 [nucl-th]} \BibitemShut
  {NoStop}%
\bibitem [{\citenamefont {Kashevarov}\ \emph {et~al.}(2017)\citenamefont
  {Kashevarov} \emph {et~al.}}]{A2:2017gwp}%
  \BibitemOpen
  \bibfield  {author} {\bibinfo {author} {\bibfnamefont {V.~L.}\ \bibnamefont
  {Kashevarov}} \emph {et~al.} (\bibinfo {collaboration} {A2}),\ }\bibfield
  {title} {\enquote {\bibinfo {title} {{Study of \ensuremath{\eta} and
  \ensuremath{\eta}' Photoproduction at MAMI}},}\ }\href {\doibase
  10.1103/PhysRevLett.118.212001} {\bibfield  {journal} {\bibinfo  {journal}
  {Phys. Rev. Lett.}\ }\textbf {\bibinfo {volume} {118}},\ \bibinfo {pages}
  {212001} (\bibinfo {year} {2017})},\ \Eprint
  {http://arxiv.org/abs/1701.04809} {arXiv:1701.04809 [nucl-ex]} \BibitemShut
  {NoStop}%
\bibitem [{\citenamefont {Collins}\ \emph {et~al.}(2017)\citenamefont {Collins}
  \emph {et~al.}}]{CLAS:2017rxe}%
  \BibitemOpen
  \bibfield  {author} {\bibinfo {author} {\bibfnamefont {P.}~\bibnamefont
  {Collins}} \emph {et~al.} (\bibinfo {collaboration} {CLAS}),\ }\bibfield
  {title} {\enquote {\bibinfo {title} {{Photon beam asymmetry $\Sigma$ for
  $\eta$ and $\eta^\prime$ photoproduction from the proton}},}\ }\href
  {\doibase 10.1016/j.physletb.2017.05.045} {\bibfield  {journal} {\bibinfo
  {journal} {Phys. Lett. B}\ }\textbf {\bibinfo {volume} {771}},\ \bibinfo
  {pages} {213--221} (\bibinfo {year} {2017})},\ \Eprint
  {http://arxiv.org/abs/1703.00433} {arXiv:1703.00433 [nucl-ex]} \BibitemShut
  {NoStop}%
\bibitem [{\citenamefont {Bartalini}\ \emph {et~al.}(2007)\citenamefont
  {Bartalini} \emph {et~al.}}]{GRAAL:2007gsc}%
  \BibitemOpen
  \bibfield  {author} {\bibinfo {author} {\bibfnamefont {O.}~\bibnamefont
  {Bartalini}} \emph {et~al.} (\bibinfo {collaboration} {GRAAL}),\ }\bibfield
  {title} {\enquote {\bibinfo {title} {{Measurement of eta photoproduction on
  the proton from threshold to 1500~MeV}},}\ }\href {\doibase
  10.1140/epja/i2007-10439-9} {\bibfield  {journal} {\bibinfo  {journal} {Eur.
  Phys. J. A}\ }\textbf {\bibinfo {volume} {33}},\ \bibinfo {pages} {169--184}
  (\bibinfo {year} {2007})},\ \Eprint {http://arxiv.org/abs/0707.1385}
  {arXiv:0707.1385 [nucl-ex]} \BibitemShut {NoStop}%
\bibitem [{\citenamefont {Adlarson}\ \emph {et~al.}(2015)\citenamefont
  {Adlarson} \emph {et~al.}}]{A2:2015mhs}%
  \BibitemOpen
  \bibfield  {author} {\bibinfo {author} {\bibfnamefont {P.}~\bibnamefont
  {Adlarson}} \emph {et~al.} (\bibinfo {collaboration} {A2}),\ }\bibfield
  {title} {\enquote {\bibinfo {title} {{Measurement of $\pi^0$ photoproduction
  on the proton at MAMI C}},}\ }\href {\doibase 10.1103/PhysRevC.92.024617}
  {\bibfield  {journal} {\bibinfo  {journal} {Phys. Rev. C}\ }\textbf {\bibinfo
  {volume} {92}},\ \bibinfo {pages} {024617} (\bibinfo {year} {2015})},\
  \Eprint {http://arxiv.org/abs/1506.08849} {arXiv:1506.08849 [hep-ex]}
  \BibitemShut {NoStop}%
\bibitem [{\citenamefont {Gottschall}\ \emph {et~al.}(2021)\citenamefont
  {Gottschall} \emph {et~al.}}]{CBELSATAPS:2019hhr}%
  \BibitemOpen
  \bibfield  {author} {\bibinfo {author} {\bibfnamefont {M.}~\bibnamefont
  {Gottschall}} \emph {et~al.} (\bibinfo {collaboration} {CBELSA/TAPS}),\
  }\bibfield  {title} {\enquote {\bibinfo {title} {{Measurement of the helicity
  asymmetry $E$ for the reaction $ \gamma p\rightarrow \pi ^0 p$}},}\ }\href
  {\doibase 10.1140/epja/s10050-020-00334-2} {\bibfield  {journal} {\bibinfo
  {journal} {Eur. Phys. J. A}\ }\textbf {\bibinfo {volume} {57}},\ \bibinfo
  {pages} {40} (\bibinfo {year} {2021})},\ \Eprint
  {http://arxiv.org/abs/1904.12560} {arXiv:1904.12560 [nucl-ex]} \BibitemShut
  {NoStop}%
\end{thebibliography}%

\end{document}